\documentclass[]{aa}
\usepackage{amsmath}
\usepackage{txfonts} 
\usepackage{subfigure}
\usepackage{color}
\usepackage{bm}
\usepackage{natbib}
\usepackage{url}
\usepackage{tabularx}
\usepackage{graphicx}
\usepackage{txfonts}
\usepackage{longtable}
\usepackage{hhline}
\usepackage{arydshln}
\usepackage{multirow}

\newcommand{\lsun}{$L_\odot$}
\newcommand{\msun}{$M_\odot$}

\newcommand{\mic}{$\mu$m}

\newcolumntype{R}[1]{>{\raggedleft\arraybackslash }b{#1}}
\newcolumntype{L}[1]{>{\raggedright\arraybackslash }b{#1}}
\newcolumntype{C}[1]{>{\centering\arraybackslash }b{#1}}

\newlength{\pointwidth}
\begin{document}

\title{Low dust emissivities and radial variations in the envelopes 
of Class 0 protostars: Possible signature of early grain growth}

  \author{M. Galametz 
               \inst{1},
              A. J. Maury
               \inst{1,2},
              V. Valdivia
               \inst{1},
              L. Testi
               \inst{3,4},
              A. Belloche
               \inst{5},
              Ph. Andr\'{e}
               \inst{1}
}

\institute{
AIM, CEA, CNRS, Universit\'{e} Paris-Saclay, Universit\'{e} Paris Diderot, Sorbonne Paris Cit\'{e}, F-91191 Gif-sur-Yvette, France\\
\email{maud.galametz@cea.fr} 
\and
Harvard-Smithsonian Center for Astrophysics, Cambridge, MA 02138, USA
\and
ESO, Karl Schwarzschild Strasse 2, 85748 Garching bei M\"{u}nchen, Germany
\and
INAF ? Osservatorio Astrofisico di Arcetri, Largo E. Fermi 5, 50125 Firenze, Italy
\and
Max-Planck-Institut f\"{u}r Radioastronomie, Auf dem H\"{u}gel 69, 53121 Bonn, Germany
}

\abstract
{Analyzing the properties of dust and its evolution in the early phases of star formation is crucial to put constraints on the collapse and accretion processes as well as on the pristine properties of planet-forming seeds.}
{In this paper, we aim to investigate the variations of the dust grain size in the envelopes of the youngest protostars.}
{We analyzed Plateau de Bure interferometric observations at 1.3 mm and 3.2 mm for 12 Class 0 protostars obtained as part of the CALYPSO survey. We performed our analysis in the visibility domain and derived dust emissivity index ($\beta_\mathrm{1-3mm}$) profiles as a function of the envelope radius at 200-2000 au scales. }
{Most of the protostellar envelopes show low dust emissivity indices decreasing toward the central regions. The decreasing trend remains after correction of the (potentially optically thick) central region emission, with surprisingly low $\beta_\mathrm{1-3mm} <$ 1 values across most of the envelope radii of NGC1333-IRAS4A, NGC1333-IRAS4B, SVS13B, and Serpens-SMM4.}
{We discuss the various processes that could explain such low and varying dust emissivity indices at envelope radii 200-2000 au. Our observations of extremely low dust emissivity indices could trace the presence of large (millimeter-size) grains in Class 0 envelopes, in which case our results would point to a radial increase of the dust grain size toward the inner envelope regions. While it is expected that large grains in young protostellar envelopes could be built via grain growth and coagulation, we stress that the typical timescales required to build millimeter grains in current coagulation models are at odds with the youth of our Class 0 protostars. Additional variations in the dust composition could also partly contribute to the low $\beta_\mathrm{1-3mm}$ we observe. We find that the steepness of the $\beta_\mathrm{1-3mm}$ radial gradient depends strongly on the envelope mass, which might favor a scenario in which large grains are built in high-density protostellar disks and transported to the intermediate envelope radii, for example with the help of outflows and winds.
}

\keywords{Stars: protostars, formation, circumstellar matter -- ISM: dust -- Techniques: interferometric -- Radio continuum: ISM}

\authorrunning{M. Galametz et al}
\titlerunning{Low dust emissivities and radial variations in Class 0 protostellar envelopes.}
\maketitle

\begin{table*}
\centering
\caption{Properties of the sample}
\begin{tabular}{ccccccccc}
\hline
\vspace{-5pt}
&\\
Name & $\alpha$$_{2000}$$^a$  &  $\delta$$_{2000}$ & Cloud      & Dist. $^b$    & Outflow P.A. $^c$  
& $T_\mathrm{bol}$ $^d$ & $L_\mathrm{int}$ $^e$ & $M_\mathrm{env}$ $^f$ \\ 
\vspace{-5pt}
& \\
& [h:m:s] & [$^{\circ}$:\arcmin:\arcsec] & & (pc) & ($^{\circ}$) & (K) & (\lsun) & (\msun)                       \\ 
\vspace{-5pt}
&\\
\hline
\vspace{-5pt}
&\\
NGC1333-IRAS2A1 & 03:28:55.57 & 31:14:37.07             & Perseus               & 293             &  +205         &       41 & 47     & 7.9       \\
NGC1333-IRAS4A1 & 03:29:10.54 & 31:13:30.98             & Perseus               & 293             &  +180         &       29 & 4.7    & 12.2      \\
NGC1333-IRAS4B  & 03:29:12.02 & 31:13:08.02             & Perseus               & 293             &  +167         &       28 & 2.3    & 4.7       \\
L1448-2A                        & 03:25:22.40 & 30:45:13.26             & Perseus                 & 293           &  -63          &       43 & 4.7    \\
L1448-C                 & 03:25:38.87 & 30:44:05.33             & Perseus                 & 293           &  -17          &       47 & 11          & 1.9   \\
L1448-NB1               & 03:25:36.38 & 30:45:14.77             & Perseus                 & 293           &  -80          &       57 & 3.9    & 4.9         \\
SVS13B                  & 03:29:03.08 & 31:15:51.74             & Perseus                 & 293           &  +167         &       20 & 3.1    & 2.8         \\
L1527                   & 04:39:53.87 & 26:03:09.66             & Taurus                         & 140           &  +60          &       59 & 0.9    & 1.2        \\
Serpens Main S68N       & 18:29:48.09 & 01:16:43.41             & Serp. Main                 & 436 $^g$      &  -45          &       58 & 11     & 11                \\
Serpens Main SMM4       & 18:29:56.72 & 01:13:15.65             & Serp. Main            & 436 $^g$        &  +30          &       26 & 2.2    & 7.7       \\
Serpens South MM18& 18:30:04.12 & -02:03:02.55          & Serp. South           & 350 $^g$        &  +188         &       35 & 29          & 5.4  \\
L1157                   & 20:39:06.27 & 68:02:15.70             & Cepheus                 & 352           &  163  &       42 & 4.0    & 3.0       \\
\vspace{-5pt}
&\\
\hline
\\
\end{tabular}
\begin{list}{}{}
\vspace{-10pt}
\item[$^a$] Peak position of CALYPSO 1.3 mm continuum emission as derived by \cite{Maury2019}.
\item[$^b$] Distance to the hosting cloud. References: \citet{Ortiz-Leon2018} for the Perseus sources, \citet{Ortiz-Leon2018_2} for the Serpens Main Cloud, \citet{Torres2009} for L1527, and \citet{Zucker2019} for L1157. 
\item[$^c$] Position angle of the outflow derived from the CALYPSO molecular line emission maps. References: Podio \& CALYPSO, in prep and \citet{Maury2019}.
\item[$^d$] Bolometric temperature. References: \citet{Motte2001} for IRAS2A1, L1527, and L1157, \citet{Enoch2011} for SerpM-S68N; \citet{Kristensen2012} for SerpM-SMM4; \citet{Maury2011} for SerpS-MM18; \citet{Tobin2016} for SVS13B; and from \citet{Sadavoy2014} for the rest of the sources.
\item[$^e$] Internal luminosity. The $L_\mathrm{int}$ were calculated from the analysis of the {\it Herschel} maps from the Herschel Gould Belt survey \citep[][; Ladjelate in prep]{Andre2010} and rescaled to the distances used in this analysis.
\item[$^f$] Envelope mass. References: \citet{Enoch2009} for L1448-2A, \citet{Chini1997} for SVS13B, \citet{Karska2013} for IRAS2A1, \citet{Motte2001} for L1527 and L1157, \citet{Kaas2004} for the Serpens Main cloud sources, \citet{Maury2011} for the Serpens South source, and \citet{Sadavoy2014} for the rest of the sample. The $M_\mathrm{env}$ were rescaled to the distances used in this analysis.
\item[$^g$] A mean distance of 436 pc for the Serpens Cloud has been determined by \citet{Ortiz-Leon2018_2} using the Gaia-DR2 results and seems to correspond to the average distance determined from the Main cloud or the star-forming region W40. Ongoing work reanalyzing the Gaia data toward Serpens South seems, however, to suggest a first layer of extinction around 300-350pc while its associated young stellar objects could be distributed in a larger extinction layer up to distances of 350 pc (Palmeirim, Andr\'{e} et al. in prep). We use this new distance for the paper. As we are working on flux ratios, this does not affect the slope we present for SerpS-MM18. This, however, has an impact on the envelope mass discussed in Sect.~6, which would be higher by a factor of 1.6 if the distance is 436 pc.
\end{list}
\label{SourceCharacteristics}
\end{table*}


\section{Introduction} 

Investigating the properties of dust in the early phases of star formation is essential to understand the physico-chemistry taking place in the collapsing core as well as to trace the pristine properties of planet-forming material building up the protoplanetary disks. From star-forming cores down to disk scales, the collisions and interactions between grain particles increase and the physical conditions and chemistry evolve, affecting the grain population, their progressive growth, and reprocessing. 
Deriving the dust properties from observations is not trivial. Submillimeter and millimeter wavelength observations of the dust thermal emission have been extensively used to put constraints on the dust opacity. In this regime indeed, the opacity has a strong frequency dependence $\kappa_{\nu}$ = $\kappa_0$ $(\nu/\nu_0)^{\beta}$, with $\beta$ the so-called dust emissivity index. 
Analyzing this dust emissivity index and its variations can help us to probe the grain size distribution \citep{Draine2006,Natta2007,Testi2014}. Works based on PRONAOS, \textit{Herschel}, the Balloon-borne Large Aperture Sub-millimetre Telescope (BLAST), or James Clerk Maxwell Telescope (JCMT) measurements have revealed that star-forming clumps present significant variations in their grain emissivity \citep{Chini1997,Dupac2003,Martin2012}. No evidence of major grain growth was discovered toward star-forming filaments such as OMC 2/3 \citep{Sadavoy2016} or the center of pre-stellar cores such as L1544 \citep{Chacon-Tanarro2017,Chacon-Tanarro2019}, at least at the angular resolution at which these studies were performed. In some Perseus star-forming clumps, however, \citet{Chen2016} found dust emissivity indices as low as 1.0, while these clumps have values of about 1.6 in the diffuse  (ISM) \citep{Planck2014,Juvela2015}. These results are in line with models predicting that grain coagulation and the formation of aggregates take place in dense molecular clouds \citep{Ossenkopf1994} and that the submillimeter and millimeter emissivity increasing with fluffiness \citep{Bazell1990,Ossenkopf1993}. Grain growth processes inside molecular clouds are also required to explain the `coreshine' phenomenon, a scattering process at 3.6 \mic\  and 4.5 \mic\ that dominates absorption \citep{Pagani2010,Steinacker2010,Andersen2013,Lefevre2014,Steinacker2015} and that was first discovered using {\it Spitzer} observations. 

At the other end of the star formation timeline, observations of pre-main-sequence objects (in particular T Tauri stars) with circumstellar disks reveal $\beta$ values ranging from 2 down to -1 \citep{BeckwithSargent1991}; these values are compatible with the presence of millimeter \citep{Perez2012,Perez2015,Tazzari2016} and even centimeter grain sizes (\citet{Testi2003}; see also models from \citet{DAlessio2001}). How and when dust grains evolve during the building of the star and disk system, i.e., during the protostellar phase, is less clear. Several analyses, mostly on individual objects, have tried to address these questions. While \citet{Agurto2019} have found no clear sign of grains larger than 100 \mic\ in the inner envelope of the Class I protostar Per-emb-50, \citet{Jorgensen2007} showed that the inner envelopes of deeply embedded protostars could present dust emissivity indices of $\sim$1 that may be indicative of grain growth in these young objects. No evidence seems to be found yet of an evolution of the grain growth from the inner regions of the envelopes to the planet-forming disks \citep[e.g.,][]{Ricci2010,Testi2014}.

Several studies have suggested that grain growth could occur even earlier than the Class I phase, i.e., at the Class 0 phase. Class 0 objects are the youngest protostars: most of their mass resides in an extended envelope (with outer radii as large as 10$^4$ au) that is being accreted onto a protostellar embryo \citep{Andre1993,Andre2000} during very short timescales \citep[t $\le$ 5$\times$10$^4$ yr;][]{Maury2011}. \citet{Kwon2009}, \citet{Chiang2012}, and \citet{Li2017} showed that a dozen of Class 0 protostellar envelopes present $\beta$ at millimeter wavelengths lower than 1.7, and for some objects even lower than 1. $\beta$ values of 1 have been obtained at 1.3 mm from models using grain growth alone \citep{Miyake1993,Ossenkopf1994,Draine2006}. Many questions remain, however, about whether  $\beta$ values lower than 1 are realistic or about how large (millimeter-size) grains can form in Class 0 envelopes despite the long timescales and high densities required to build them. Current radiative transfer models of young protostellar envelopes and disks already have tried to incorporate variations in the grain composition (e.g., ices) and size distribution \citep[][]{Whitney2003,Robitaille2006,Facchini2017}. More observations are, however, needed to put constraints on the dust properties within the protostellar envelopes during the disk-forming stage, that is the Class 0 stage, and to identify the environmental parameters (protostar mass, temperature, and hosting cloud conditions) leading to these variations to probe the pristine material building up the planet-forming disks and build a coherent end-to-end scenario of the grain evolution. 

Class 0 protostars are ideal candidates to investigate the potential grain growth processes taking place during the main accretion phase and establish the pristine dust properties from which the future disk originates. In this work, we aim to perform a resolved and multifrequency millimeter analysis of the variations of the dust properties in a sample of 12 Class 0 protostellar envelopes observed as part of the Continuum And Lines in Young ProtoStellar Objects survey (CALYPSO; PI: Ph. Andr\'{e}; \url{http://irfu.cea.fr/Projets/Calypso/}).
Interferometric observations at millimeter wavelengths are ideal to perform this analysis: most of the envelope/disk emission is optically thin in this regime, limiting the biases in accessing the dust opacity and studying its spatial variations.
We provide details on the sample and data reduction in Sect.~2 and analyze the 1 to 3 mm $\beta$ radial profiles in Sect.~3. We correct our $\beta_\mathrm{1-3mm}$ profiles for the central regions in Sect.~4 and describe the implications of these corrections for our results. We analyze the dust emissivity values and radial gradients and their dependence on the global source properties in Sect.~5. Our results are summarized in Sect.~6.

\begin{figure*}
\begin{tabular}{m{5.6cm}m{5.6cm}m{5.6cm}}
\hspace{5pt} \vspace{-11pt} \includegraphics[width=5.8cm]{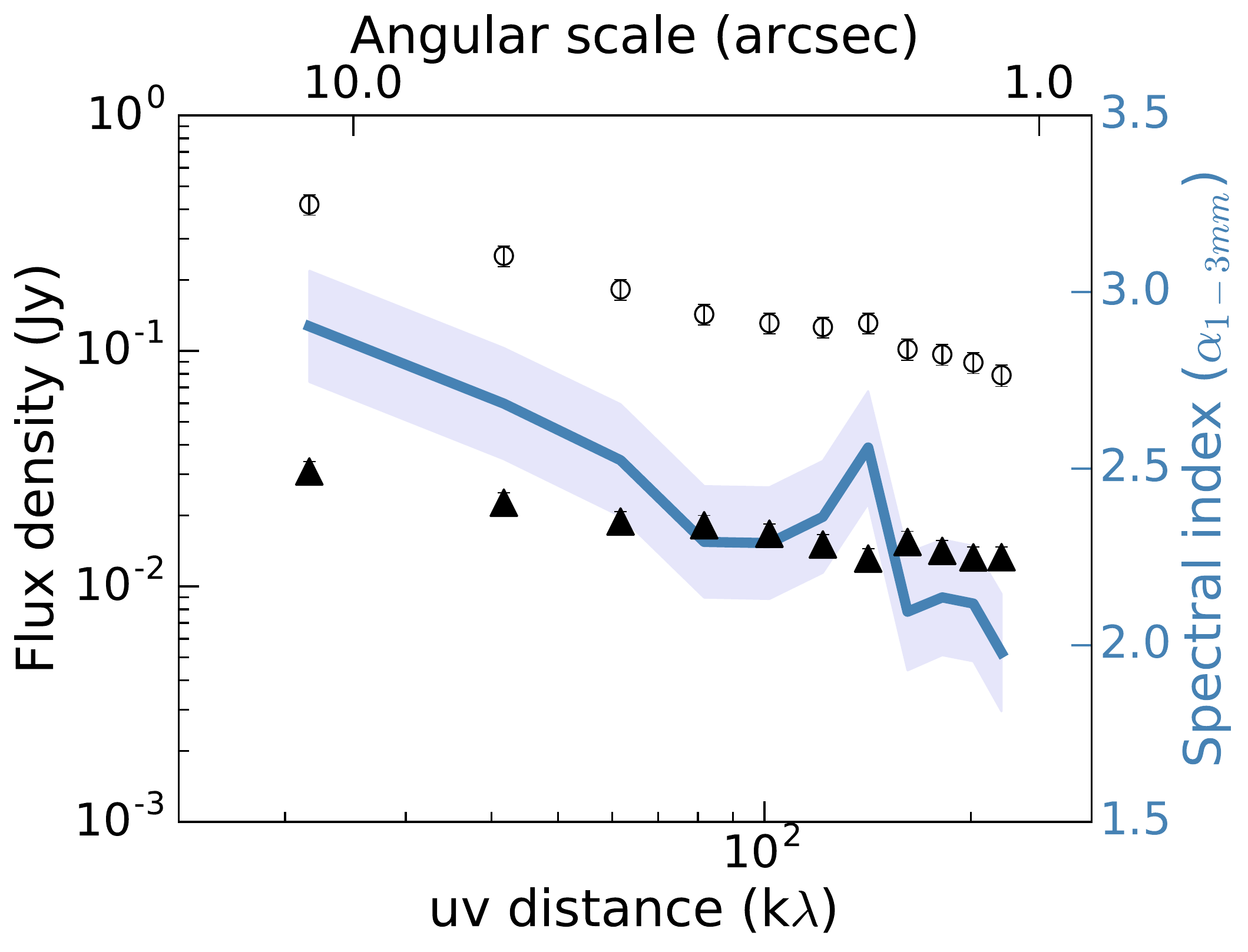} &
\hspace{3pt} \vspace{-11pt} \includegraphics[width=5.8cm]{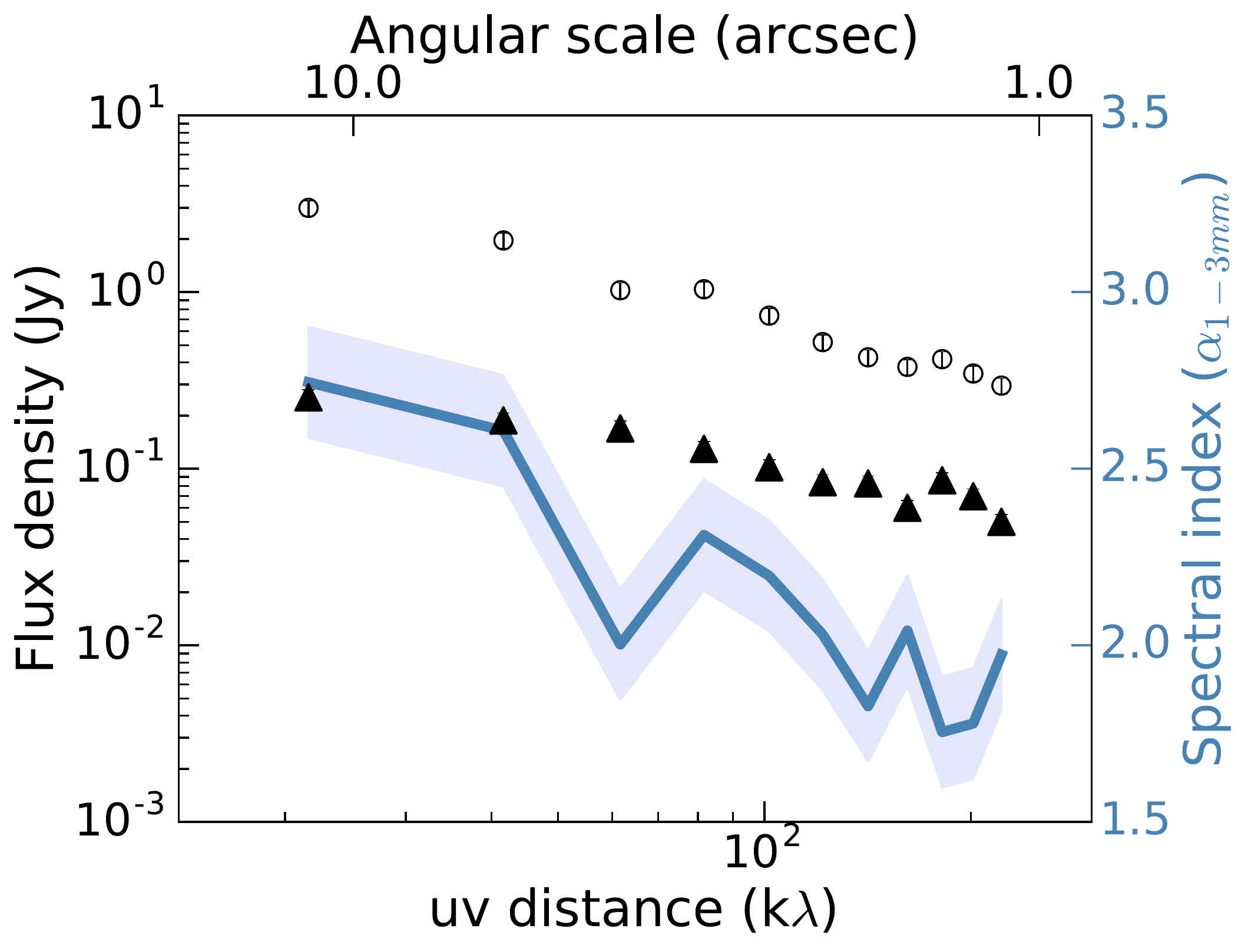} &
\hspace{3pt} \vspace{-11pt} \includegraphics[width=5.8cm]{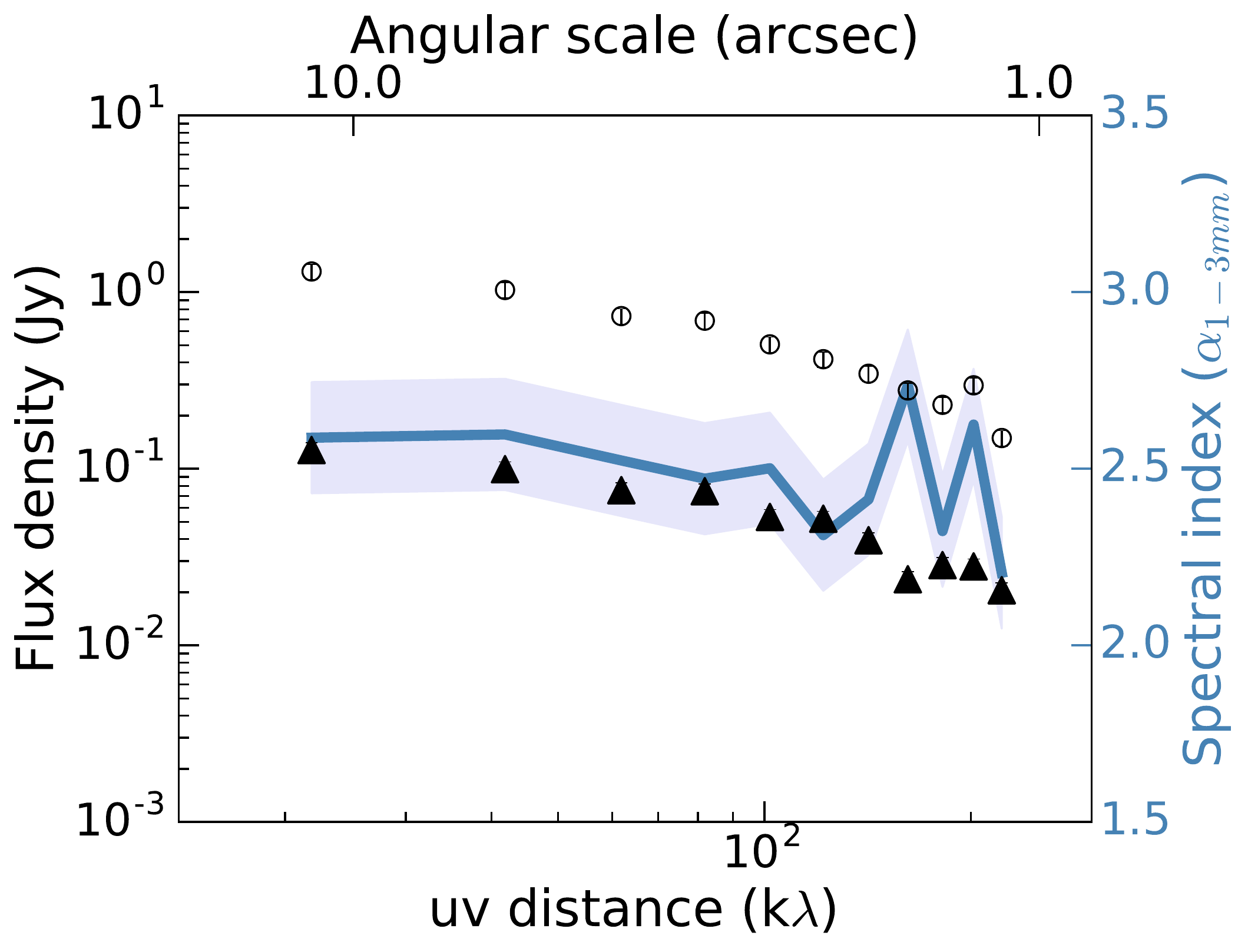} \\
\vspace{-50pt}\hspace{35pt}{\bf IRAS2A1} & 
\vspace{-50pt}\hspace{35pt}{\bf IRAS4A1} & 
\vspace{-50pt}\hspace{35pt}{\bf IRAS4B} \\
\hspace{5pt}  \vspace{-11pt} \includegraphics[width=5.8cm]{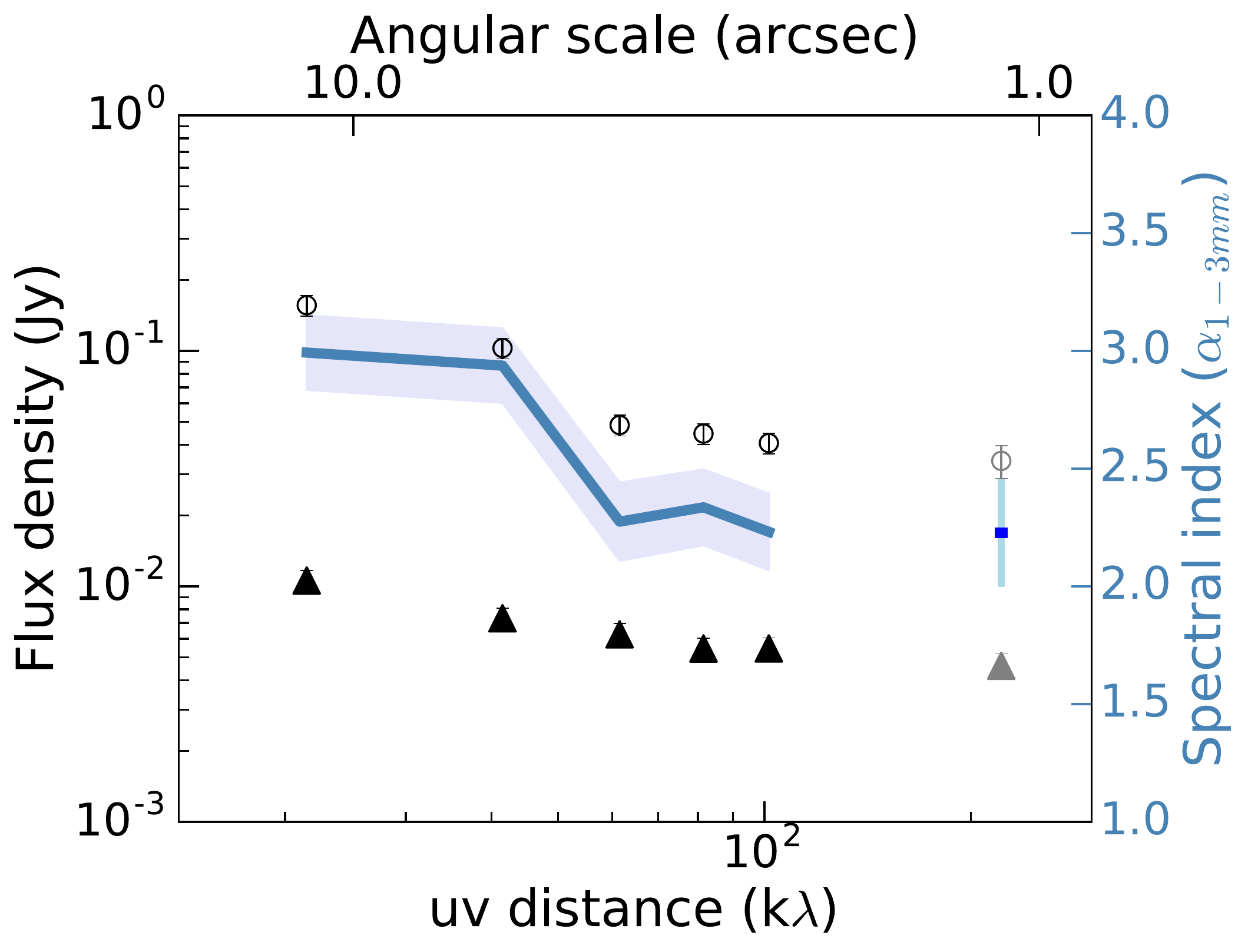} &
\hspace{3pt}  \vspace{-11pt} \includegraphics[width=5.8cm]{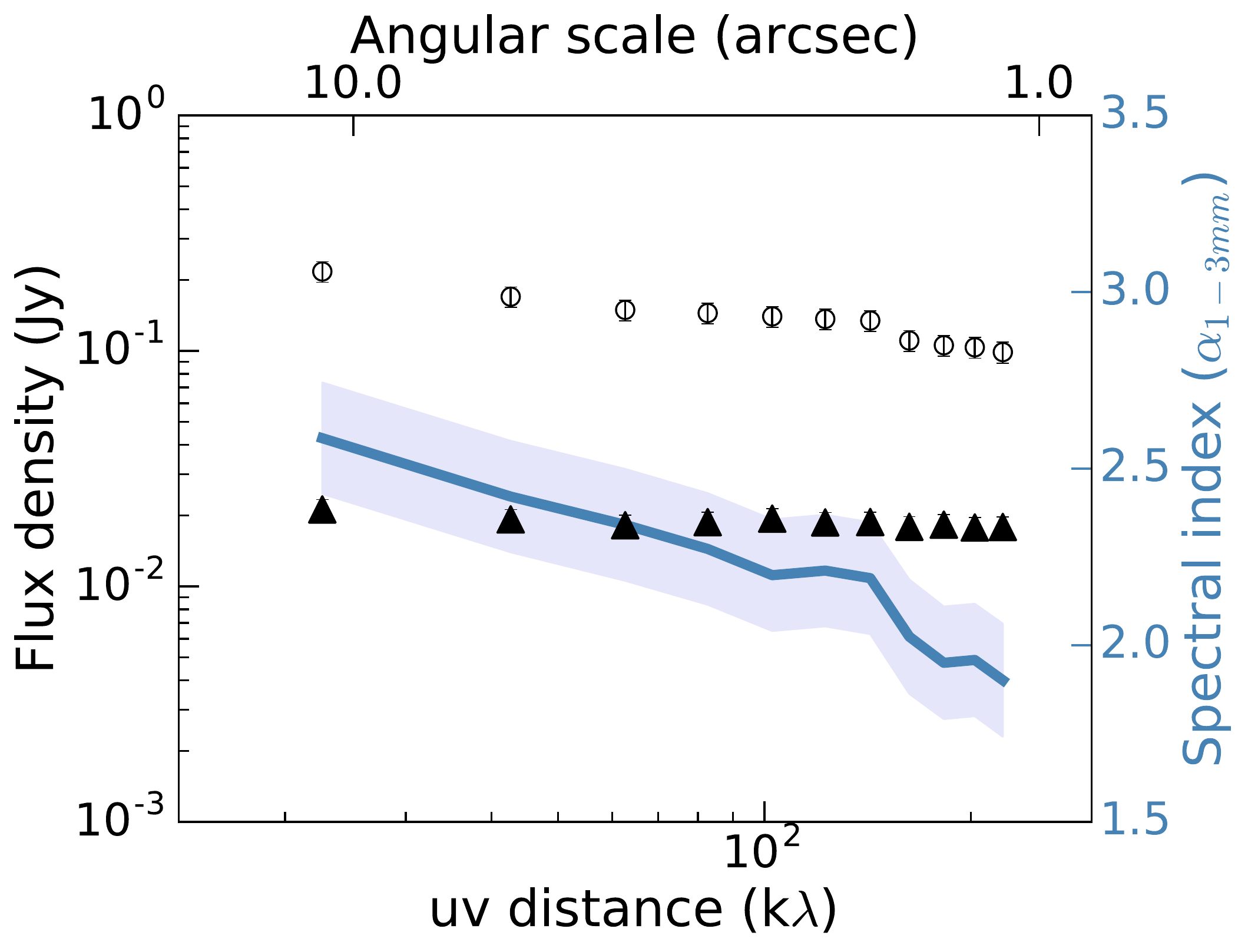} &
\hspace{3pt}  \vspace{-11pt} \includegraphics[width=5.8cm]{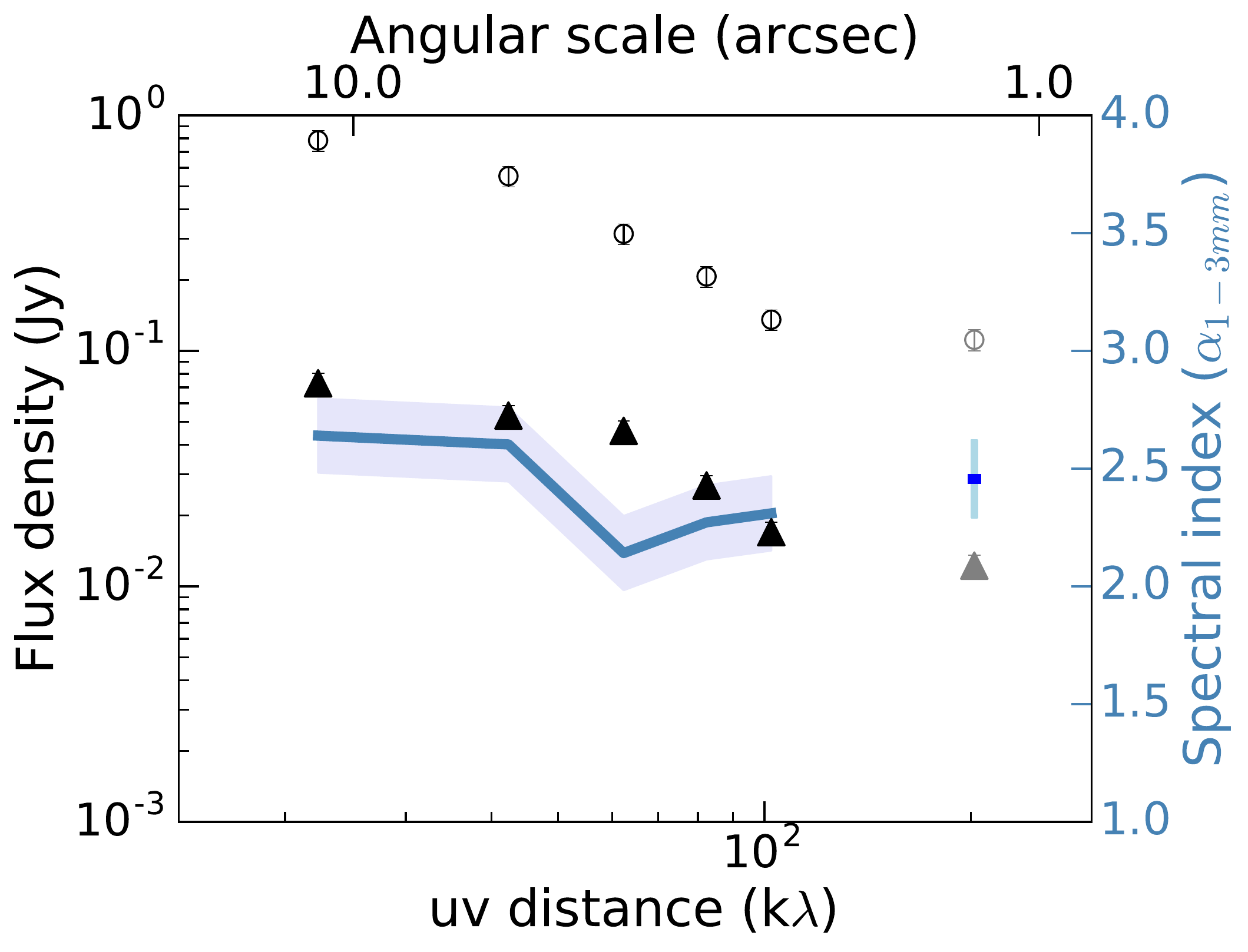} \\
\vspace{-50pt}\hspace{35pt}{\bf L1448-2A} & 
\vspace{-50pt}\hspace{35pt}{\bf L1448-C} & 
\vspace{-50pt}\hspace{35pt}{\bf L1448-NB1} \\
\hspace{5pt} \vspace{-11pt} \includegraphics[width=5.8cm]{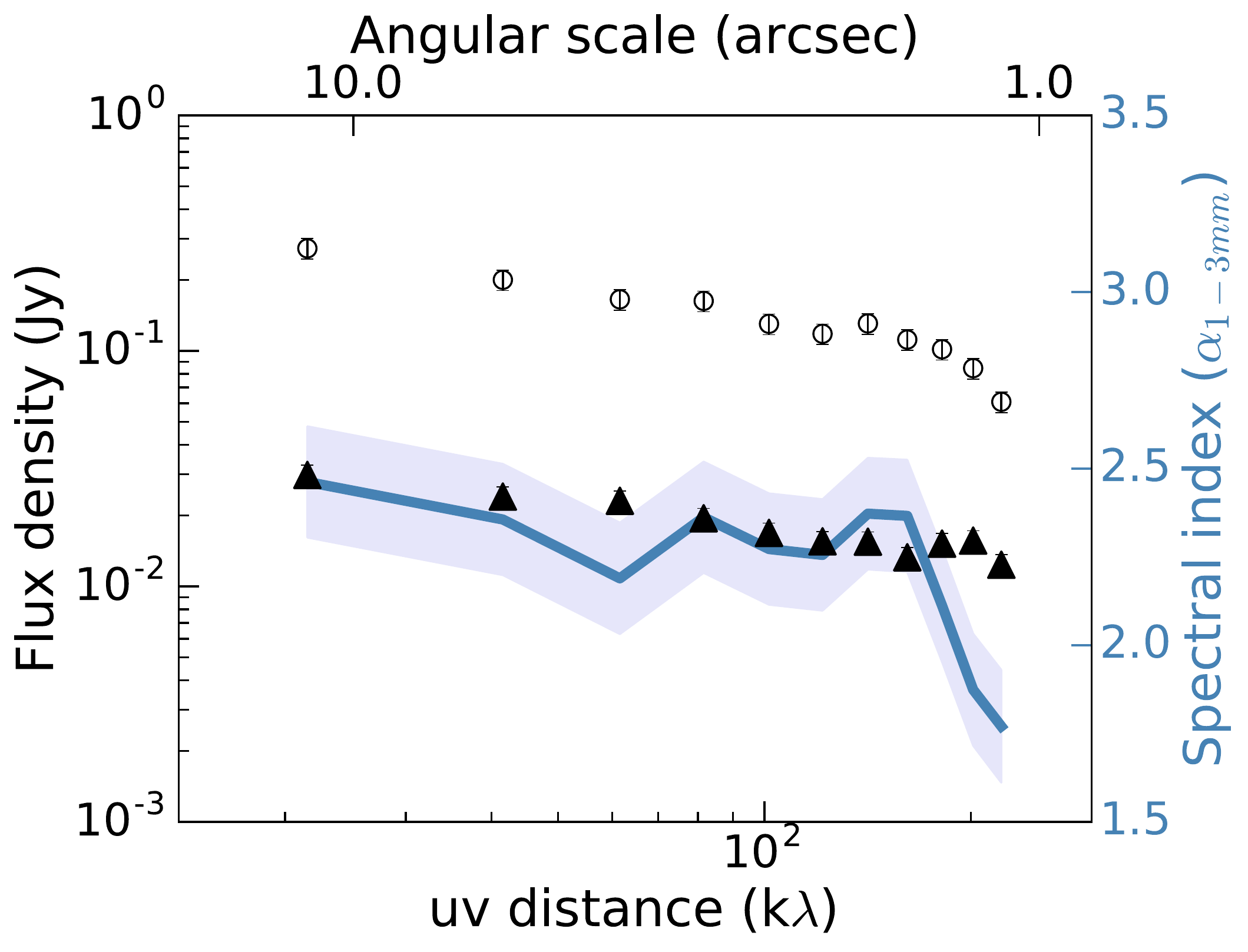} &
\hspace{3pt} \vspace{-11pt} \includegraphics[width=5.8cm]{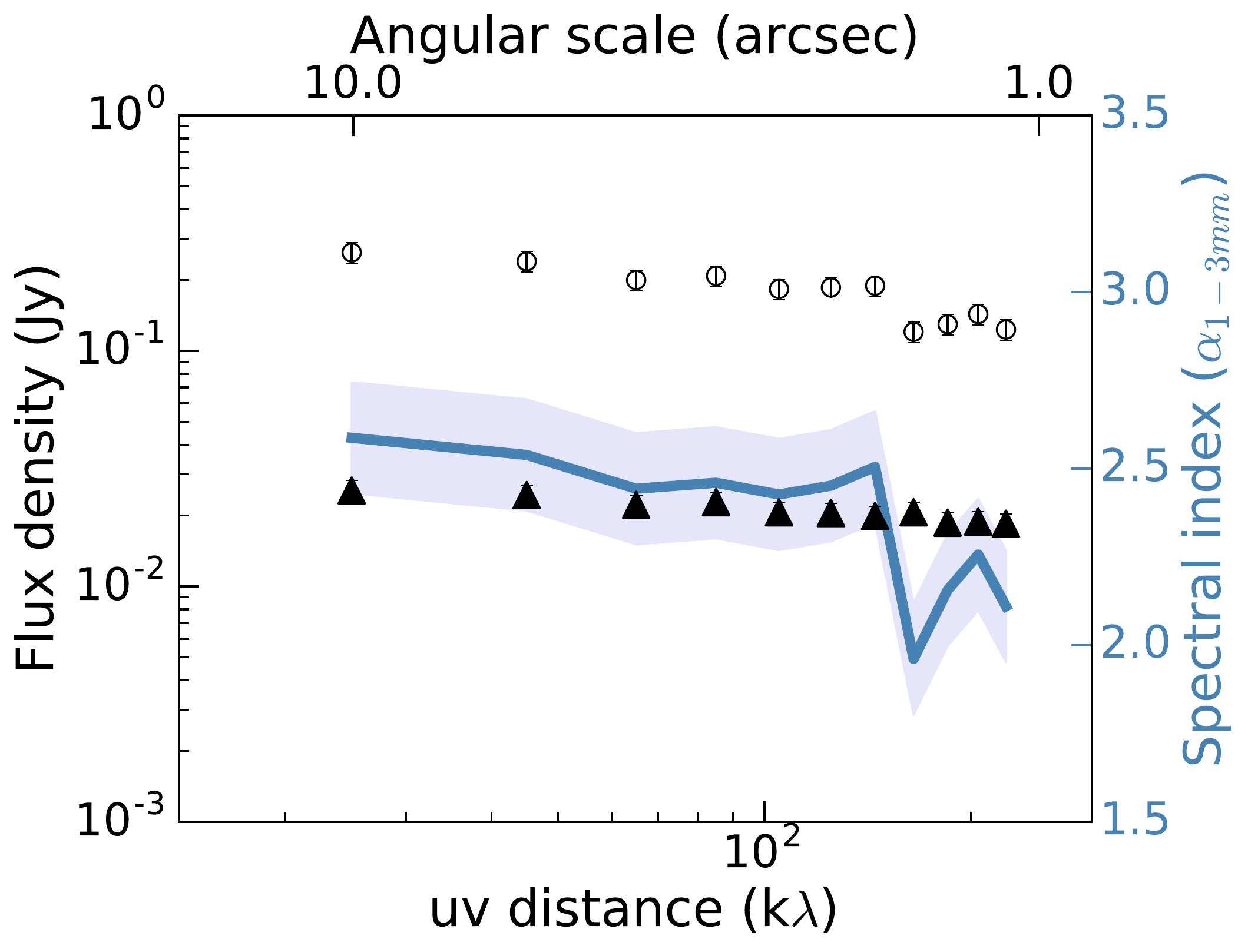} &
\hspace{3pt} \vspace{-11pt} \includegraphics[width=5.8cm]{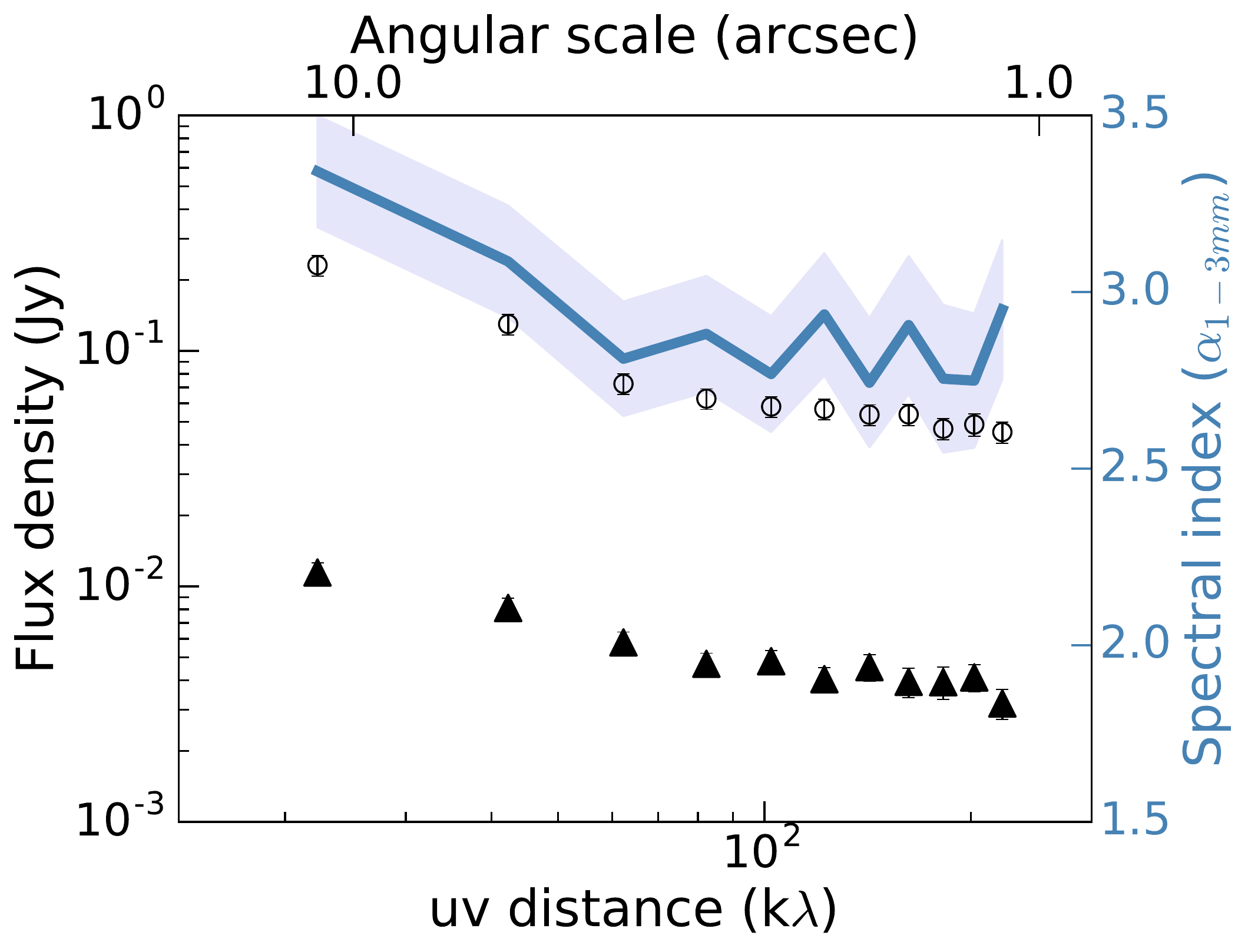} \\
\vspace{-50pt}\hspace{35pt}{\bf SVS13B} & 
\vspace{-50pt}\hspace{35pt}{\bf L1527} & 
\vspace{-50pt}\hspace{35pt}{\bf SerpM-S68N} \\
\hspace{5pt} \vspace{-11pt} \includegraphics[width=5.8cm]{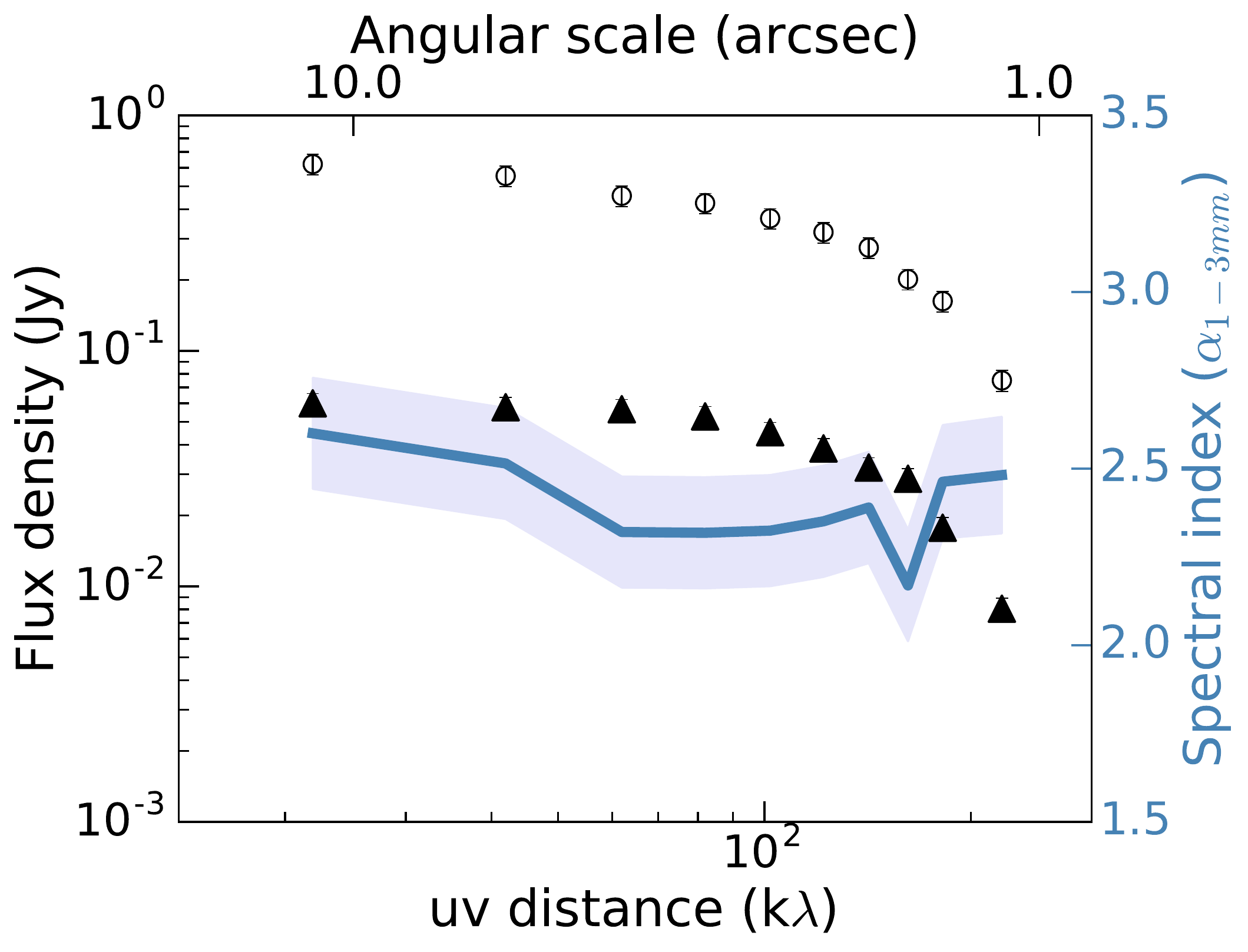} &
\hspace{3pt} \vspace{-11pt} \includegraphics[width=5.8cm]{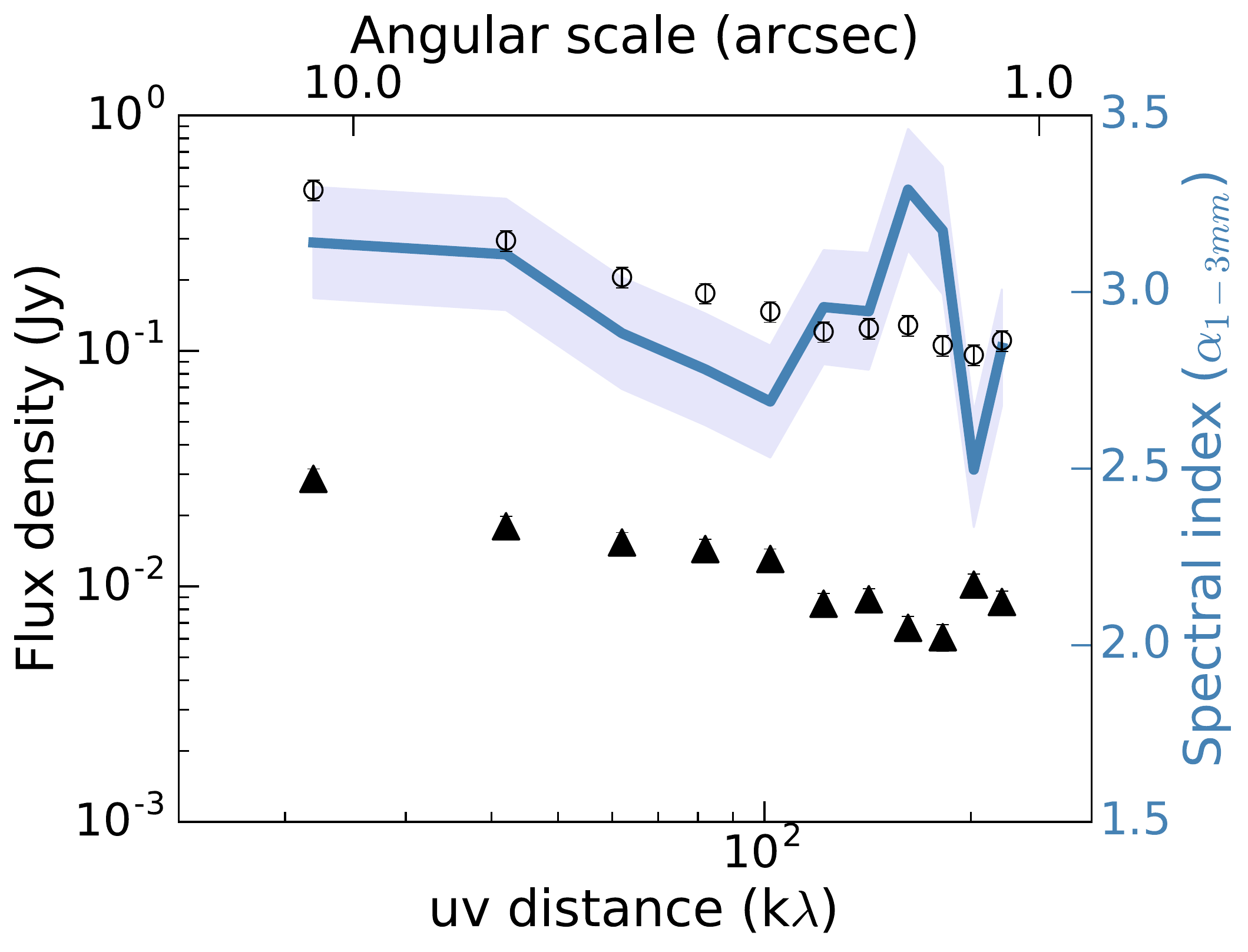} &
\hspace{3pt} \vspace{-11pt} \includegraphics[width=5.8cm]{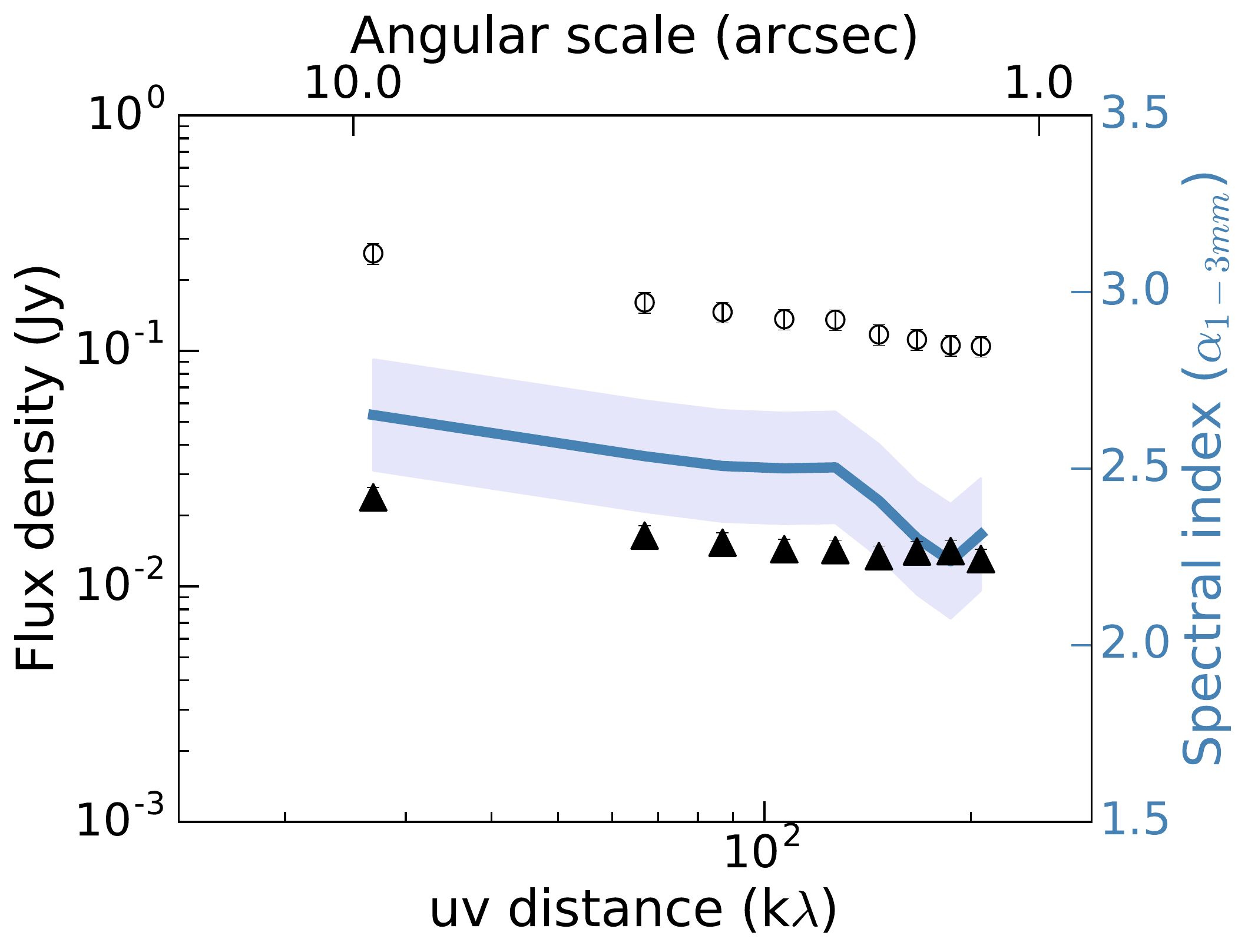} \\
\vspace{-50pt}\hspace{35pt}{\bf SerpM-SMM4} & 
\vspace{-50pt}\hspace{35pt}{\bf SerpS-MM18} & 
\vspace{-50pt}\hspace{35pt}{\bf L1157} \\
\end{tabular}
\caption{Continuum flux density at 1.3 mm (open circles) and 3.2 mm (filled triangles) as {a function of} the baselines averaged every 20 k$\lambda$. The variations of the observed spectral index $\alpha_\mathrm{1-3mm}$ (see Eq.~\ref{equalpha}) are overlaid (blue line). Their values are shown on the right y-axis. The error bars indicate the flux uncertainties and include the calibration errors. These errors are used to derive the uncertainties on $\alpha_\mathrm{1-3mm}$ shown as the shaded blue area. As L1448-N and L1448-2A host a secondary source, we only show the continuum flux densities at the shortest baselines but also indicate the flux at $\sim$200 k$\lambda$ obtained using the primary source as the phase center (see Sect.~2.2 for more details).
}
\label{Visibilities}
\end{figure*}


\section{Data}

\subsection{The sample}

The CALYPSO sources were selected from various nearby ($d=140-436$ pc) star-forming clouds and observed at 1.3 mm and 3.2 mm using the IRAM/PdBI (Institut de Radioastronomie Millimétrique / Plateau de Bure Interferometer) \citep{Maury2019}. Both the most extended configuration (A array) and the intermediate antenna configuration (C array) were used to probe the dust continuum emission over a wide range of baselines. 
For this analysis, we selected the 12 brightest protostars of the CALYPSO sample, namely sources with a 1.3 mm PdBI integrated flux higher than 100 mJy \citep[according to Table 4 in][]{Maury2019}. The sample is diverse in terms of morphologies (including single objects as well as close and wide binaries), bolometric luminosities (from 1 up to 30 \lsun), or bolometric temperatures (from 25 to 60 K). The source characteristics are summarized in Table~\ref{SourceCharacteristics}. For further details on the morphology of the dust continuum emission in this sample of sources, we refer to the maps shown in Appendix B of \citet{Maury2019}. Maps of the gas kinematics are shown in \citet{Gaudel2019}.

\begin{figure*}
\begin{tabular}{m{5.6cm}m{5.6cm}m{5.6cm}}
\hspace{5pt} \vspace{-11pt} \includegraphics[width=5.9cm]{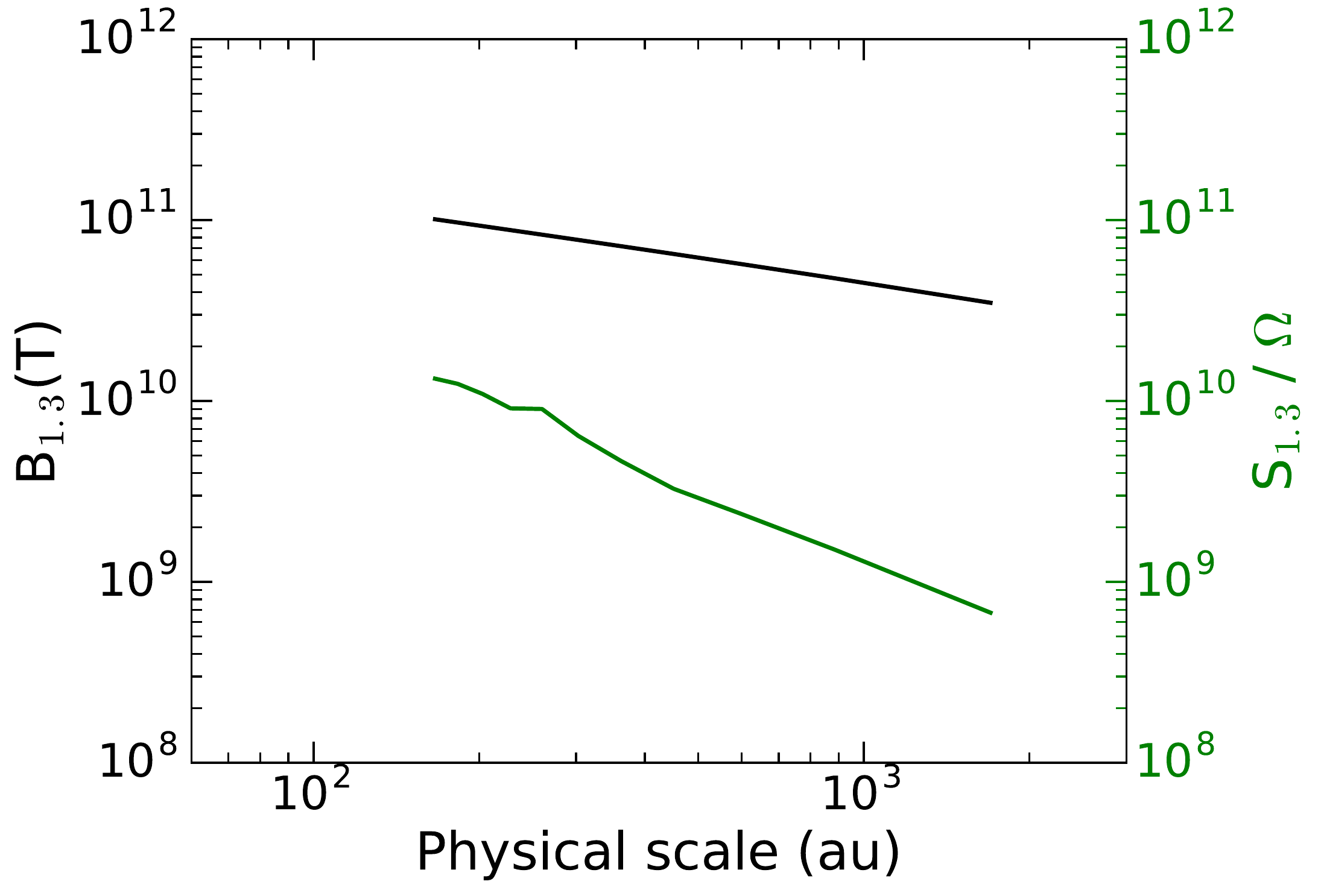} &      
\hspace{3pt} \vspace{-11pt} \includegraphics[width=5.9cm]{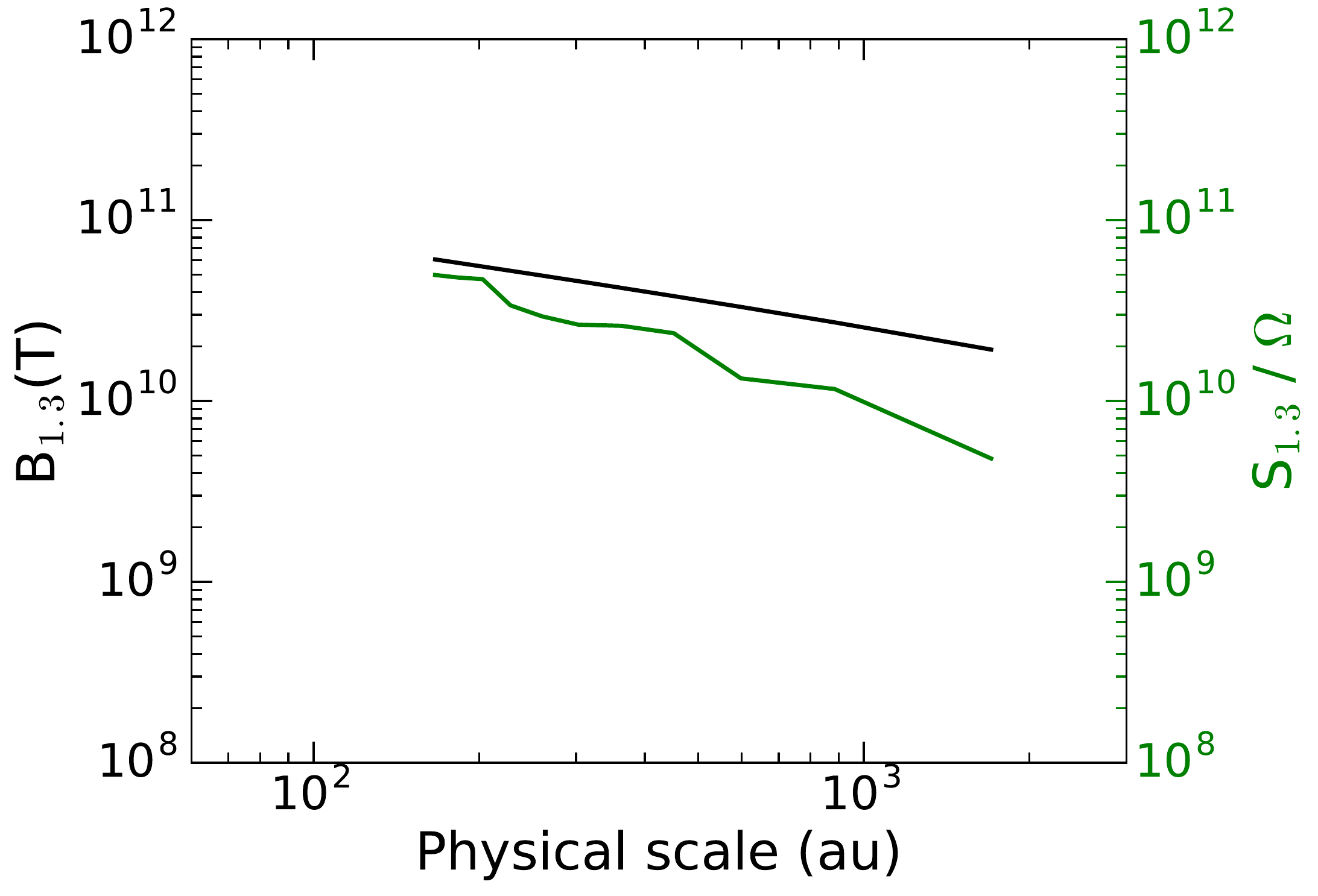} &
\hspace{3pt} \vspace{-11pt} \includegraphics[width=5.9cm]{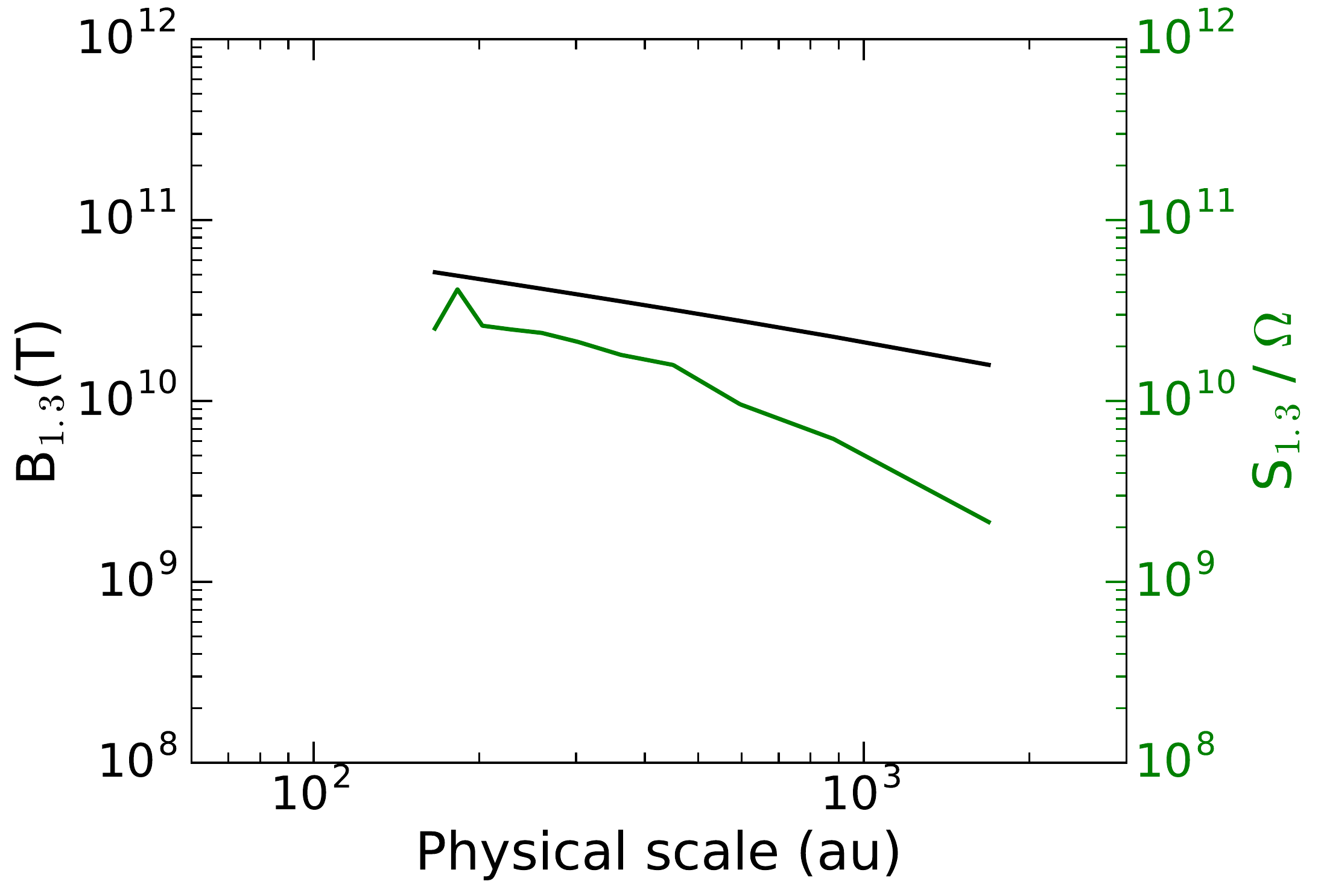} \\
\vspace{-50pt}\hspace{35pt}{\bf IRAS2A1} & 
\vspace{-50pt}\hspace{35pt}{\bf IRAS4A1} & 
\vspace{-50pt}\hspace{35pt}{\bf IRAS4B} \\
\hspace{5pt}  \vspace{-11pt} \includegraphics[width=5.9cm]{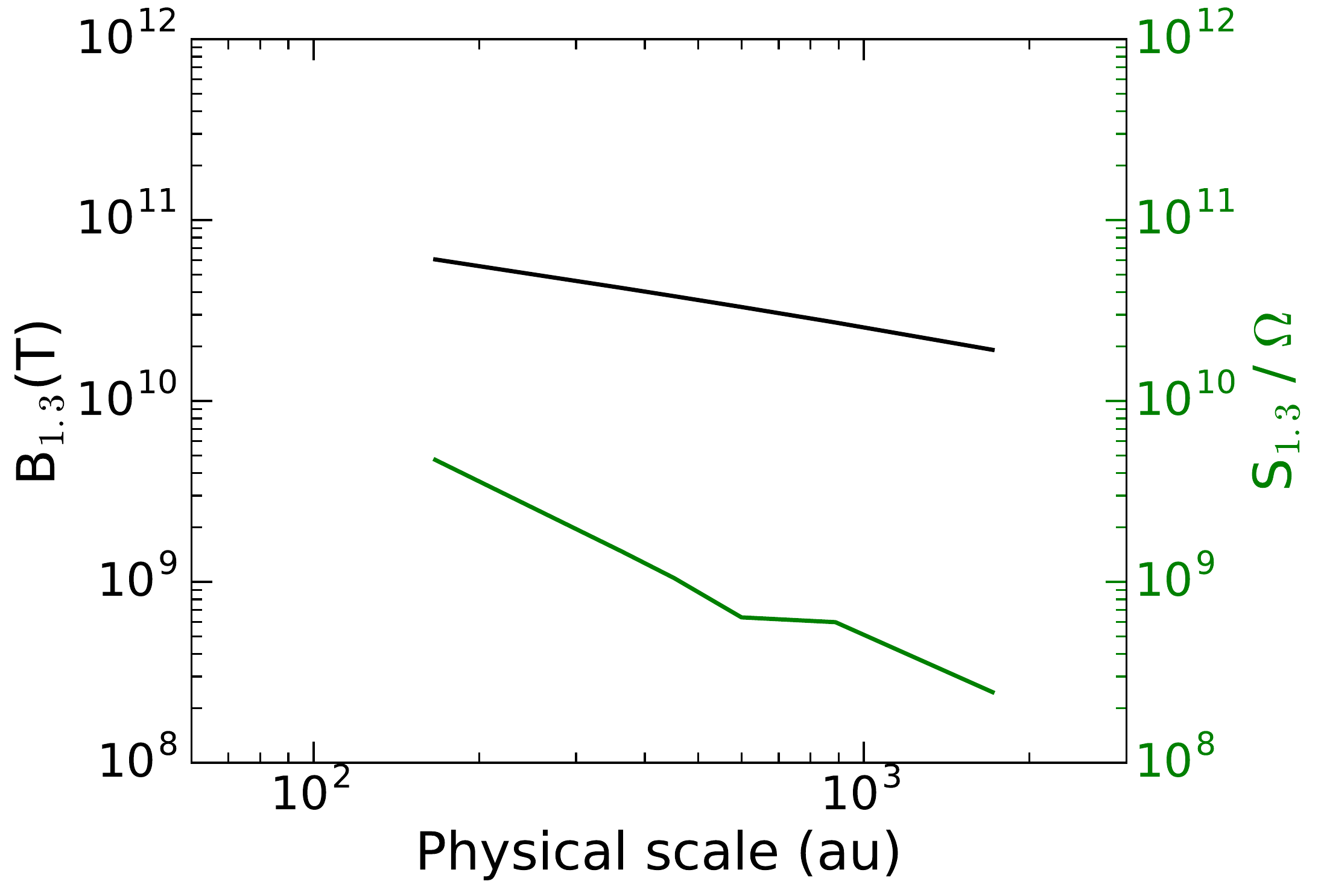} &
\hspace{3pt}  \vspace{-11pt} \includegraphics[width=5.9cm]{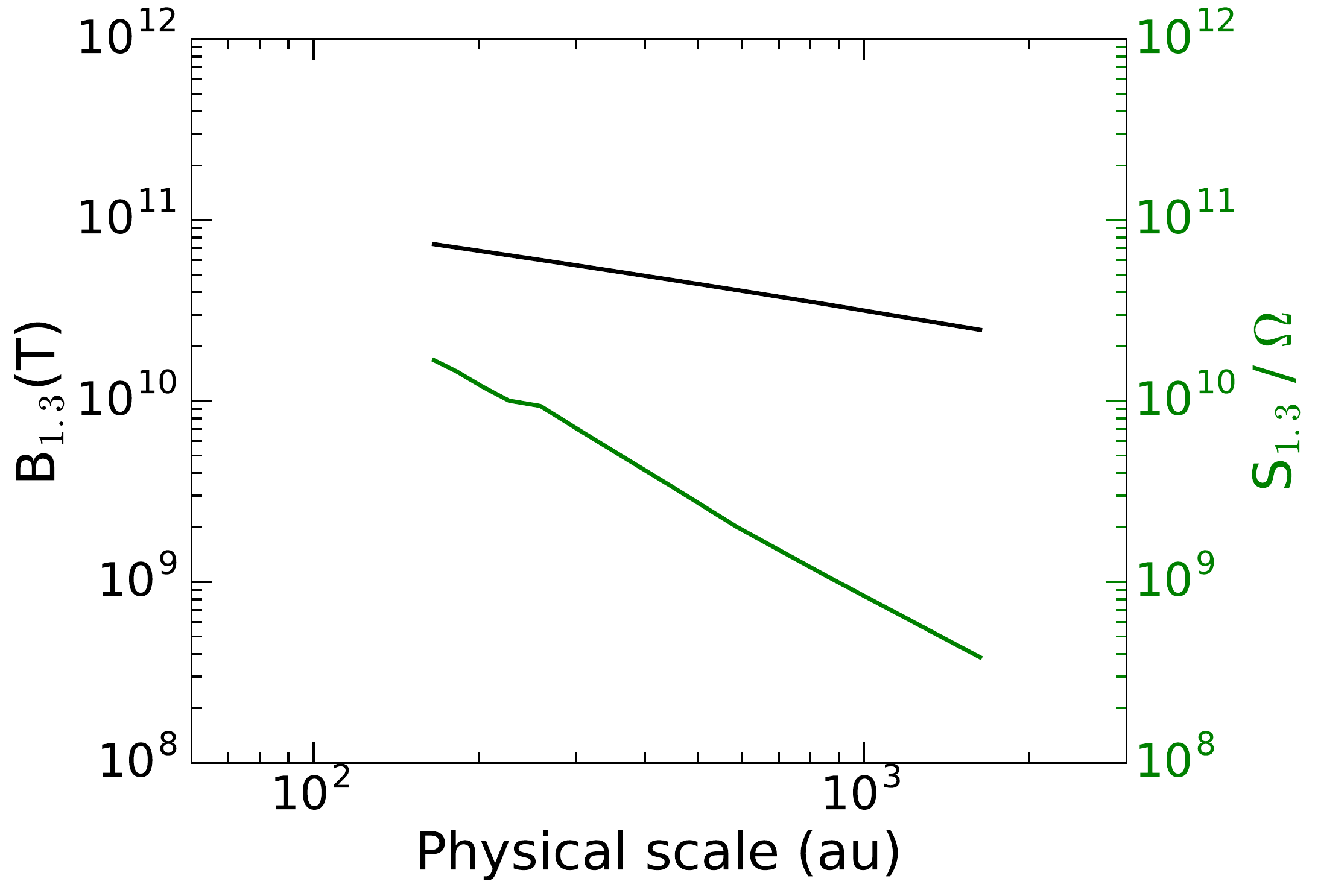} &
\hspace{3pt}  \vspace{-11pt} \includegraphics[width=5.9cm]{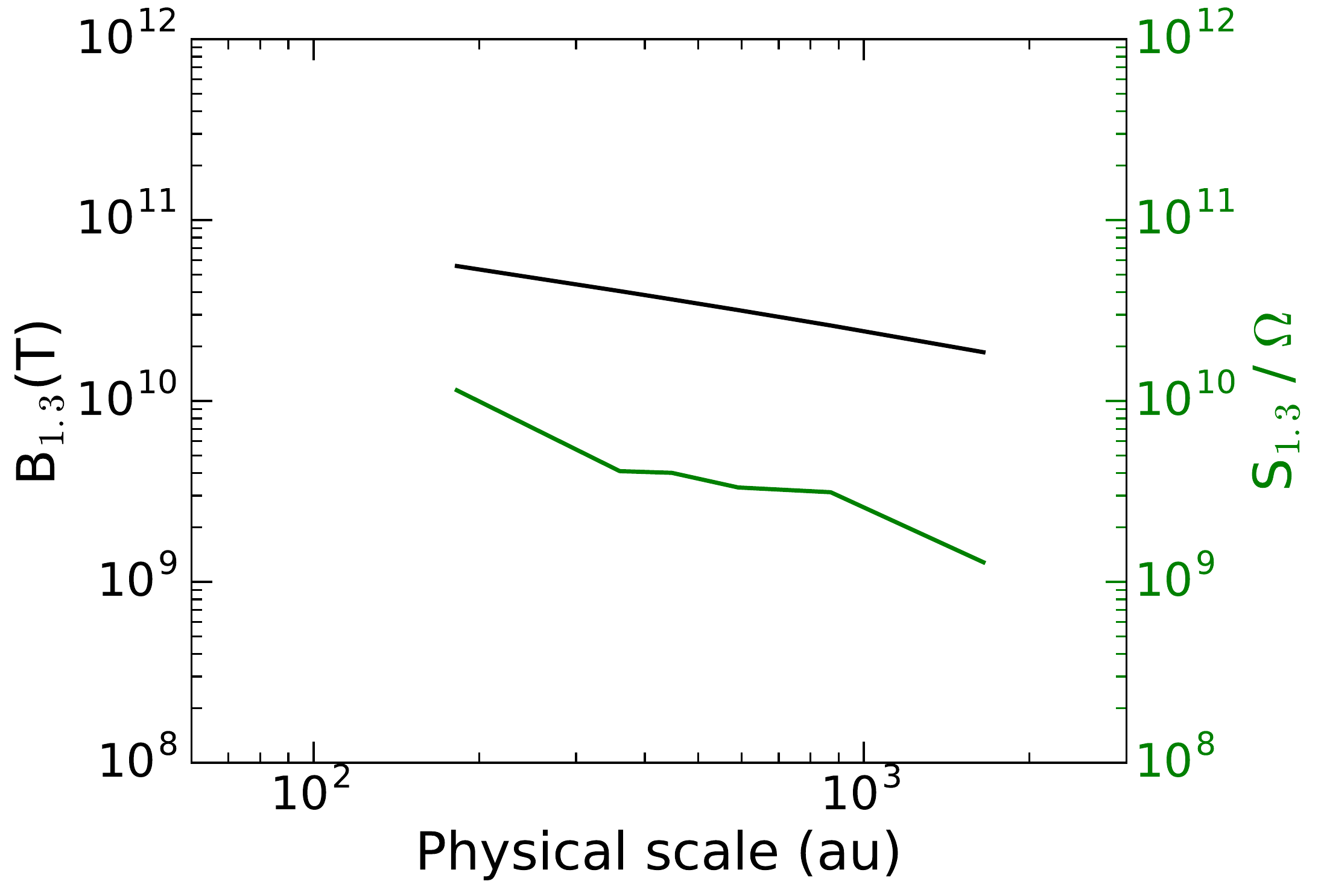} \\
\vspace{-50pt}\hspace{35pt}{\bf L1448-2A} & 
\vspace{-50pt}\hspace{35pt}{\bf L1448-C} & 
\vspace{-50pt}\hspace{35pt}{\bf L1448-NB1} \\
\hspace{5pt} \vspace{-11pt} \includegraphics[width=5.9cm]{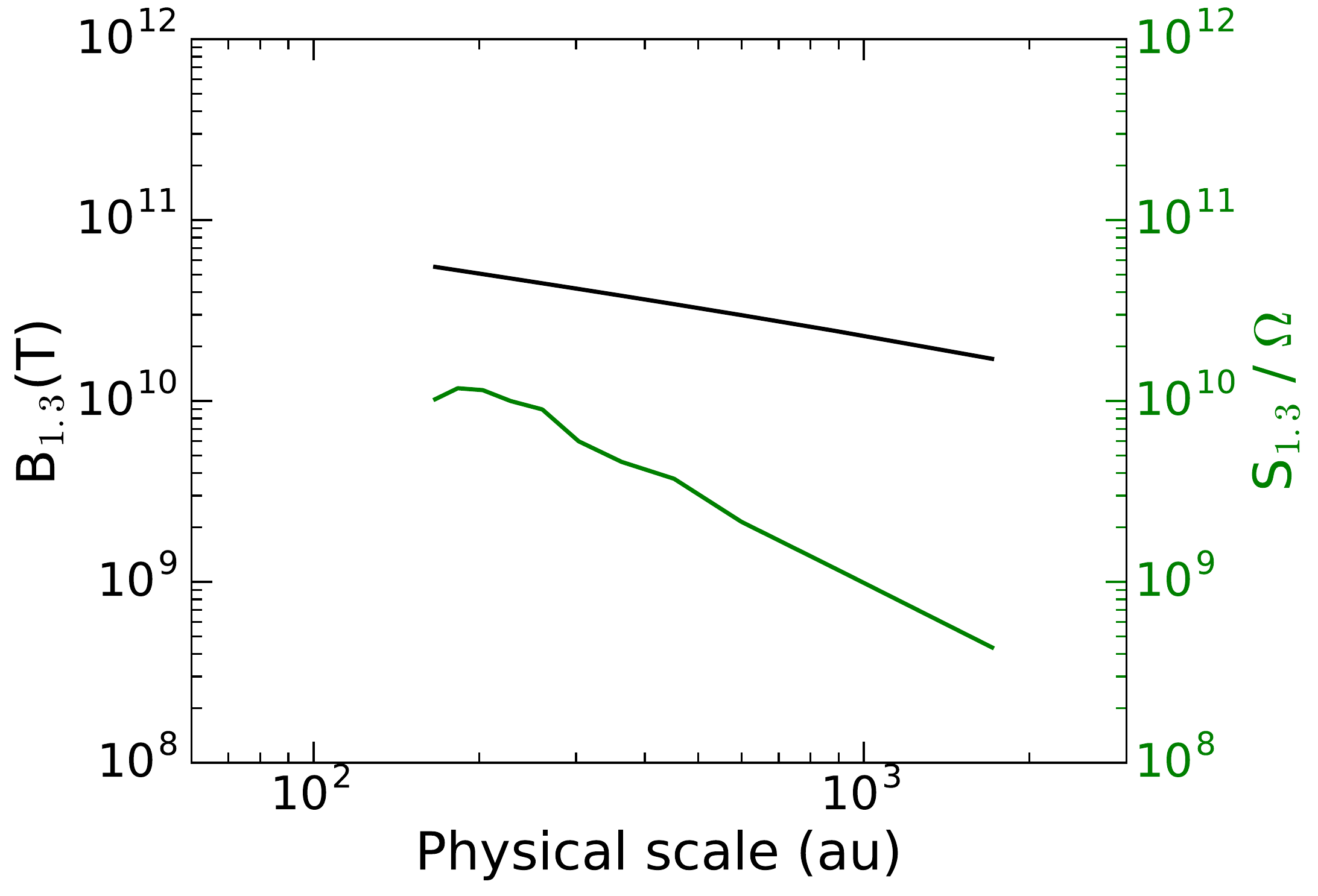} &
\hspace{3pt} \vspace{-11pt} \includegraphics[width=5.9cm]{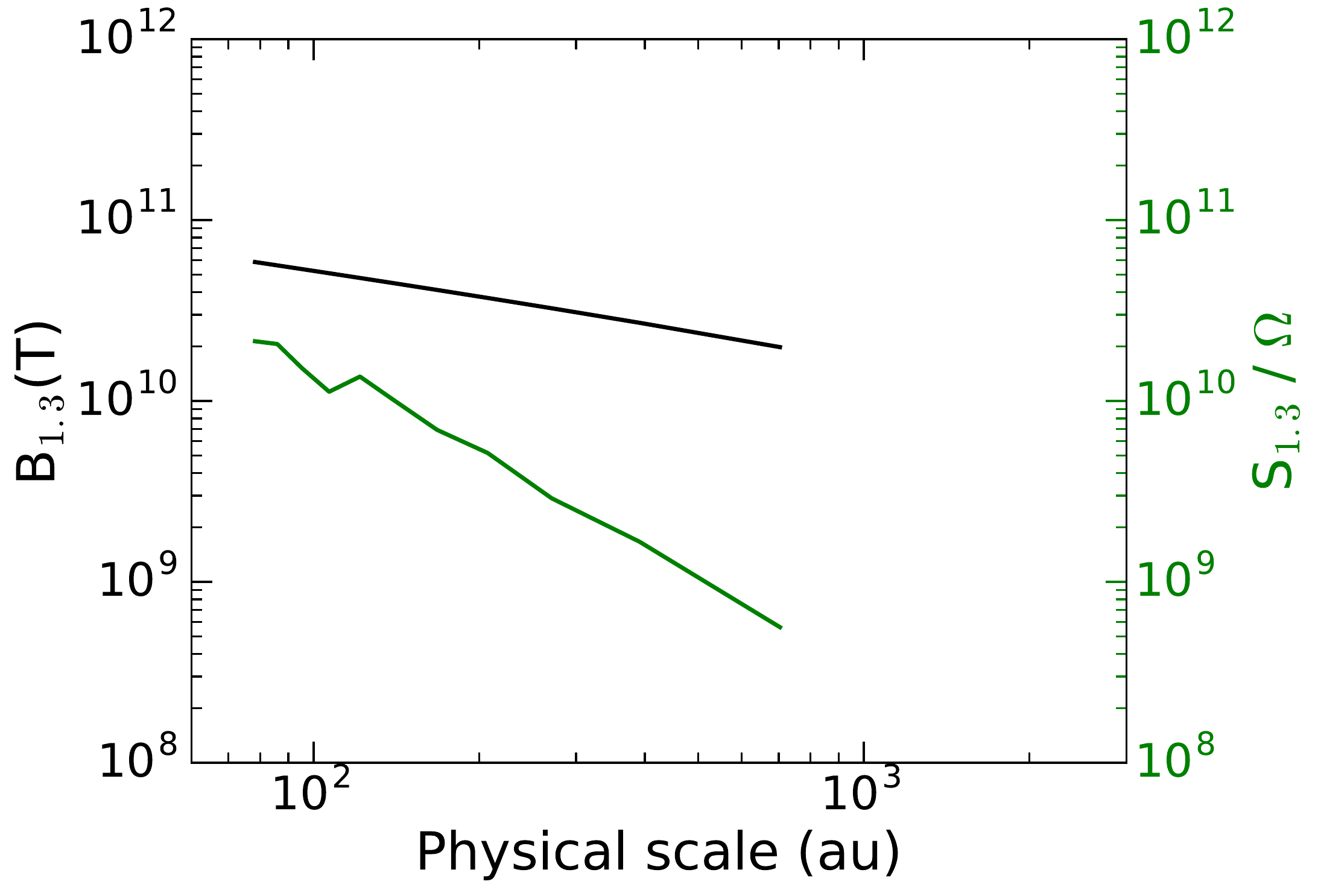} &
\hspace{3pt} \vspace{-11pt} \includegraphics[width=5.9cm]{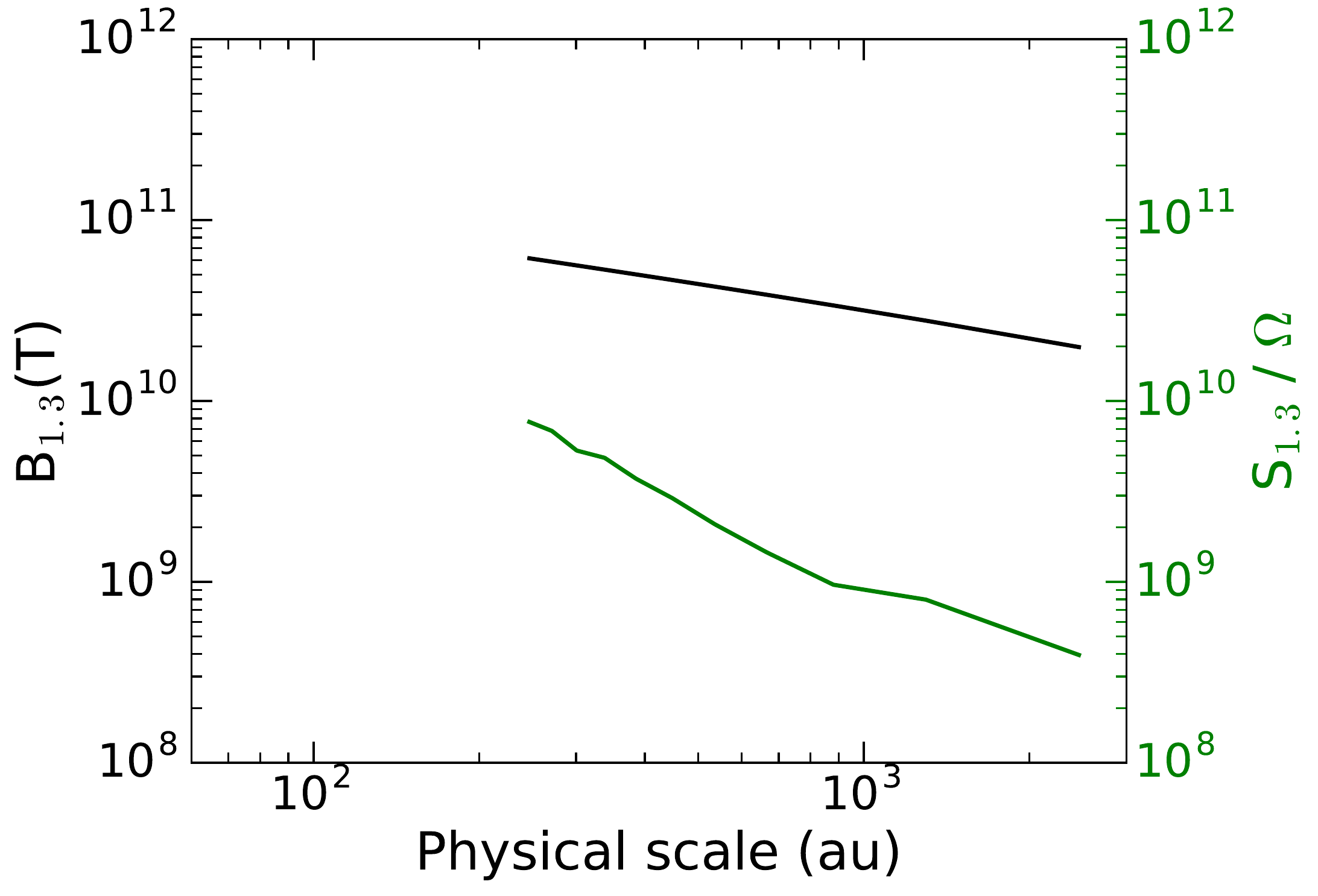} \\
\vspace{-50pt}\hspace{35pt}{\bf SVS13B} & 
\vspace{-50pt}\hspace{35pt}{\bf L1527} & 
\vspace{-50pt}\hspace{35pt}{\bf SerpM-S68N} \\
\hspace{5pt} \vspace{-11pt} \includegraphics[width=5.9cm]{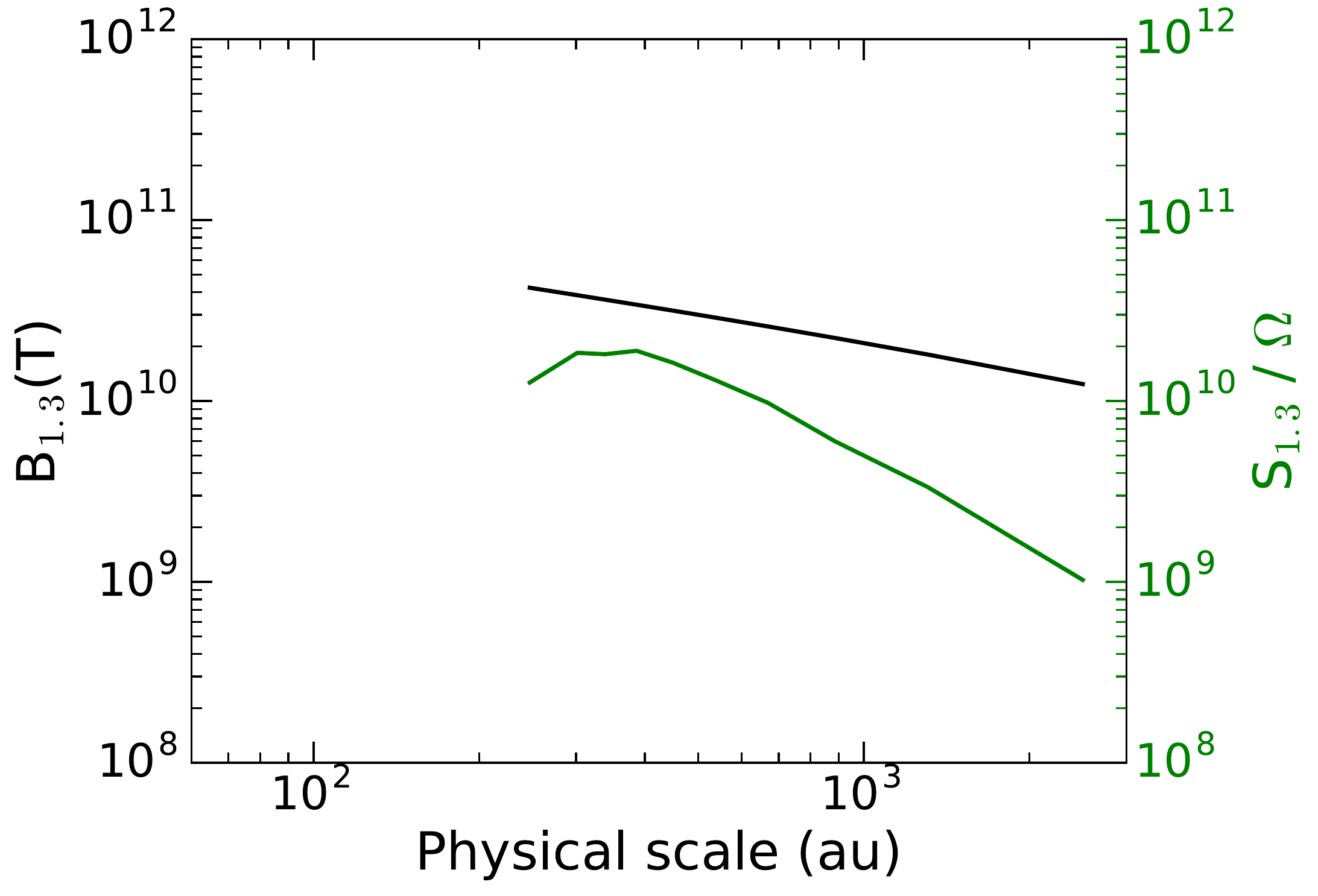} &
\hspace{3pt} \vspace{-11pt} \includegraphics[width=5.9cm]{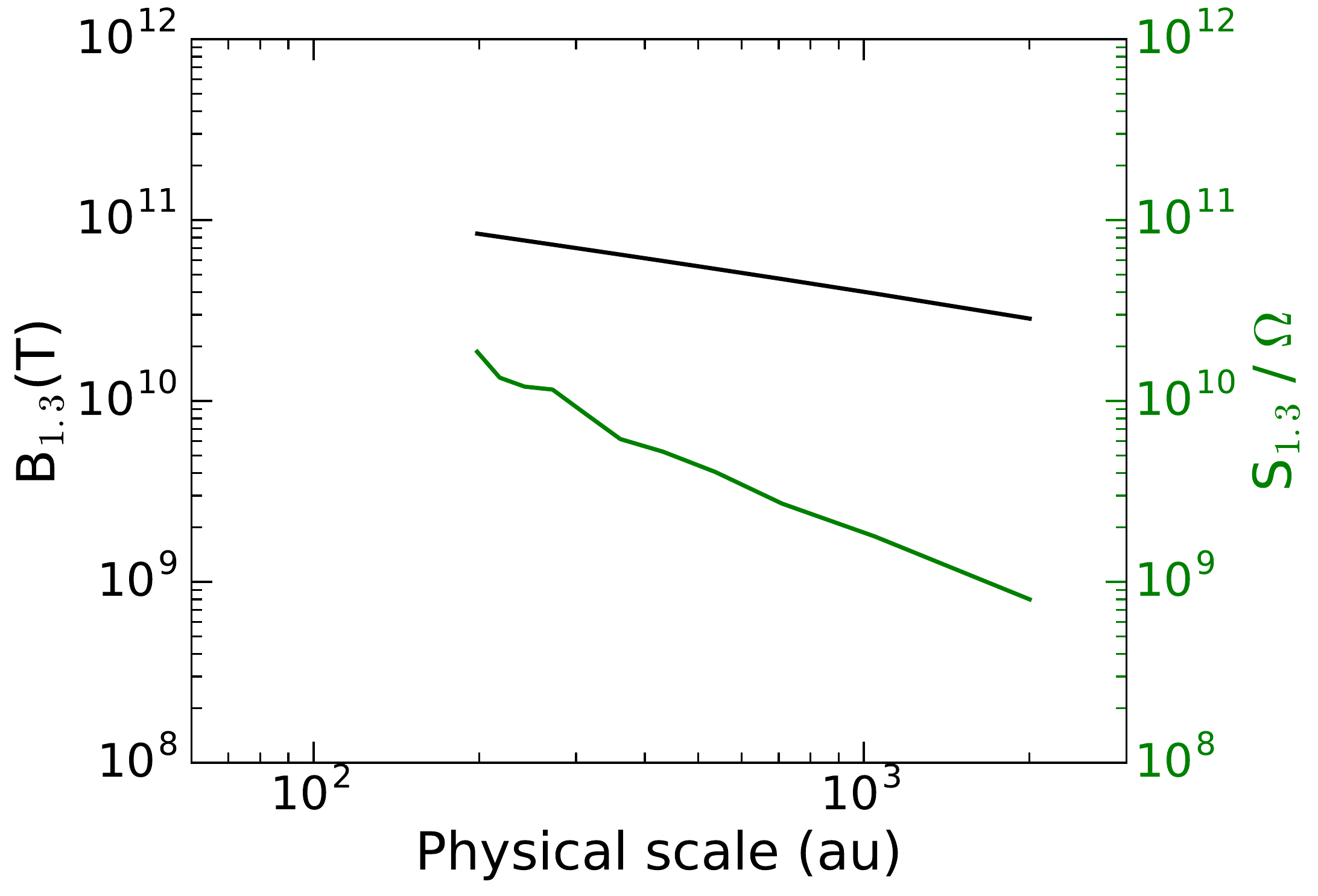} &
\hspace{3pt} \vspace{-11pt} \includegraphics[width=5.9cm]{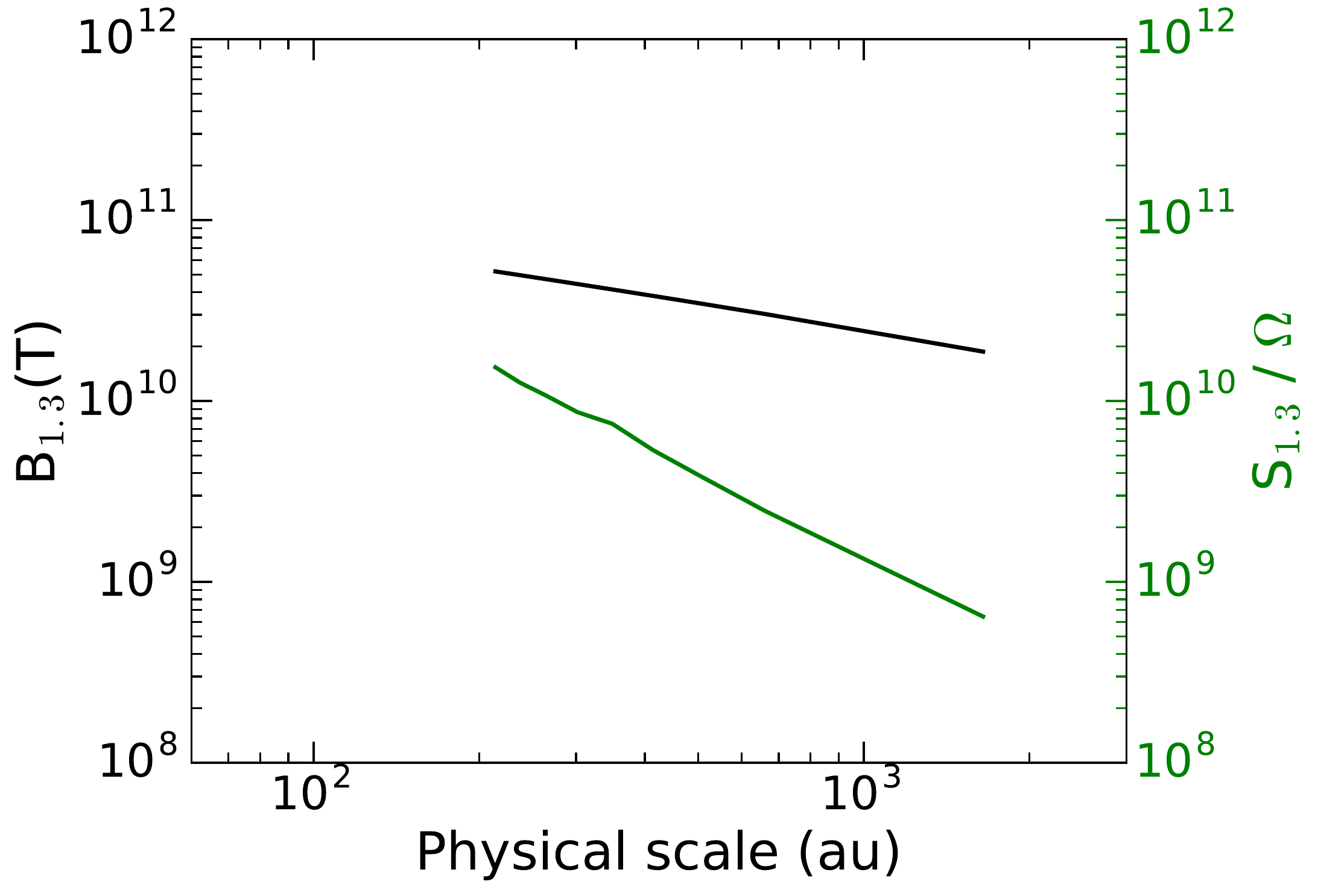} \\
\vspace{-50pt}\hspace{35pt}{\bf SerpM-SMM4} & 
\vspace{-50pt}\hspace{35pt}{\bf SerpS-MM18} & 
\vspace{-50pt}\hspace{35pt}{\bf L1157} \\
\end{tabular}
\caption{Comparison between $S_{1.3}/\Omega$ (with $S$ the flux density at 1.3 mm and $\Omega$ the solid angle associated with the physical scale probed) and the Planck function profile associated with the dust temperature profile of the envelope. The fact that $S_{1.3}/\Omega < B_{1.3}(T)$ in the objects indicates that the envelopes (at the radii probed by our analysis) can be considered as optically thin. 
}
\label{Tb}
\end{figure*}

\subsection{Data processing}
We worked on the common range of baselines sampled at 1.3 mm and 3.2 mm, thus ranging from 20 k$\lambda$ to $~$220 k$\lambda$. The data calibration (and self-calibration for some of the sources) was performed with \texttt{GILDAS}\footnote{https://www.iram.fr/IRAMFR/GILDAS/} using the Continuum and Line Interferometer Calibration (CLIC) package. \citet{Maury2019} have provided details on the observations, data reduction, and Plummer models fitted to the dust continuum visibility profiles. Figure~\ref{Visibilities} shows the real part of the visibilities at 1.3 mm and 3.2 mm (circles and triangles, respectively); the baselines are averaged every 20 k$\lambda$. Errors are estimated by adding in quadrature the calibration uncertainties ($\sim$10\% at 94 GHz and $\sim$15\% at 231 GHz) and the rms dispersion around the weighted mean in each bin. Our analysis is restricted to the properties of the dust continuum emission around the primary protostars, as defined in Table~2 of \citet{Maury2019}. The two sources L1448-N and L1448-2A host a secondary source that greatly contributes to the visibility profiles. For these sources, we chose to show only the amplitudes of the shortest baselines derived using the center of the binary system as the phase center and thus tracing the general dust properties in the outer envelope of the binary system. We additionally indicated the fluxes at the $\sim$200 k$\lambda$ baseline obtained using the primary source as the phase center to show how the dust properties connect between these two scales.

\subsection{Free-free + synchrotron contribution}
\label{ffc}
In protostars, thermal (free-free) emission produced by a jet or a photo-evaporative wind \citep[see][for observations and theoretical models]{Anglada1996,Lugo2004,Pascucci2012,Tychoniec2018_III} and to a lesser extent, from nonthermal (synchrotron) emission \citep{Andre1996} originating from the presence of magnetic fields, can contribute to the 1-3 mm emission within the central 50 au around the protostar. We estimated the contribution to the 1.3 mm and 3.2 mm fluxes not coming from thermal dust emission from radio measurements beyond 2 cm. Details for each source are provided in Appendix \ref{Nonthermaldust}.
The contribution is systematically lower than 10\% at 1.3 mm, but can be non-negligible at 3.2 mm. 
Correcting for this compact contamination is one motivation for the `central region correction' (c.r.c) we apply and discuss in Sect.~4.

\subsection{Optically thick emission at 1.3 and 3.2 mm} 

If regions of our Class 0 envelopes are optically thick, the intensity should only depend on the temperature of the emitting material and the brightness temperature should be equal to the actual temperature in the region. Following \citet{Motte1998}, we have
\begin{equation}
    \frac{S_{1.3}}{\Omega} \propto (1-\exp(-\tau))~B_{1.3}(T_\mathrm{dust})
\end{equation}

\noindent where $S$ is the flux density (here taken at 1.3 mm), $\Omega$ the solid angle associated with the physical scale probed, $\tau$ the optical depth, and $B$ the Planck function derived from the dust temperature. The dust temperature is estimated with Eq.~\ref{equtemperature} (cf. Sect. 3). If the medium is optically thick and if tau$>$1, (1-$\exp(-\tau$))*B tends toward B. If the medium is optically thin, we should observe S/$\Omega$ $<$ $B(T)$. We thus use the radially determined fluxes at 231 GHz maps divided by $\Omega$ and compare them to the $B(T)$ profiles. 

Figure~\ref{Tb} shows how the two quantities vary with the scales in astronomical units. The absence of overlap between the two profiles indicates that most of the continuum emission of the envelope scales we probe is optically thin, except perhaps for the 1.3 mm emission coming from the $<$ 200 au regions of IRAS4A, IRAS4B, or SerpM-SMM4. In this case again the central region correction applied in Sect.~4 allows us to correct for the potential central optical thickness effects affecting the visibility profiles.

\section{Observational results}

An interferometer only samples the uv plane at discrete spatial frequencies. Voids in this under-sampled Fourier space are a challenge for interferometric image reconstruction. We thus decided to perform our analysis of the grain size distribution variations directly in the visibility domain. As described in many analyses \citep[][among others]{Harvey2003,Maury2019}, if the density and temperature distributions in the envelope follow a power law, $\propto r^{-p}$ and $\propto r^{-q}$ \citep{Adams1991,Motte2001}, respectively, then in the Rayleigh-Jeans approximation, the intensity will follow a power law $I(r)$ $\propto r^{-(p+q-1)}$. The resulting visibility distribution will be a power law as well, expressed as $V(b)$ $\propto b^{(p+q-3)}$. 
The ratio of the amplitude visibility measured at two wavelengths at a common baseline thus traces the spectral index of the dust composing the emitting material within a given area corresponding to the scale probed by the specific baseline. The variations of the millimeter spectral index $\beta_{mm}$ can then be used as a signature of variation of the grain size distribution with lower $\beta$ values associated with the presence of more emissive grains, as expected for larger grains \citep{Kruegel1994}.

\subsection{Observed spectral index profiles with uv distance}

The dust continuum emission fluxes at 1.3 mm and 3.2 mm, as measured in bins of baselines, are shown in Fig.~\ref{Visibilities}. The figure also shows the variations of the observed spectral index $\alpha_\mathrm{1-3mm}$ defined as
\begin{equation}
\alpha_\mathrm{1-3mm} = \frac{log(F(\nu_0) / F(\nu_1))}{log(\nu_0 / \nu_1)}
\label{equalpha}
,\end{equation}
\noindent where $\nu_0$ and $\nu_1$ are equal to 231 GHz and 94 GHz,  respectively. 
Figure~\ref{Visibilities} indicates, overlaid on the visibilities, the variations of the observed spectral index $\alpha_\mathrm{1-3mm}$ (blue lines, scale provided on the right axis) for each object. 

In order to express the uv distances in terms of angular separations, we use the relation linking the brightness distribution I of an object to its complex visibility given by the Fourier transform 

\begin{equation}
I(x, y)=\int_{-\infty}^{\infty} \int_{-\infty}^{\infty} V(u, v) \exp (2 \pi i(x u + y v)) d u d v
,\end{equation}

\noindent where (x, y) represent the angular coordinates on the sky (in radians) and (u, v) the coordinates describing the baseline. 
For an assumed circularly symmetric object, we can use the zeroth-order Bessel function J$_0$ to write

\begin{equation}
V(b)= \int_{0}^{\infty} 2 \pi I(\theta) J_0(2 \pi \theta b) d \theta \sim \int_{0}^{x_0/2 \pi b} 2 \pi I(\theta) \theta d \theta
,\end{equation}

\noindent where x$_0$ is the first zero of J$_0$. The visibility reaches its first zero when b$_{V=0}$=1.22($\lambda/\theta$), resulting in a baseline - angular scale relation given by

\begin{equation}
\label{eqtheta}
\theta = 1.22 \frac{\lambda}{b}
.\end{equation}

These angular scales $\theta$, over which the dust continuum emission is integrated to compute the spectral index are indicated on the top axis of Fig.~\ref{Visibilities}.
We observe that most of the objects show a clear decrease of the observed spectral index $\alpha_\mathrm{1-3mm}$ toward larger uv distances, thus smaller envelope radii. Several objects of the sample have already been studied using other millimeter interferometric observations with similar resolutions, allowing us to confront our observations with previous works. L1157 has been studied using CARMA by \citet{Kwon2009} and \citet{Chiang2012} who derived the profile of the dust emissivity index as a function of uv distance using the Rayleigh-Jeans approximation and the optically thin assumption. Our results are consistent with the relatively flat profile they observe on 10-40 au and 25-2500 au scales, respectively: L1157 has one of the weakest gradient of our sample.
For L1527, we find a very similar flat profile of the observed spectral index as a function of uv distance than that found by \citet{Tobin2013_2}. 

\begin{figure}
\includegraphics[width=9cm]{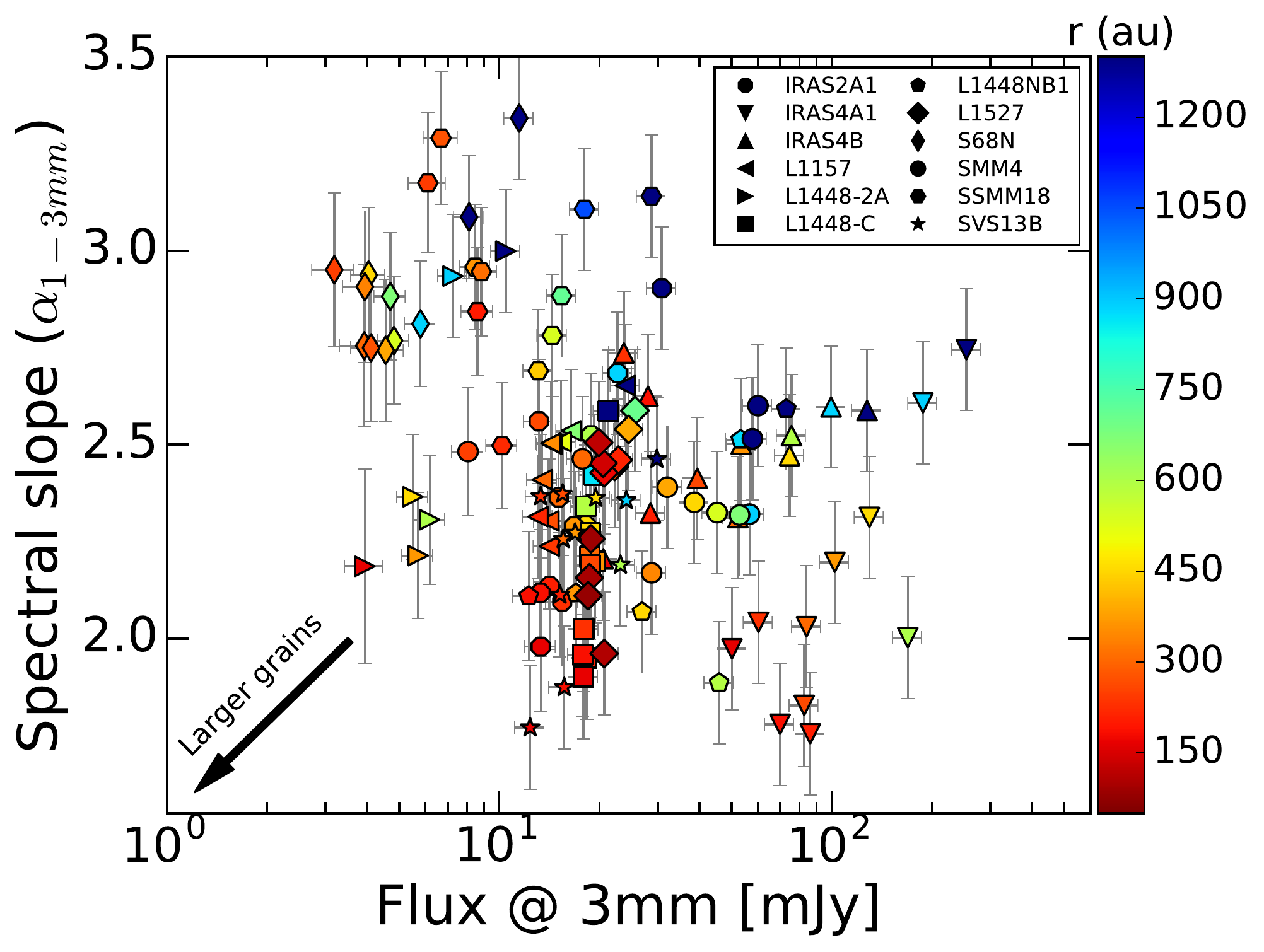} 
\caption{Variation of the observed spectral index $\alpha_\mathrm{1-3mm}$ derived in the baseline bins of Fig.~\ref{Visibilities} as a function of its associated 3.2 mm flux density. Each symbol represents a different protostellar envelope and symbols are color coded with respect to the physical distances probed (in au). The bottom left arrow indicates the trend predicted when the grain size is increasing \citep[see models from][]{Miotello2014}.}
\label{AlphaVersusF3mm}
\end{figure}

\vspace{5pt}
The observed spectral indices derived in baseline bins as a function of of their associated 3.2 mm flux densities are shown in Fig.~\ref{AlphaVersusF3mm}. Each point is color coded as a function of the radius over which the integrated quantities are computed, in astronomical units (using eq.~\ref{eqtheta} and r = $\theta \times d/2$). \citet{Miotello2014} generated a series of families of disk models following the method developed in \citet{Testi2001} to interpret their millimeter observations of the two Class I protostars Elias29 and WL12. We overlay in the bottom left corner of Fig.~\ref{AlphaVersusF3mm} the evolution of the grain size expected from these models, even if those do not directly apply to our Class 0 sources. Observationally, the outer regions of our protostellar envelopes populate the top right part of the plot,while the central regions have lower $\alpha_\mathrm{1-3mm}$ and fluxes and thus are located in the bottom left part. The \citet{Miotello2014} models predict a progressive increase of the grain size as one moves from the top right to the bottom left corner of the plot. We conclude that our observations are consistent with the presence of larger grains when probing smaller and smaller scales.

\begin{figure}
\includegraphics[width=9cm]{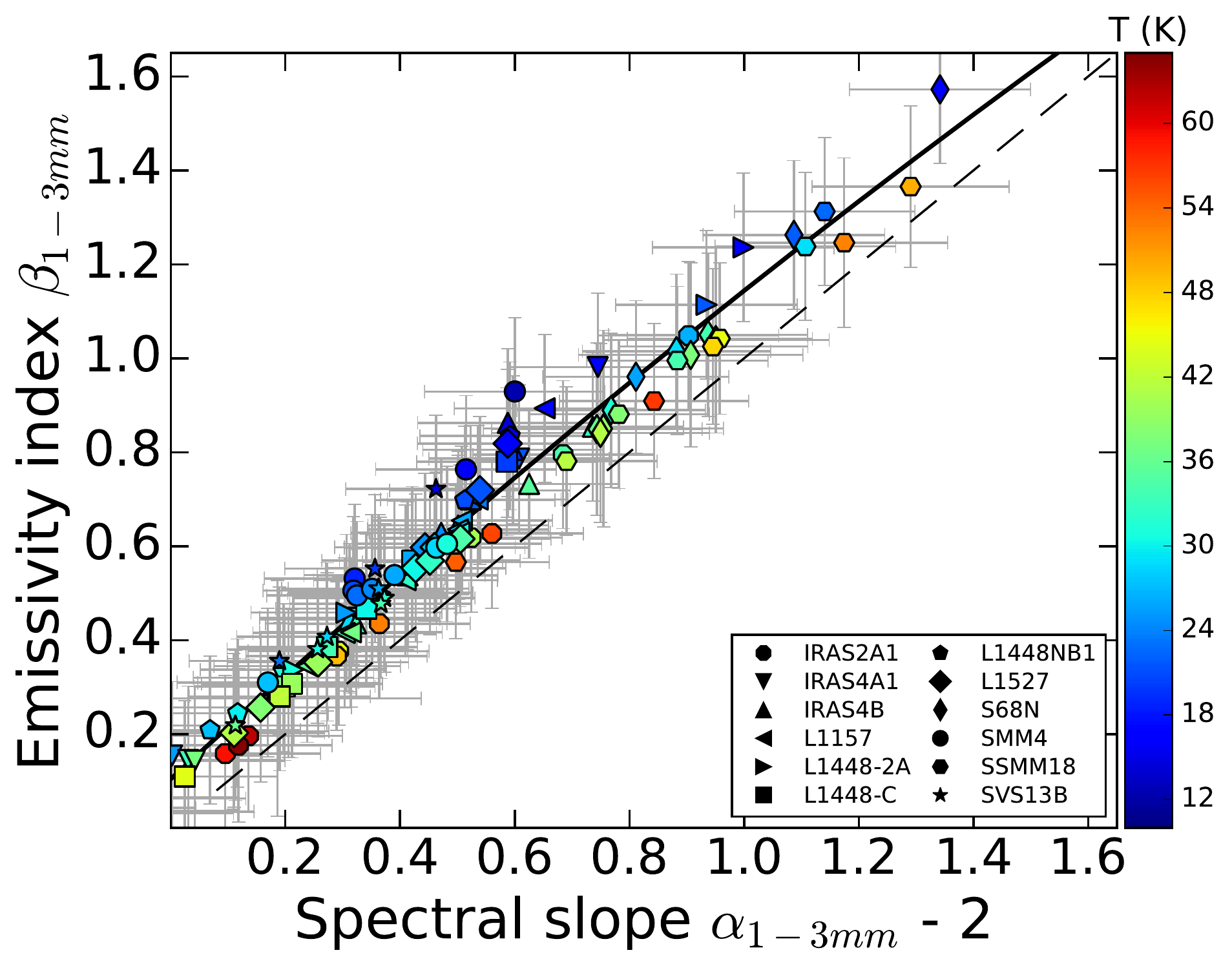} 
\caption{Relation between the dust emissivity index $\beta_\mathrm{1-3mm}$ and the observed spectral index $\alpha_\mathrm{1-3mm}$-2. The dashed line represents the 1:1 relation, while the plain line represents the polynomial regression to the data ($\beta_\mathrm{1-3mm}$ = -0.07 ($\alpha_\mathrm{1-3mm}$-2)$^2$ + 1.12 ($\alpha_\mathrm{1-3mm}$-2) + 0.10). Each symbol represents a different protostellar envelope and symbols are color coded with respect to our modeled dust temperature.}
\label{BetaVersusAlpha}
\end{figure}

\begin{figure*}
\begin{tabular}{m{5.6cm}m{5.6cm}m{5.6cm}}
\includegraphics[width=5.8cm]{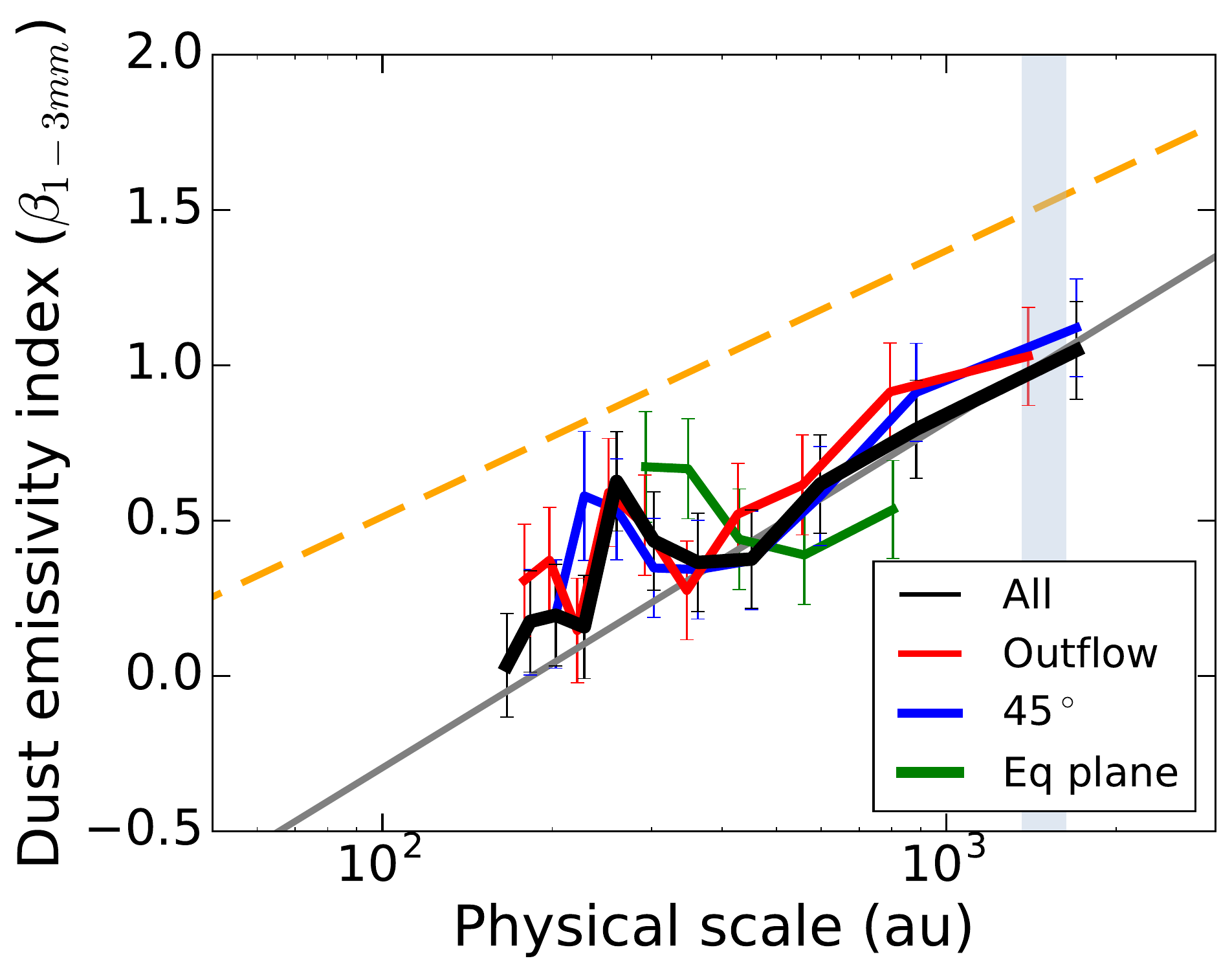} & \includegraphics[width=5.8cm]{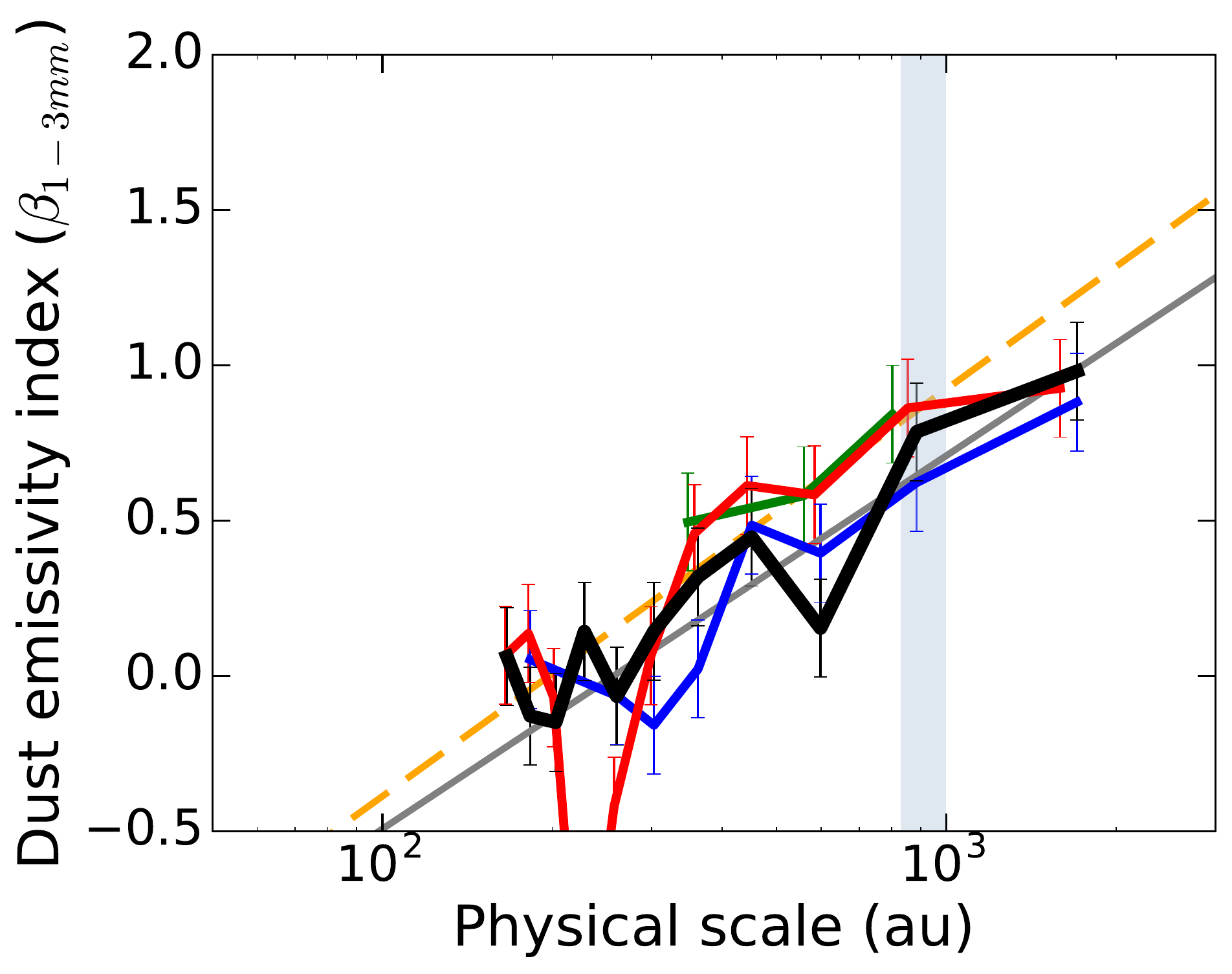} & \includegraphics[width=5.8cm]{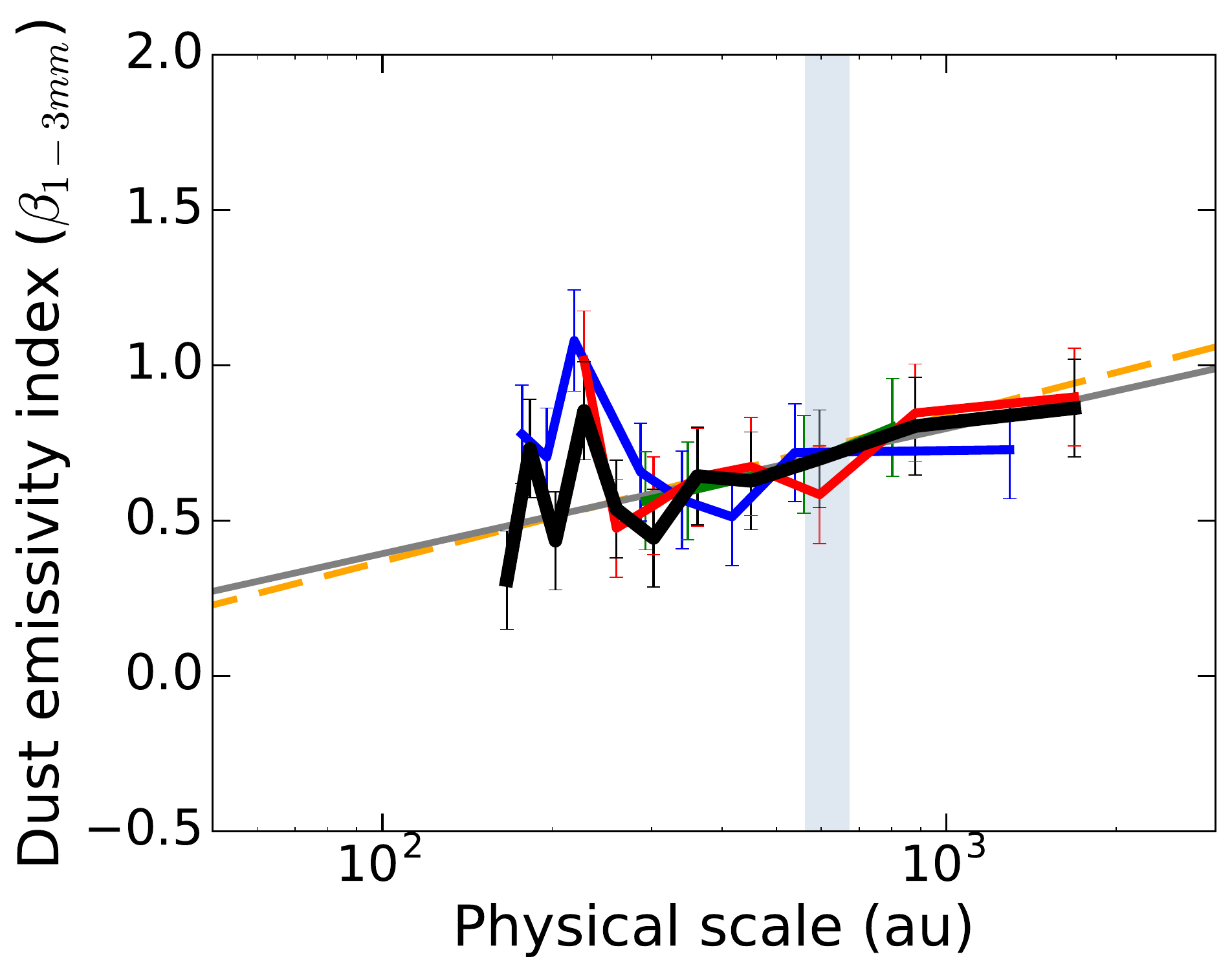} \\
\vspace{-245pt}\hspace{35pt}{\bf IRAS2A1} & 
\vspace{-245pt}\hspace{35pt}{\bf IRAS4A1} & 
\vspace{-245pt}\hspace{35pt}{\bf IRAS4B} \\
\vspace{-10pt} \includegraphics[width=5.8cm]{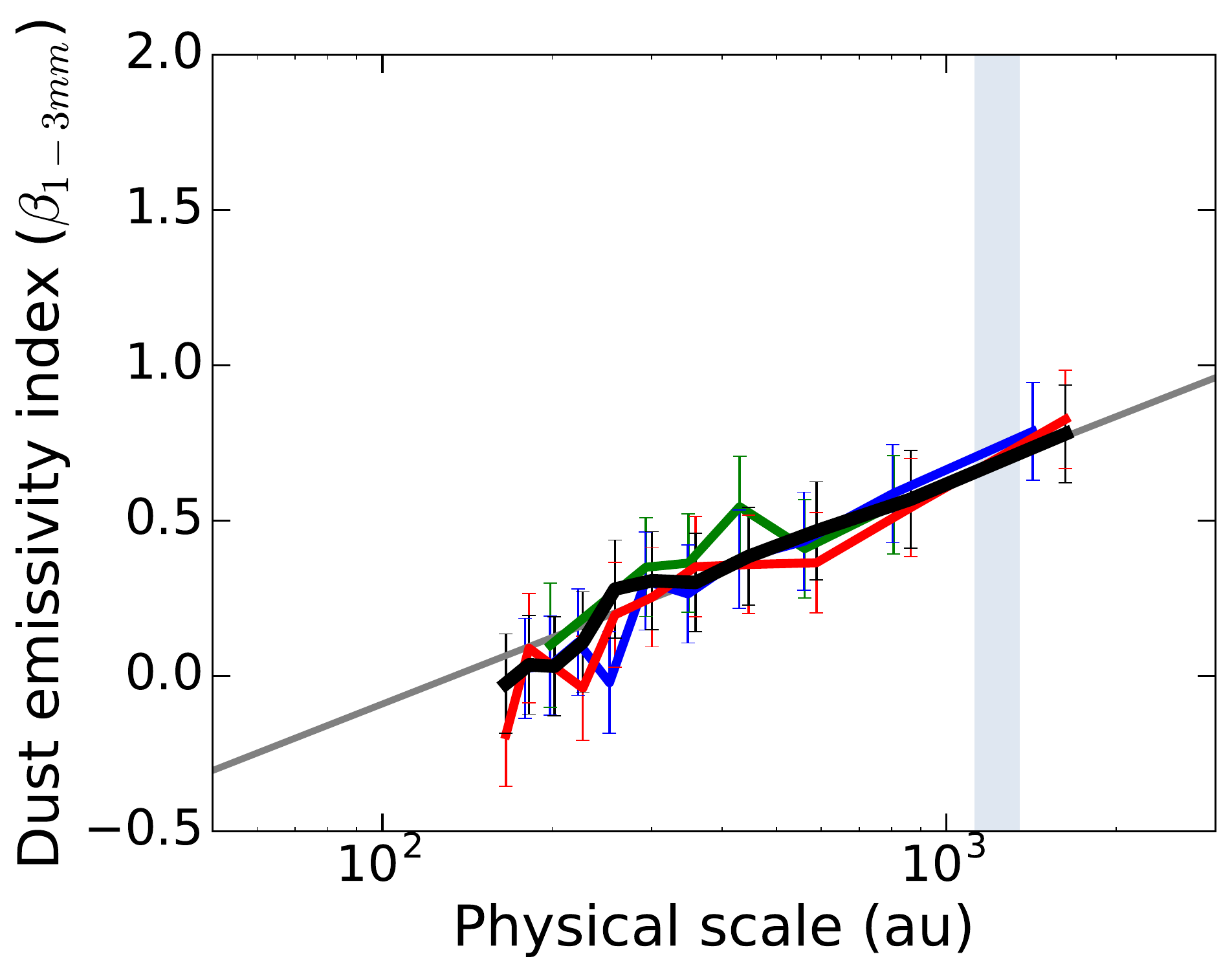} & \vspace{-10pt} \includegraphics[width=5.8cm]{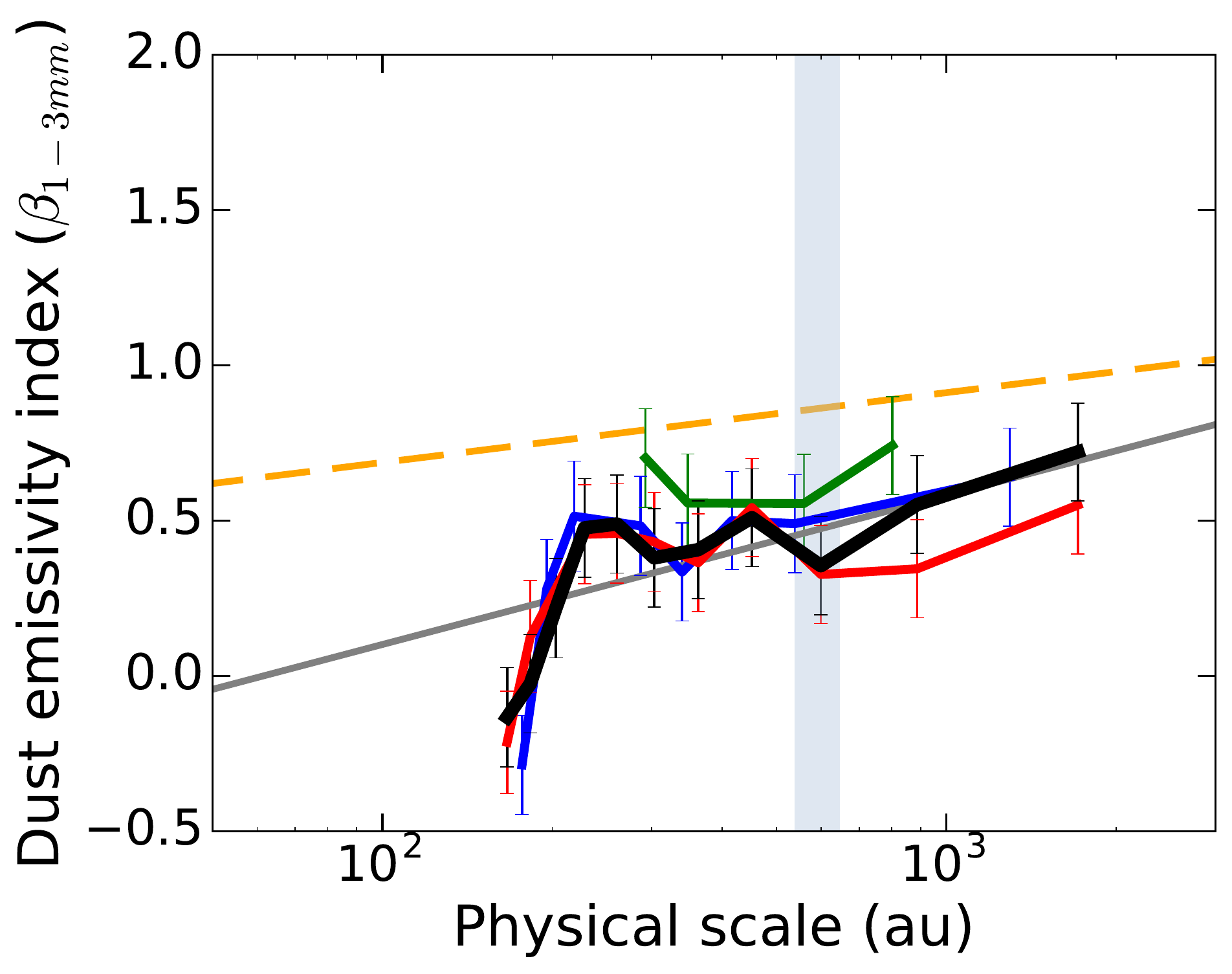} &
\vspace{-10pt} \includegraphics[width=5.8cm]{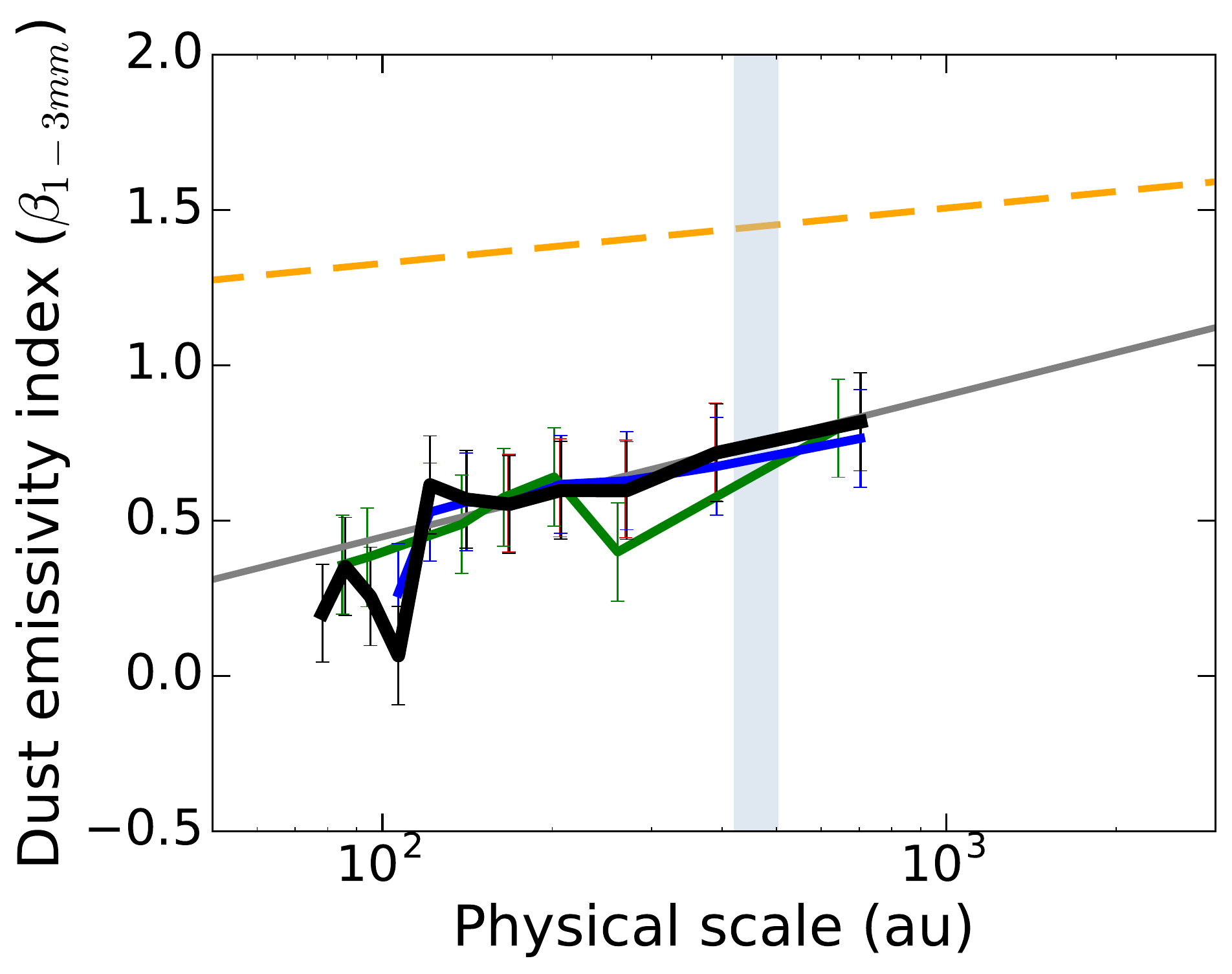} \\
\vspace{-245pt}\hspace{35pt}{\bf L1448-C} & 
\vspace{-245pt}\hspace{35pt}{\bf SVS13B} & 
\vspace{-245pt}\hspace{35pt}{\bf L1527} \\ 
\vspace{-10pt} \includegraphics[width=5.8cm]{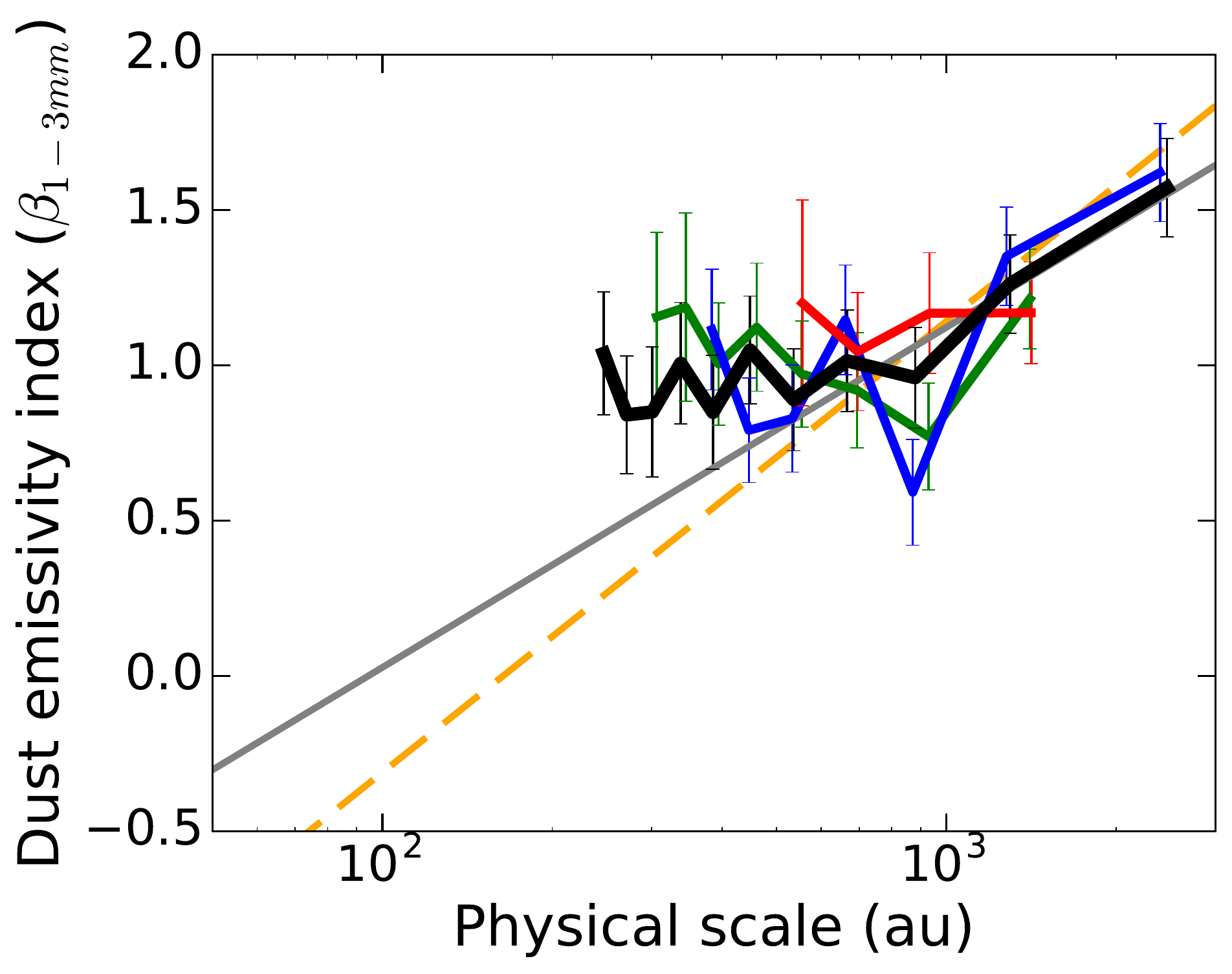} &
\vspace{-10pt} \includegraphics[width=5.8cm]{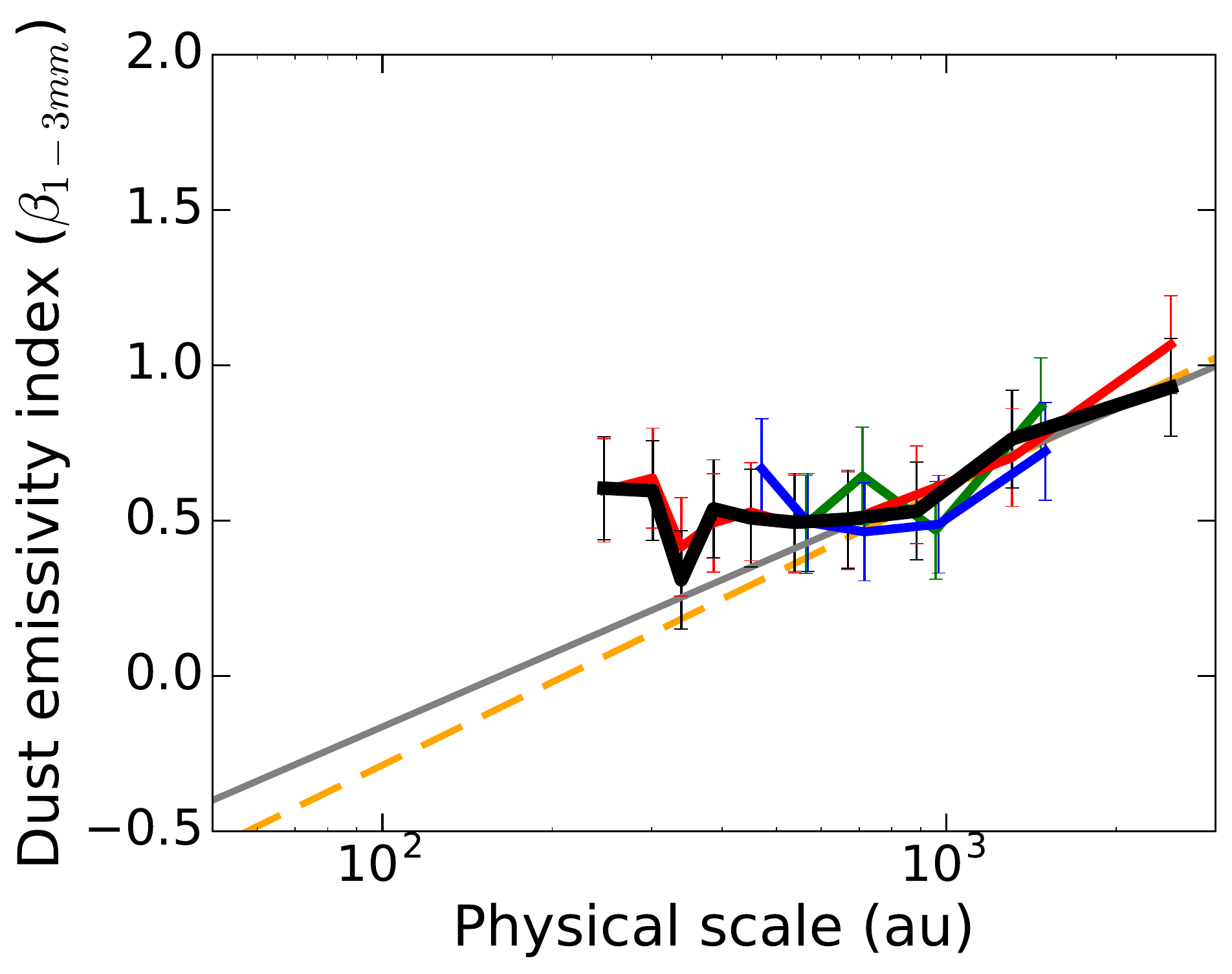} &
\vspace{-10pt} \includegraphics[width=5.8cm]{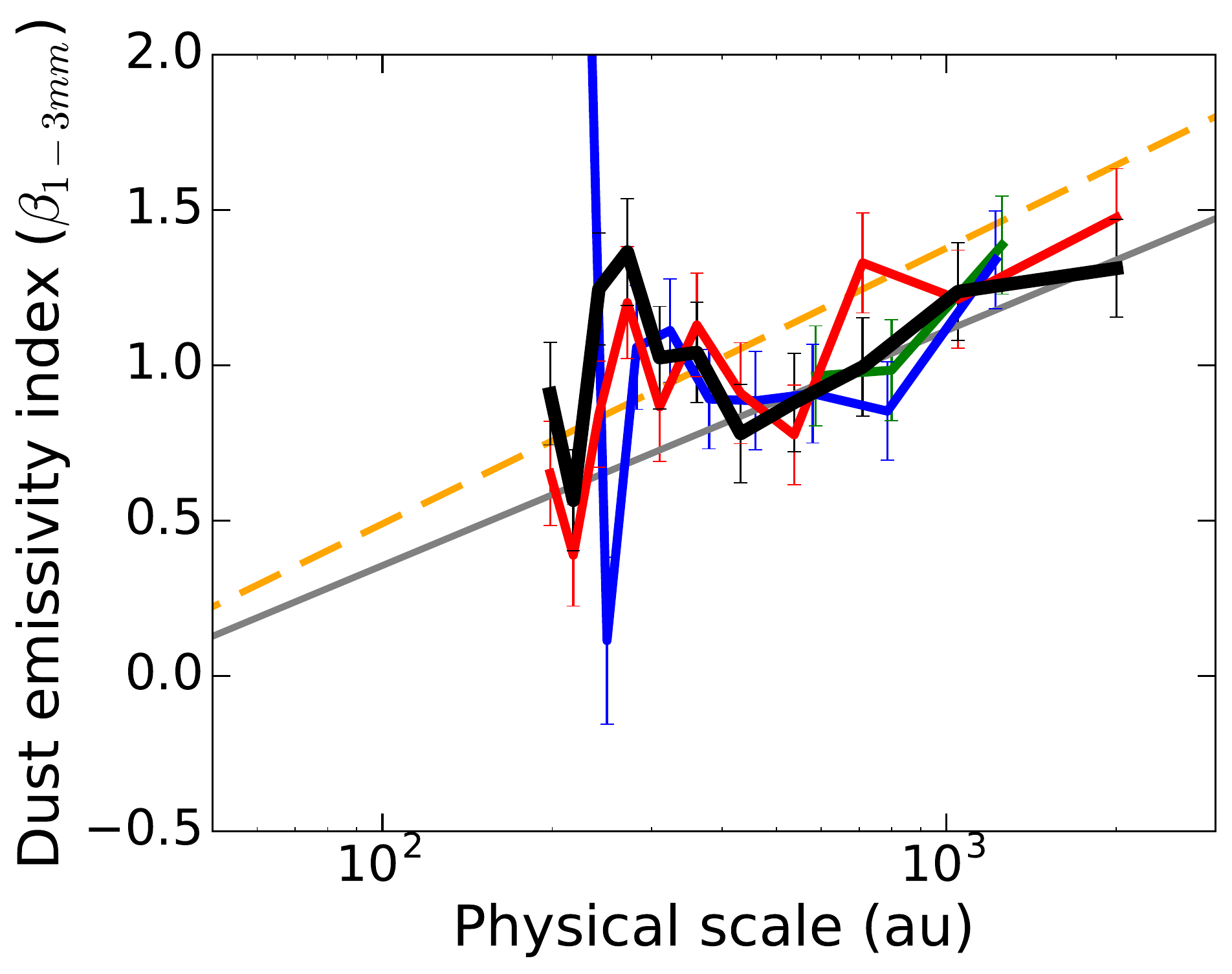} \\
\vspace{-245pt}\hspace{35pt}{\bf SerpM-S68N} &
\vspace{-245pt}\hspace{35pt}{\bf SerpM-SMM4} &
\vspace{-245pt}\hspace{35pt}{\bf SerpS-MM18} \\
&\vspace{-10pt} \includegraphics[width=5.8cm]{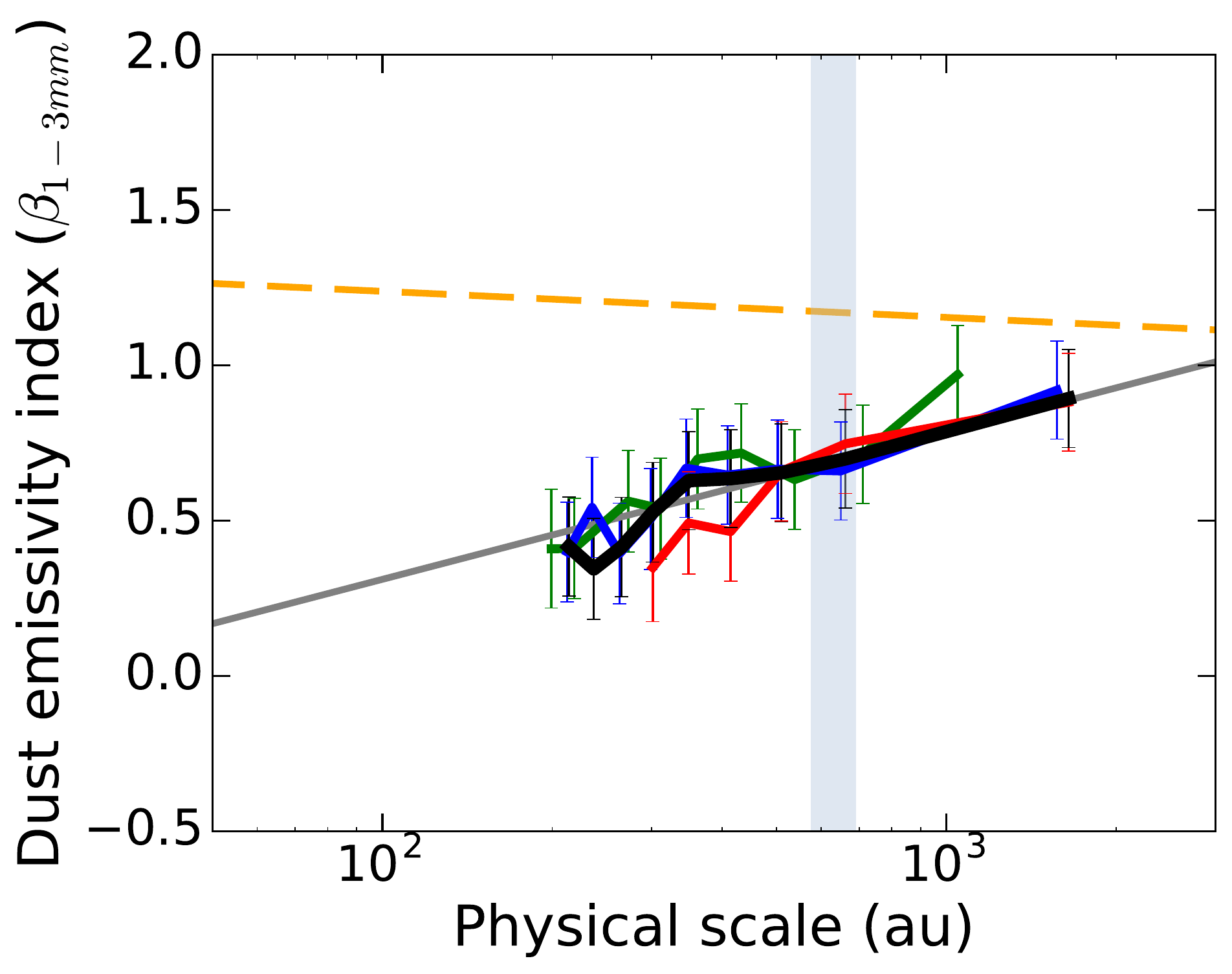} & \\
&\vspace{-245pt}\hspace{35pt}{\bf L1157} &\\
\end{tabular}
\vspace{-15pt}
\caption{Variation of the dust emissivity index $\beta_\mathrm{1-3mm}$ as a function of the physical scale integrated in the protostellar envelope. The thick black line indicates the trend when all directions are considered while the gray line indicates the (log-lin) fit to the trend (see Table~\ref{BetaTrendTable}). The colored lines indicate the trend in various angular bins with respect to the outflow direction. If PA is the position angle with respect to the outflow axis, the red lines indicate the trends for regions located close to the outflow direction (-22.5$^\circ<PA<22.5^\circ$).
The blue lines indicate the trends for regions located close to the 45$^\circ$ PA (22.5$^\circ<PA<67.5^\circ$ and -67.5$^\circ<PA<22.5^\circ$).
The green lines indicate the trends for regions located close to the equatorial plane (67.5$^\circ<PA<122.5^\circ$).
The orange dotted line indicates the fit to the $\beta_{1-3mm}$ values corrected for the central region (see Sect.~\ref{otc}). We do not apply the correction for L1448-C, as its 3 mm flux is dominated at most scales by the compact component, with barely any flux coming from the envelope, leading to large uncertainties in the $\beta_{1-3mm}$ determination. Finally, the light blue shaded line indicates the position of the CO snow line determined from the N$_2$H$^+$ and C$^{18}$O observations \citep{Gaudel2019,Anderl2016}. 
}
\label{RadialProfiles_individual}
\end{figure*}

\subsection{Dust emissivity index radial profiles}

In the Rayleigh-Jeans approximation (and assuming optically thin emission at millimeter wavelengths), $\alpha$ values translate to a dust emissivity index $\beta$~=~$\alpha$-2. This is, however, only a proxy for $\beta$ as the Rayleigh-Jeans approximation does not hold across protostellar envelopes because of their low dust temperatures. To offer a more robust approximation, we recalculate $\beta_\mathrm{1-3mm}$ assuming that our objects are centrally illuminated spherical envelopes in which the temperature varies with radius \citep{Terebey1993,Motte2001} as

\begin{equation}
T = 38~L_\mathrm{int}^{0.2}~\left(\frac{\rm r}{100~\mathrm{au}}\right)^{-0.4}
\label{equtemperature}
.\end{equation}

\noindent The protostellar internal luminosities $L_\mathrm{int}$ (in \lsun) are tabulated in Table~\ref{SourceCharacteristics}. We then use the temperature profiles to calculate the binned $\beta$ values using\begin{equation}
\beta = \frac{log((F(\nu_0)/B_{\nu_0}(T))/(F(\nu_1)/B_{\nu_1}(T)))}{log(\nu_0 / \nu_1)}
\label{equbeta}
,\end{equation}
\noindent where $\nu_0$ and $\nu_1$ are equal to 231 and 94 GHz,  respectively, and $B_{\nu}(T) $ is the Planck function. 
The relation linking the dust emissivity index $\beta$ to the observed spectral index $\alpha_\mathrm{1-3mm}$-2 is shown in Fig.~\ref{BetaVersusAlpha}. We immediately observe that the relation deviates from the 1:1 relation as one points toward colder regions (with $\alpha_\mathrm{1-3mm}$-2 underestimating $\beta$ at colder temperatures). This has a direct impact on the $\beta_\mathrm{1-3mm}$ radial gradients, steepening most of them. 
For a visual illustration, maps of $\beta_\mathrm{1-3mm}$ have been generated from the 1.3 mm and 3.2 mm dust continuum cleaned maps (see Table~\ref{BeamSizes} and Appendix \ref{BetaMaps}) but we performed the following analysis on the visibility datasets rather than on the maps. 

Figure \ref{RadialProfiles_individual} (black lines) shows how the dust emissivity index $\beta_\mathrm{1-3mm}$ varies within the envelopes for each object of our sample. Our $\beta_\mathrm{1-3mm}$ values can be reconnected with those presented in \citet{Bracco2017} on larger scales derived with {\it Herschel} and the Neel-IRAM-KID-Array (NIKA). These authors found that the emissivity $\beta$ decreases radially in the two protostellar cores studied in Taurus while staying flat in the pre-stellar core they analyzed; values of $\beta$=1.0-1.5 nicely match the range of $\beta_\mathrm{1-3mm}$ we observe in most of our sources beyond 2000 au. We provide the $\beta_\mathrm{1-3mm}$ values we derive at 500 au in Table~\ref{BetaTrendTable}. These values range from 0.38 to 0.92. We note that the typical value found in the diffuse ISM is $\beta$ $\sim$ 1.6 \citep{Planck2014,Juvela2015}. The $\beta_\mathrm{1-3mm}$ values observed at 500 au are thus extremely low for the whole sample. If absolute flux calibration affects the $\beta_\mathrm{1-3mm}$ measured, the fact that all sources present low $\beta_\mathrm{1-3mm}$ values would imply a systematically underestimated 1.3 mm flux (or overestimated 3.2 mm flux) that is not expected since the data comes from $>$ 30 independent tracks observed over a period of less than years. Phase noise could produce a loss of 1.3 mm flux at long baselines, but this does not affect our analysis since we limit ourselves to baselines $<$ 200 k$\lambda$, thus recovering the dust continuum emission at envelope scales.

\vspace{5pt}
Clear radial variations appear in the envelopes of several objects. To quantify these variations, we fit the $\beta_\mathrm{1-3mm}$ radius trend in lin-log scale and call `$\beta_\mathrm{1-3mm}$ gradient' the slope of the relation linking beta with the logarithm of the radius (in au) as follows: 
\begin{equation}
\beta_\mathrm{1-3mm} = A~log(r) + B\end{equation}

\noindent where A is the $\beta_\mathrm{1-3mm}$ gradient. With this definition, A = 2 indicates that $\beta_\mathrm{1-3mm}$ doubles its value over an order of magnitude in radius. Gradients are calculated using baselines below 100 k$\lambda$ to limit the noise present at longer baselines and only probe the dust emissivity variations at envelope scales (i.e., 300 to 4000 au, depending on the source distance). We provide the $\beta_\mathrm{1-3mm}$ gradients in Table~\ref{BetaTrendTable}. Some objects show flat $\beta_\mathrm{1-3mm}$ profiles, in particular IRAS4B, SVS13B, and L1157. In L1157, \citet{Kwon2015} found that the dust emissivity index changes with radius and has a mean $\beta_\mathrm{1-3mm}$ value of 0.76, but values as low as 0.1 at 80 k$\lambda$. We do not detect such low values at these scales. Our results are closer to those of \citet{Chiang2012} in that respect, namely that $\beta_\mathrm{1-3mm}$ does not vary much depending on the envelope radius in L1157.

\begin{table}[]
\centering
\caption{Emissivity index $\beta_\mathrm{1-3mm}$ at 500 au and radial gradients $^a$}
\begin{tabular}{cccc}
\hline
\vspace{-5pt}
&\\
Name    && $\beta_\mathrm{1-3mm}$ @ 500 au       & $\beta_\mathrm{1-3mm}$ gradient  \\ 
\vspace{-5pt}
&\\
\hline
\vspace{-5pt}
&\\
IRAS2A1&&        0.38 $\pm$       0.16 &       1.12 $\pm$       0.37 \\
&c.r.c&         0.82 $\pm$       0.17 &       0.86 $\pm$       0.37 \vspace{5pt}\\
IRAS4A1&&        0.44 $\pm$       0.16 &       1.21 $\pm$       0.36 \\
&c.r.c&         0.54 $\pm$       0.16 &       1.31 $\pm$       0.36 \vspace{5pt}\\
IRAS4B&&        0.64 $\pm$       0.16 &       0.40 $\pm$       0.36 \\
&c.r.c&         0.62 $\pm$       0.16 &       0.46 $\pm$       0.36 \vspace{5pt}\\
L1448-C&&       0.41 $\pm$       0.16 &       0.70 $\pm$       0.37 \\
&c.r.c&         -     &            -     \vspace{5pt}\\
SVS13B&&        0.51 $\pm$       0.16 &       0.48 $\pm$       0.37 \\
&c.r.c&         0.99 $\pm$       0.16 &       0.22 $\pm$       0.36 \vspace{5pt}\\
L1527&&         0.73 $\pm$       0.16 &       0.45 $\pm$       0.40 \\
&c.r.c&         1.41 $\pm$       0.16 &       0.18 $\pm$       0.39 \vspace{5pt}\\
SerpM-S68N&&    0.92 $\pm$       0.17 &       1.09 $\pm$       0.38 \\
&c.r.c&         0.66 $\pm$       0.27 &       1.47 $\pm$       0.44 \vspace{5pt}\\
SerpM-SMM4&&    0.50 $\pm$       0.16 &       0.79 $\pm$       0.36 \\
&c.r.c&         0.44 $\pm$       0.16 &       0.90 $\pm$       0.37 \vspace{5pt}\\
SerpS-MM18&&    0.81 $\pm$       0.16 &       0.77 $\pm$       0.37 \\
&c.r.c&         0.74 $\pm$       0.17 &       0.90 $\pm$       0.37 \vspace{5pt}\\
L1157&&         0.65 $\pm$       0.16 &       0.48 $\pm$       0.42 \\
&c.r.c&         1.17 $\pm$       0.18 &      -0.09 $\pm$       0.44 \vspace{5pt}\\
\hline
\end{tabular}
\begin{list}{}{}
\item[$^a$] $\beta_\mathrm{1-3mm}$ gradients are defined as A so that $\beta_\mathrm{1-3mm}$ = A~log(r) + B. $\beta_\mathrm{1-3mm}$ values at 500 au and gradients are estimated by generating 2000 sets with $\beta_\mathrm{1-3mm}$ values varying randomly within their individual error bars (following a normal distribution around their nominal value). Only baselines below 100 k$\lambda$ are taken into account for the calculation of the gradients. The central region correction (c.r.c)  rows indicate the values when the central regions are removed (see Sect.~\ref{otc}).  
\end{list}
\label{BetaTrendTable}
\end{table}

\subsection{Investigating dust emissivity variations with respect to source morphology}

The source physical properties (high densities in the equatorial plane, outflows) 
could have an impact on the grain populations and thus affect the structure and evolution of the central regions. 
To analyze potential local variations of the dust emissivity index, we chose to analyze the visibilities corresponding to different angular bins separately. The outflow opening angles are still poorly constrained and can vary significantly from one source to another. In \citet{Hsieh2017}, the median opening angle for the outflow of their low-mass objects is 42$^{\circ}$ (their Table 1). We thus chose to study the visibilities in 45$^o$ angular bins separately, namely {\it i)} close to the outflow direction, {\it ii)} close to the equatorial plane, or {\it iii)} close to the $\pm45^\circ$ PA.
The separation of the visibilities with respect to their uv position angle is performed using the \texttt{GILDAS/MAPPING}\footnote{\url{https://www.iram.fr/IRAMFR/GILDAS/doc/pdf/map.pdf}} software. The outflow position angles are provided in Table~\ref{SourceCharacteristics}. 
We plot the radial trends of the dust emissivity located in these different regions in Fig.~\ref{RadialProfiles_individual}. We did not perform the analysis for L1448-2A and L1448NB1 because they are close binaries driving nonaligned protostellar jets, which confuse the assumption of a simple geometry on the envelope for these sources. We observe that for most of the objects, the trends are similar in all directions. Our data therefore does not point to significantly different dust properties in the outflow direction or the equatorial plane, at least at the 80-2500 au scales we trace in this analysis. While this is surprising at first sight, we stress that such analysis should be repeated on sensitive data probing the dust continuum emission both at the very small and very large envelopes scales, so the outflow cavity walls can be detected and explored individually for example. A better knowledge of the outflow opening angles and orientation would help us refine our crude separation in outflow or equatorial plane regions.
We finally note that in L1157, the 3.2 mm emission is more extended than the 1.3 mm emission in the outflow direction \citep{Maury2019}, an elongation also observed at 2.7 mm with PdBI \citep{Beltran2004} and 3 mm with CARMA \citep{Chiang2010}\footnote{In both cases, the emission is resolved out by the interferometers and not linked with a beam elongation}. \citet{Gueth1997} suggested that the asymmetry could be linked with an extended component arising from the compressed outflow cavity edges (their Fig.~10). In the visibility plots, we observe that the dust emissivity index seems to be slightly lower in the outflow direction. The trend is also observable in the $\beta_\mathrm{1-3mm}$ map in Fig.~\ref{Maps} and could also be the signature of different dust properties in the outflow compared to the rest of the envelope for this source.

\section{Correcting for the central regions} \label{otc}

\subsection{Applying a sub-arcsecond component correction}

The central sub-arcsecond emission can affect the observed spectral index estimated at larger scales when the analysis is performed in the uv domain, as shown by the work of \citet{Miotello2014} in which unresolved components were shown to affect the emission at all uv distances. Their results are reinforced by the tests performed by our team using predictions from synthetic observations of MHD simulations that show that the $\alpha_\mathrm{1-3mm}$ values of the simulation are recovered for most of the envelope once our central region correction is applied (see Appendix \ref{SyntheticObservations}). If the brightness temperature profiles derived for our sources suggest that the Class 0 envelopes studied in this work are mostly optically thin at millimeter wavelengths (see Sect. 2.4 and Fig.\ref{Tb}), the low $\beta_\mathrm{1-3mm}$ values we observe could be due to unresolved optically thick emission in the central regions of some of our sources, but also partly driven by the presence of a central compact component with a different grain population and thus a different emissivity. Moreover, we saw in Sect.~\ref{ffc} that on small scales (beyond 200 k$\lambda$), the emission could be non-negligibly contaminated by nonthermal dust emission, especially at 3.2 mm, which would also bias our $\beta_\mathrm{1-3mm}$ values. To correct for these contaminations, we subtracted the $>$200 k$\lambda$ average amplitude from the shorter visibilities at both wavelengths to recalculate corrected $\beta_\mathrm{1-3mm}$ values. These average amplitudes are tabulated in Table~\ref{FluxCorrected}. 
Because of the asymmetry of the systems at small scales, we again did not perform the analysis for the two binary sources L1448-N and L1448-2A. We also did not include L1448-C whose 3 mm flux is dominated at most scales studied in this work by the compact component and for which our correction would lead to extremely uncertain $\beta_\mathrm{1-3mm}$ values. 
To take the uncertainties of the correction into account, the corrected $\beta_\mathrm{1-3mm}$ values at 500 au and gradients are estimated by generating 2000 sets of radial $\beta_\mathrm{1-3mm}$ values varying randomly within their individual error bars following a normal distribution around their nominal value. The corrected $\beta_\mathrm{1-3mm}$ values and gradients are provided in Table~\ref{BetaTrendTable}. The orange dashed lines in Fig.~\ref{RadialProfiles_individual} indicate the $\beta_\mathrm{1-3mm}$ corrected radial trends when the central region correction is applied.

\noindent Robustness of the correction - We performed a series of tests before choosing our final 200 k$\lambda$ threshold to ensure that our findings were robust. When we subtracted 150 k$\lambda$ average amplitude for instance, we find that most of the sources have the same $\beta_\mathrm{1-3mm}$ radial trends. Two sources have a modified slope compared to the 200 k$\lambda$ correction. SerpS-MM18 has a steeper slope for the 150 k$\lambda$ correction but the gradient stays within the 40\% error bar quoted in Table 2. The second source is L1527 for which a 150 k$\lambda$ correction leads to a flat beta profile. We note however that the error bar for this source is one of the largest of our sample: the flat $\beta$ gradient for the 200 k$\lambda$ correction quoted in Table 2 is also consistent with no gradient at all in this source, within the error bars.

\subsection{Impact on the $\beta_\mathrm{1-3mm}$ values and radial gradients} 

After correction, the $\beta_\mathrm{1-3mm}$ values at 500 au range from 0.4 to 1.2. The $\beta_\mathrm{1-3mm}$ values and radial gradients are not affected in the same way for all the sampled sources. Since a significant correction could be the sign of an optically thick compact dust emission, we first explored the link with the presence of a disk. \citet{Maury2019} have detected candidate disks in all but four CALYPSO sources (L1448-2A, L1448-NB1, IRAS4A1, and SerpM-S68N). The disk is resolved for IRAS4B, L1448-C, Serp-MM4, and L1527. In the two sources presenting the largest disk-like structures in the dust emission, namely Serp-MM4 and IRAS4B (radii of 290 and 125 au, respectively), we observe that the sub-arcsecond correction does not significantly affect our previous estimates. This suggests that there is not a significant difference between the dust properties in the envelope and those obtained integrating over the inner 1\arcsec\ region. On the contrary, in L1527 \citep[estimated disk radius of 54 au in][]{Maury2019}, the correction is relatively large, suggesting that the small disk could be partially optically thick or host more emissive (thus lower $\beta_\mathrm{1-3mm}$) grains than the envelope. The interpretation is very similar for IRAS2A1: the presence of an unresolved disk \citep[suggested by][]{Maury2019} with different grain properties could explain the lower $\beta_\mathrm{1-3mm}$ values derived before the correction is applied.
The 1.3 mm and 3.2 mm profiles of L1157 and the sub-arcsecond correction are very similar to those of L1527. Our results even suggest that the $\beta_\mathrm{1-3mm}$ gradient could be reversed once the correction is applied, i.e., where $\beta_\mathrm{1-3mm}$ values decrease toward larger scales, reconciling the gradient estimated in the uv domain with the $\beta_\mathrm{1-3mm}$ map derived in the L1157 envelope (see Appendix \ref{BetaMaps} and the previously mentioned 3.2 mm extension observed in the outflow direction). For sources with no disk detected in \citet{Maury2019}, namely SerpM-S68N and IRAS4A1, we find that those are only weakly affected by our central region correction.

\section {Analysis and discussion}

\begin{figure}
\begin{tabular}{m{8.5cm}}
\includegraphics[width=8.4cm]{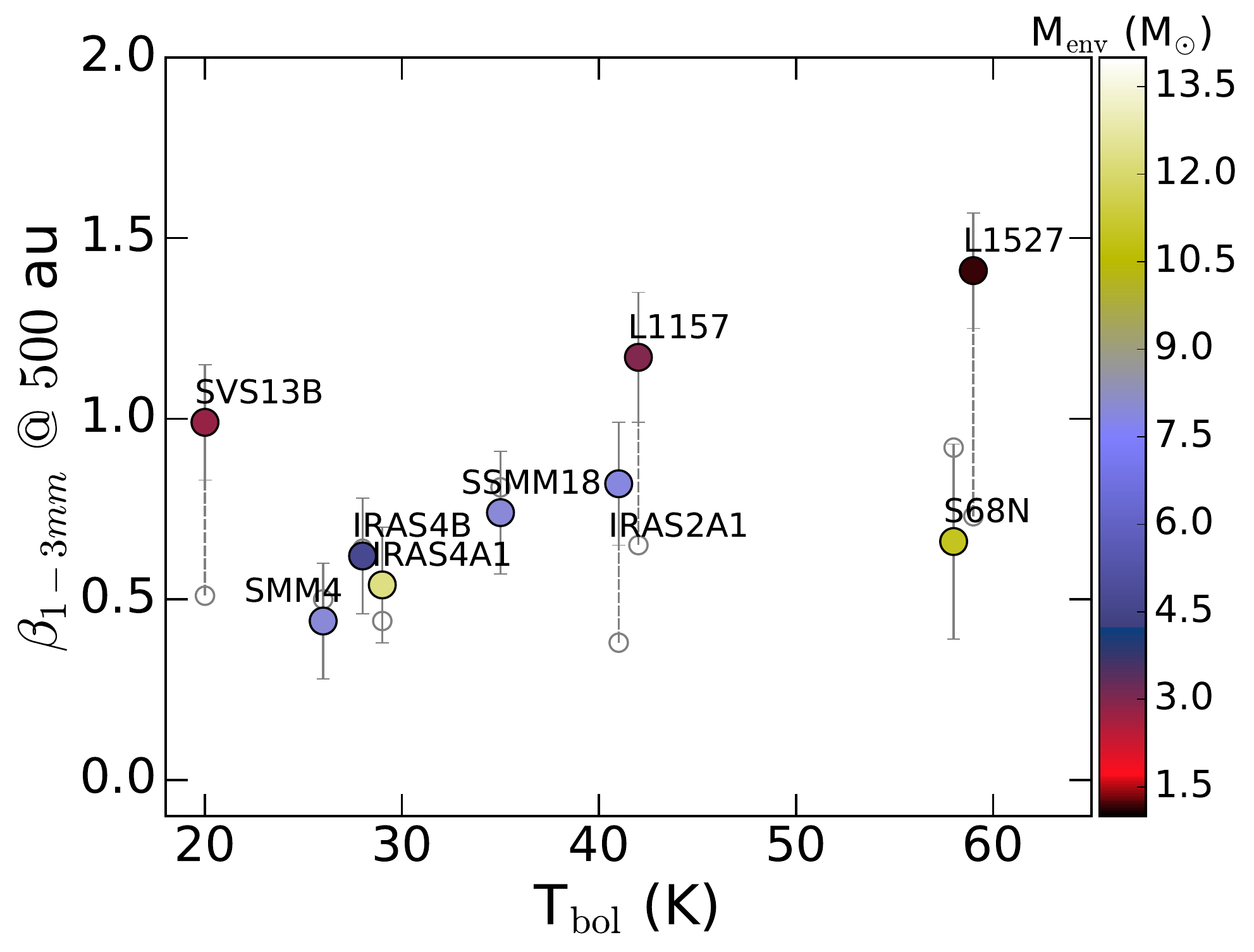} \\
\includegraphics[width=8.4cm]{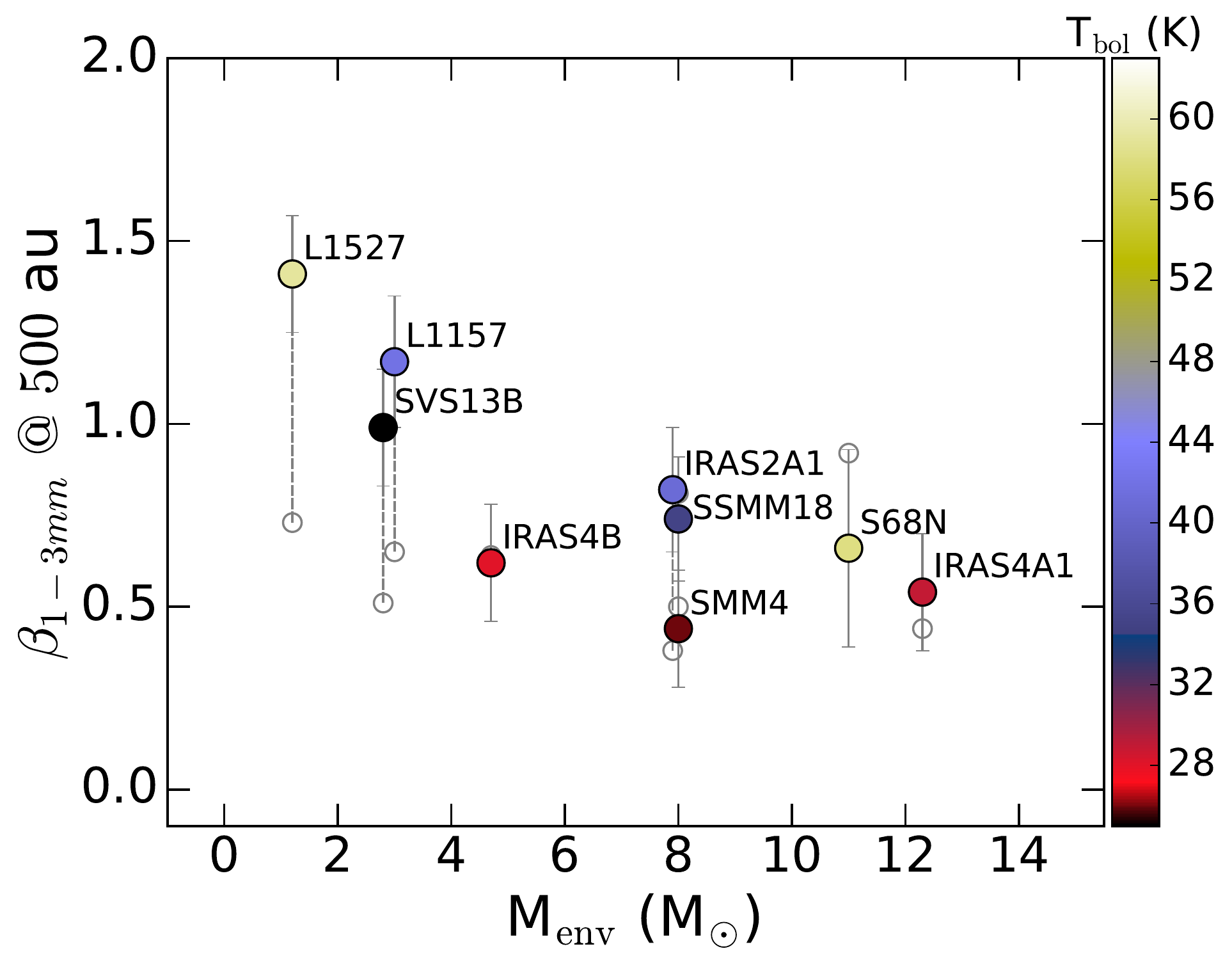} \\ 
\end{tabular}
\caption{Relation between the $\beta_\mathrm{1-3mm}$ value at 500 au and the bolometric temperature (top) and envelope mass (bottom).
The $\beta_\mathrm{1-3mm}$ values at 500 au are tabulated in Table~\ref{BetaTrendTable}. Colored/empty symbols represent 
the values after/before the central region correction (c.r.c). 
}
\label{Correlations1}
\end{figure}

\subsection{Emissivity index dependence on protostar properties}
\label{D1}

The observed $\beta_\mathrm{1-3mm}$ values stay low in most of our sources despite the central region correction. If the (corrected) $\beta_\mathrm{1-3mm}$ values at 80-2500 au scales mostly stay above 1 in the envelopes of L1157, L1527, and SerpM-S68N, the sources IRAS4A, IRAS4B, SVS13, or SerpM-SMM4 have, on average, $\beta_\mathrm{1-3mm}$ values below 1 up to 2000 au scales. No correlation was found between the average $\beta_\mathrm{1-3mm}$ value and age in a sample of Class I and II objects studied by \citet{Ricci2010}. We investigate other correlations in this work. In Fig.~\ref{Correlations1}, we compare the $\beta_\mathrm{1-3mm}$ value at 500 au with the bolometric temperatures $T_{\rm bol}$ for each object: the correlation is weak (Pearson correlation coefficient $R = 0.5$) between the two parameters. We note that this result is consistent with the absence of correlation found between the two parameters in the analysis of \citet{Froebrich2005}. Like these authors, we also find an anticorrelation between $\beta_\mathrm{1-3mm}$ and $M_\mathrm{env}$ ($R = -0.79$), which partly suggests smaller grains on average in protostars with smaller envelope mass (Fig.~\ref{Correlations1}). We note that $M_\mathrm{env}$ estimates strongly depend on the assumed distance and are impacted by the difficulty to separate the envelope from the parent cloud. The $M_\mathrm{env}$ estimates are also usually calculated assuming a fixed dust opacity; for instance with $\kappa_{1.3}$, the dust opacity at 1.3 mm is fixed to a commonly used value of 0.01 cm$^2$g$^{-1}$. This could contribute to the anticorrelation we observed between $M_\mathrm{env}$ (or N$_\mathrm{H2}$) and $\beta_\mathrm{500au}$.

\subsection{Possible steeper emissivity radial evolution in more massive envelopes}

\begin{figure}
\begin{tabular}{m{8.5cm}}
\includegraphics[width=8.4cm]{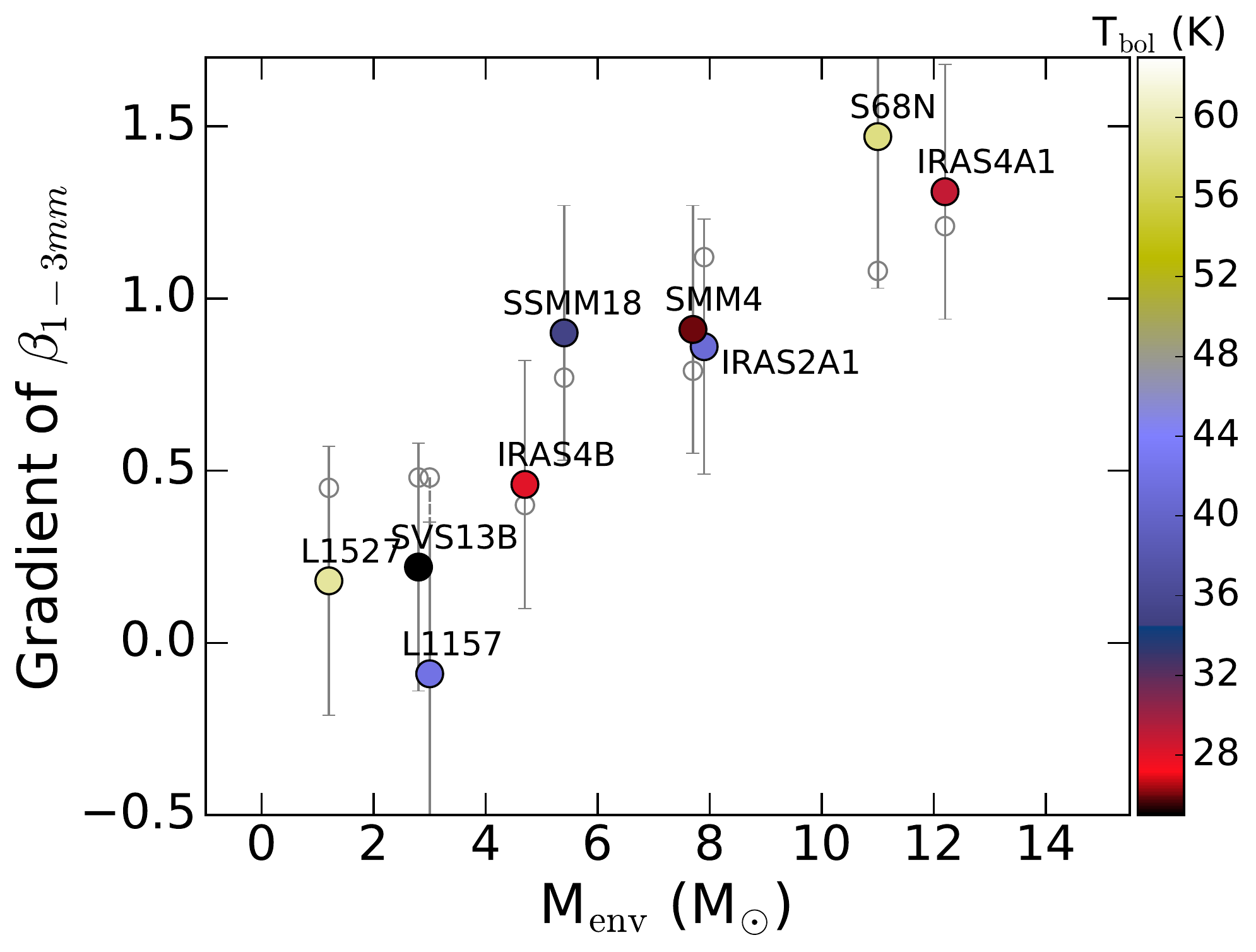} \\
\includegraphics[width=8.4cm]{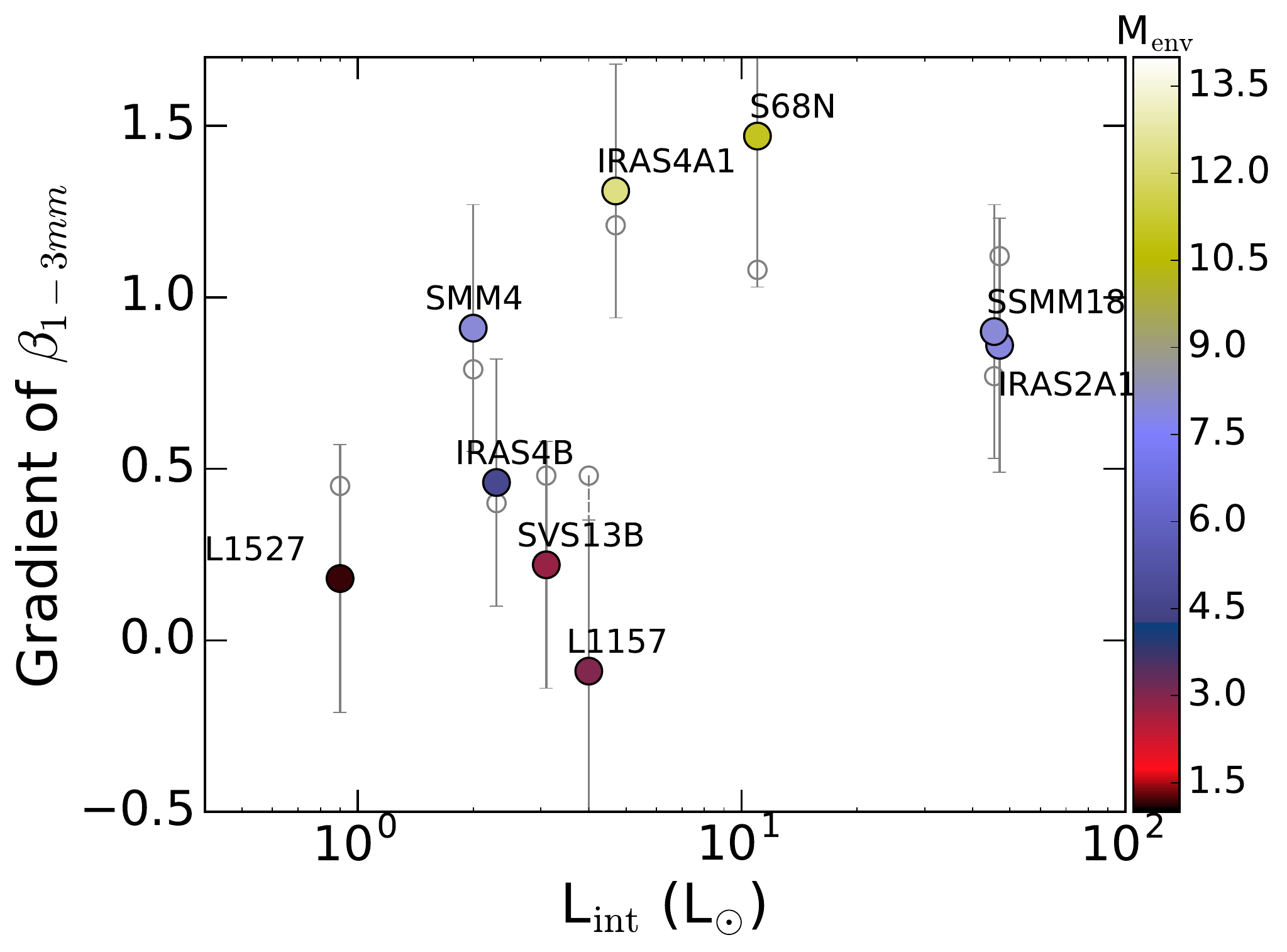} \\
\end{tabular}
\caption{Relation between $\beta_\mathrm{1-3mm}$ radial gradient and envelope mass (top) and internal luminosity (bottom). The $\beta_\mathrm{1-3mm}$ radial gradients 
are tabulated in Table~\ref{BetaTrendTable}. Colored/empty symbols represent the values after/before the central region correction. }
\label{Correlations2}
\end{figure}

As far as the $\beta_\mathrm{1-3mm}$ variations with envelope radius are concerned, IRAS2A1, IRAS4A1, SerpM-SMM4, SerpM-S68N, and SerpS-MM18 present a significant evolution of $\beta_\mathrm{1-3mm}$ in their envelopes. All these sources have a relatively large envelope mass ($>$ 6\msun). We note in particular that the two Serpens sources SerpS-MM18 and SerpM-SMM4 show the same $\beta_\mathrm{1-3mm}$ gradient. The sources IRAS4B, SVS13B, L1157, and L1527 have more moderate gradients ($<$ 0.5). Those sources have lower envelope masses than the previous group. In Fig.~\ref{Correlations2}, we show that the steepness of the gradient linking $\beta_\mathrm{1-3mm}$ to the envelope scales probed by our observations seems to be strongly correlated to the envelope mass of the object ($R = 0.93$) by over an order of magnitude. The relation suggests more homogeneous dust properties in the lowest mass sources of our sample. We note that the more evolved object L1527 presents an homogeneous $\beta_\mathrm{1-3mm}$ in its envelope. One explanation could be that the grain populations have been completely mixed over longer timescales. 

Investigating more potential relations with environmental properties, we do not find correlations between the $\beta_\mathrm{1-3mm}$ gradients with the internal luminosity (Fig.~\ref{Correlations2}); neither do we find steeper $\beta_\mathrm{1-3mm}$ gradients in objects presenting larger snowlines nor a correlation between $\beta_\mathrm{1-3mm}$ values and the position of the CO snowline itself \citep[Fig.~\ref{RadialProfiles_individual}[]{Gaudel2019,Anderl2016}. 

Complex organic molecules (COMs) are supposedly formed on the surfaces of dust grains and are released in the gas phase when the dust icy mantles sublimate \citep{Rawlings2013}. Belloche et al (in prep) performed a COM analysis for the CALYPSO sample, studying the abundance of molecules such as CH$_3$OH, C$_2$H$_5$OH, CH$_3$OCH$_3$, CH$_3$OCHO, CH$_3$CHO, NH$_2$CHO, CH$_3$CN, C$_2$H$_5$CN, a-(CH$_2$OH)$_2$, CH$_2$(OH)CHO, HNCO or NH$_2$CN. IRAS2A1, IRAS4B, and SerpS-MM18 have clear detections of most of the cited COMs but do not particularly share a common trend in terms of $\beta_\mathrm{1-3mm}$ values or gradients. Both SerpM-S68N and L1157 only have CH$_3$OH, CH$_3$OCHO, and CH$_3$CN detected but the former has the steepest $\beta_\mathrm{1-3mm}$ gradient while the latter the lowest. The absence of correlation between COM abundances and the radial variations of the dust emissivity index we observe thus suggests that the decrease of the emissivity index at small envelope radii is not driven by the loss of the icy mantles.

\begin{figure}
\begin{tabular}{m{8.5cm}}
\includegraphics[width=8.4cm]{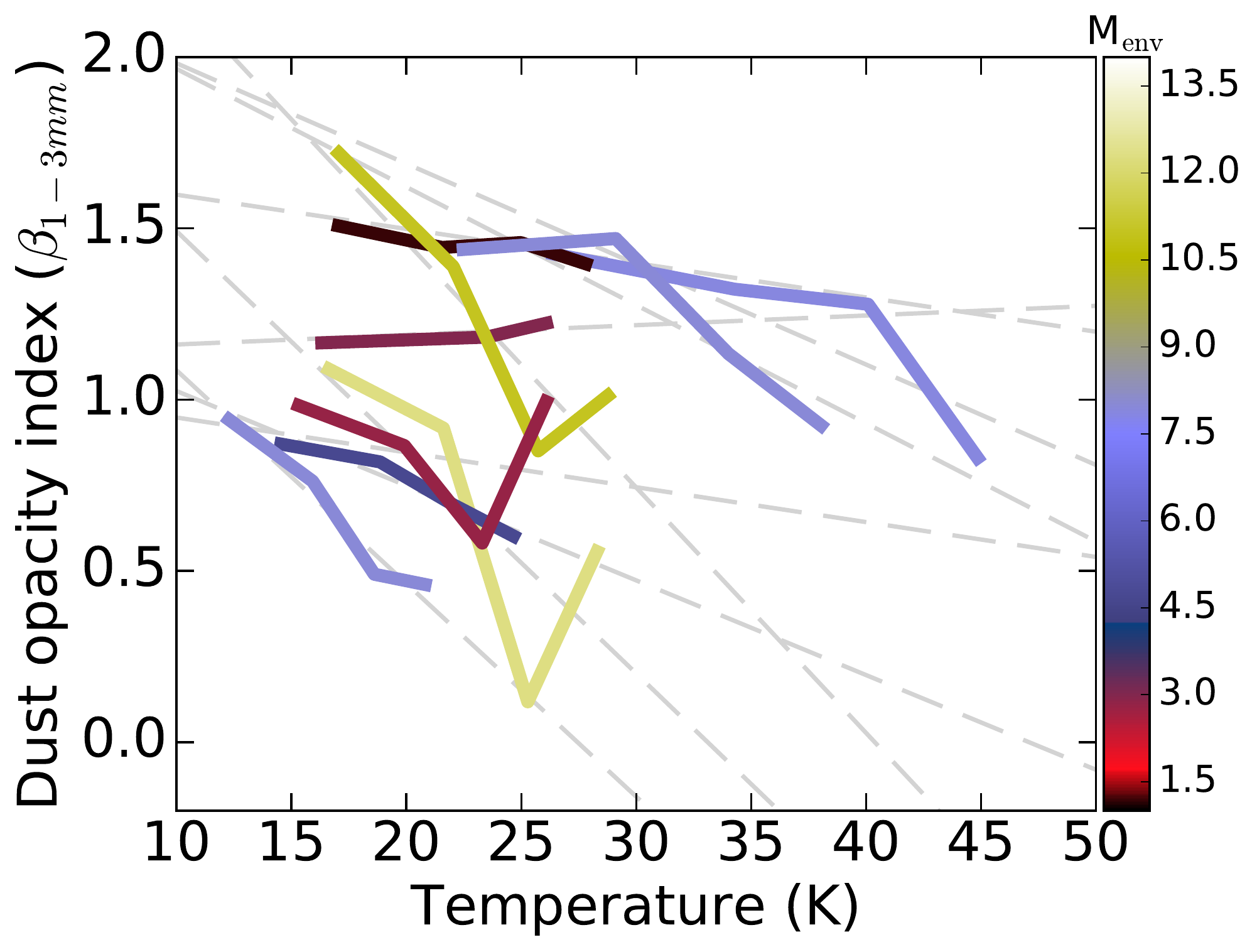} \\
\end{tabular}
\vspace{-10pt}
\caption{Local variation of the dust emissivity index $\beta_\mathrm{1-3mm}$ as a function of the temperature 
in the envelope of each source. Lines are color coded depending on the source envelope mass.  The dashed gray lines
indicate the fit to the relation for each source to guide the eye.}
\label{Beta_temperature}
\end{figure}

Finally, we plot in Fig.~\ref{Beta_temperature} the variation of $\beta_\mathrm{1-3mm}$ as a function of the dust temperature at the scale it is measured (Eq.~\ref{equtemperature}), allowing a source-to-source local comparison. The large range of $\beta_\mathrm{1-3mm}$ values obtained at similar dust temperatures across the sample envelopes suggests that the local dust temperature does not have a major influence on the dust emissivity itself.

\subsection{Millimeter-size grains: Grown in situ or transported}

Several sources have $\beta_\mathrm{1-3mm}$ values below 1. \citet{Miyake1993}, \citet{Ossenkopf1994}, or \citet{Draine2006} have shown that grain growth only could already produce $\beta_\mathrm{1-3mm}$ $\le$ 1 at millimeter wavelengths. A recent study by \citet{Ysard2019} has quantified the size distribution effects on the shape (peak, $\beta$) of the spectral energy distribution (SED) of dense interstellar regions including young stellar objects. Their results (see their Fig.~10) clearly show that the millimeter spectral indices decrease with an increasing size distribution. For most of the material mixtures probed, $\beta$ values $<$1 require grain sizes larger than 100 \mic.

Many questions remain regarding how these large grains can form in the youngest protostellar envelopes despite the long timescales and high densities required to build them. \citet{Ormel2009} showed that long support mechanisms are needed for coagulation to happen and impact the $\beta_\mathrm{1-3mm}$ observed in molecular clouds. These authors found that millimeter-size particles require ice-coated grains and coagulation times of 10$^7$ yr to be formed from ISM-like grains. This timescale shrinks when considering hydrogen number densities $>$ 10$^6$ cm$^{-3}$ (densities that are reached in our protostellar envelopes) or if the coagulation processes start on already grown grains, as expected to explain the coreshine effect observed in cold cores. The denser the medium, the stickier the grains: this enhances the coagulation processes in the inner envelopes during the protostellar collapse. 

More recently however, \citet{Wong2016} proposed new coagulation models and found that higher densities (n$_\mathrm{H}$ = 10$^{10}$ cm$^{-3}$) would be required to grow grains above 300 \mic\ in reasonable (a few free-fall time) timescales. To explain the presence of millimeter-size grains in the Class 0 envelopes, these authors suggested that large grains could be built up in the high-density central region (in the protostellar disk), then could be `launched' into the inner envelope via the gas dragged by the outflow. The transportation conditions would depend on many criteria: for instance, the central mass of the stellar embryo, i.e., the larger the mass, the less efficient the transportation; the gas velocity, i.e., low v$_\mathrm{gas}$ is needed to transport millimeter grains; the opening angle of the outflow; its mass loss rate; and  the sputtering and shattering rate, namely, the ability of a grain to survive gas friction and grain-grain collisions. 

Since the outflow momentum flux correlates well with the circumstellar envelope mass \citep{Bontemps1996}, the correlation we observe between $M_\mathrm{env}$ and the $\beta_\mathrm{1-3mm}$ gradient could be linked with stronger outflow energetics in high-mass protostellar envelopes, leading to a more efficient transport of large grains from the inner disk to the inner/intermediate envelope scales. 
Interestingly, the detection of strongly polarized (from a few percent to up 10\%) dust emission at millimeter wavelengths following precisely protostellar outflow cavity walls \citep{Maury2018,LeGouellec2019} suggests a population of large grains \citep{Valdivia2019} in these specific locations, which strengthens this scenario. The scales on which dust grains could be reinjected into the envelopes using this mechanism and outflow dynamics (gas velocity and grain motions) need to be constrained observationally, especially at the small scales at which the outflow is launched and the cavity develops. Current efforts in the theoretical aspects of these investigations, in particular the ongoing implementation of the dust grain dynamics into dusty collapse simulations \citep[e.g.,][]{Lebreuilly2019} along with further tests on the decoupling of the dust and gas and potential grains segregation sorting in the equatorial plane \citep{Bate2017}, might provide valuable predictions to carry out additional observational tests to constrain the full dust evolution from the core to disk scales.

\subsection{Grain composition effects}

The grain chemical composition is expected to vary during the collapse of protostellar envelopes as a result of, for example, grain-grain collisions and thermal desorption. Composition effects could thus also participate in the low $\beta$ values we observed at millimeter wavelengths. So far, the very low values of $\beta_\mathrm{1-3mm}$ ($<<$1) have been difficult to explain from optical properties derived in laboratory studies of interstellar dust analogs. \citet{Kohler2015} showed that additional carbonaceous mantles or ices would not lead to such low beta. Studies on magnesium-rich or iron-rich amorphous silicates \citep{Demyk2017,Demyk2017_b} have shown that the mass absorption coefficient strongly varies with wavelength but the derived millimeter $\beta$ values were still higher than 1. Recent work using the optical constants from the The Heterogeneous dust Evolution Model for Interstellar Solids \citep[THEMIS; ][]{Jones2013} dust model has highlighted the strong influence of the chemical composition on the emission of various mixtures of amorphous hydrogenated carbon grains and amorphous silicates \citep{Ysard2019}. Their study, however, shows that large grains seem to be a necessary condition to explain $\beta_{mm}$ $<$ 1.
\citet{Coupeaud2011} suggested that astrosilicate grains with a stoichiometry close to that of pyroxene could have $\beta$ $<1$ in the millimeter range, at which pyroxenes are more abundant, for a given pressure, at lower temperatures \citep{Kessler-Silacci2005}. Several studies of massive protostars and protostars from the Orion molecular cloud complex seem to confirm that their amorphous silicate population is more dominated by amorphous pyroxene than the typical ISM dust \citep{Demyk1999,Poteet2012}. The object-to-object variations of $\beta_\mathrm{1-3mm}$ could finally
be partly linked with variations in the grain porosity itself \citep[see for instance][]{Kataoka2014}. 
In conclusions, while a varying grain size distribution remains our favored hypothesis to explain the radial trends we observe, a larger spectrum of laboratory experiments would help predict the observational signatures expected for a variety of interstellar dust analogs and from the collisions of these grains. More observations of the dust polarized emission (and their use as a constraint for polarized dust models) could also provide important additional information on the dust composition \citep{Guillet2018}.

 \subsection{Other potential caveats}
 
Studies of 3 mm dust continuum measurements in nearby filaments \citep{Schnee2014,Mason2019} recently suggested that the emission at 3 mm deviates from dust emission extrapolated from a simple modified blackbody SED fitted to shorter wavelengths. The emission is inconsistent with free-free emission or spinning dust models and the enhanced emission at 3 mm has been attributed to the presence of amorphous dust grains. The question regarding whether 3 mm dust continuum emission can be safely used to study grain size variations in protostellar environments remains, however, open. Observations at additional (sub)mm wavelengths but also at higher resolution would enable us to sample better the local submillimeter SEDs  to disentangle different grain models. Future James Webb Space Telescope (JWST) instruments will also provide the opportunity to observe gases, silicate features \citep[e.g., analyzing the 10-to-18 \mic\ ratio as done in][]{Demyk1999}, or ices in absorption, probing more deeply into the warm interior of the protostellar envelopes.

\citet{GalvanMadrid2018} have recently explored the effects of self-obscuration in protostellar disks with a radially decreasing temperature gradient in producing extremely low spectral indices. Given that most of the emission arising from the Class 0 protostellar envelopes studied in this work is optically thin, our results are probably not affected by these self-obscuration effects. 

Finally, our team is currently producing a series of synthetic observations to quantify the impact of various parameters (inclination, size distribution, luminosity, and envelope mass) on the observed millimeter continuum and polarized emission (Valdivia et al. in prep). This work has also been initiated to investigate the instrumental effects linked with interferometric observations and get a better handle on whether they could partly bias the $\beta_\mathrm{1-3mm}$ trends we observe.


\section{Conclusions}

We study the dust properties in a sample of 12 Class 0 protostellar envelopes observed at 94 and 231 GHz with PdBI as part of the CALYPSO survey. We perform our analysis in the visibility domain. We derive dust emissivity index profiles as a function of the envelope scales probed. Since the central arcsecond emission can affect the observed spectral index estimated at larger scales, we correct the visibility profiles for this potential contamination. Our analysis leads to the following conclusions:

\vspace{5pt}
\noindent {\it (i)} 
Most of the Class 0 envelopes present low dust emissivity indices $\beta_\mathrm{1-3mm}$: the observed $\beta_\mathrm{1-3mm}$ often reach values lower than 1, in particular in the envelopes of IRAS4A, IRAS4B, SVS13B, and SerpM-SMM4. Many sources also present a strong decrease of $\beta_\mathrm{1-3mm}$ toward small envelope radii. Low values of $\beta_\mathrm{1-3mm}$ are found preferentially in sources presenting these large radial variations. Both results suggest that grain growth could already be at work in Class 0 inner envelopes.

\vspace{5pt}
\noindent {\it (ii)} The steepness of the $\beta_\mathrm{1-3mm}$ gradient depends strongly on the envelope mass, suggesting less grain evolution in low-mass objects and a more diverse grain mixture from the outer to the inner regions in more massive envelopes. 

\vspace{5pt}
\noindent {\it (iii)} If accretion and coagulation models predict the presence of large grains in protostellar envelopes, the timescales required seem to be too large for millimeter-size grains to be grown directly in the envelopes from ISM-like grains. Potential transportation mechanisms of large grains formed in the high-density inner envelopes and protostellar disks back to the intermediate scales of the envelopes might  explain the observations and should be observationally investigated in the future.

\vspace{5pt}
\noindent {\it (iv)} Variations in the grain composition, stoichiometry, or porosity also strongly influence the dust emission at millimeter wavelengths and could partially contribute to the radial trends observed at envelope scales. Recent predictions from a range of interstellar grain mixture analogs suggest however that a certain level of grain growth would still be necessary to explain $\beta_\mathrm{1-3mm}$ values $<$1 \citep{Ysard2019}.

\vspace{5pt}
Higher resolution observations, with JWST and ALMA, will help probe deeper into the core and connect the dust properties down to the disk/envelope transition to analyze locally if observations fit with the different accretion, coagulation, and transport model predictions.

\section*{Acknowledgments}
We would first like to thank the referee for  useful comments that help improved 
the clarity of the paper and the robustness of the results.
We thank Beno\^{i}t Commer\c{c}on, Karin Demyk, Ugo Lebreuilly, Nathalie Ysard, and Vincent Guillet 
for many fruitful discussions on the analysis, results, and interpretation, and Jean-Philippe Berger
for his help on interferometry related questions. We thank Patrick Hennebelle for producing the MHD 
simulations used to test our correction method on synthetic observations. 
We also thank the CALYPSO collaboration, in particular S\'{e}bastien Maret for our discussions on the source distances.
This project has received funding from the European Research Council (ERC) 
under the European Union Horizon 2020 research and innovation programme 
(MagneticYSOs project, grant agreement N$\degr$ 679937). LT acknowledges support 
from the Italian Ministero dell' Istruzione, Universit\`{a} e Ricerca through the 
grant Progetti Premiali 2012 - iALMA (CUP C52I13000140001), by the Deutsche 
Forschungs-gemeinschaft (DFG, German Research Foundation) - Ref no. FOR 2634/1 TE 1024/1-1, 
by the DFG cluster of excellence `Origins' (\url{https://www.origins-cluster.de/}), and 
from the European Union's Horizon 2020 research and innovation programme under the 
Marie Sklodowska-Curie grant agreement N$\degr$823823 (RISE DUSTBUSTERS project).

\vspace{-10pt}
\bibliographystyle{aa.bst}
\bibliography{/Users/mgalamet/Documents/Work/Papers/mybiblio_MY.bib}

\begin{thebibliography}{114}
\expandafter\ifx\csname natexlab\endcsname\relax\def\natexlab#1{#1}\fi

\bibitem[{{Adams}(1991)}]{Adams1991}
{Adams}, F.~C. 1991, \apj, 382, 544

\bibitem[{{Agurto-Gangas} {et~al.}(2019){Agurto-Gangas}, {Pineda}, {Szucs},
  {Testi}, {Tazzari}, {Miotello}, {Caselli}, {Dunham}, {Stephens}, \&
  {Bourke}}]{Agurto2019}
{Agurto-Gangas}, C., {Pineda}, J.~E., {Szucs}, L., {et~al.} 2019, arXiv:
  1901.05021

\bibitem[{{Anderl} {et~al.}(2016){Anderl}, {Maret}, {Cabrit}, {Belloche},
  {Maury}, {Andr{\'e}}, {Codella}, {Bacmann}, {Bontemps}, {Podio}, {Gueth}, \&
  {Bergin}}]{Anderl2016}
{Anderl}, S., {Maret}, S., {Cabrit}, S., {et~al.} 2016, \aap, 591, A3

\bibitem[{{Andersen} {et~al.}(2013){Andersen}, {Steinacker}, {Thi}, {Pagani},
  {Bacmann}, \& {Paladini}}]{Andersen2013}
{Andersen}, M., {Steinacker}, J., {Thi}, W.~F., {et~al.} 2013, \aap, 559, A60

\bibitem[{{Andr\'{e}}(1996)}]{Andre1996}
{Andr\'{e}}, P. 1996, in Astronomical Society of the Pacific Conference Series,
  Vol.~93, Radio Emission from the Stars and the Sun, ed. A.~R. {Taylor} \&
  J.~M. {Paredes}, 273--284

\bibitem[{{Andr{\'e}} {et~al.}(2010){Andr{\'e}}, {Men'shchikov}, {Bontemps},
  {K{\"o}nyves}, {Motte}, {Schneider}, {Didelon}, {Minier}, {Saraceno}, \&
  {Ward-Thompson}}]{Andre2010}
{Andr{\'e}}, P., {Men'shchikov}, A., {Bontemps}, S., {et~al.} 2010, \aap, 518,
  L102

\bibitem[{{Andre} {et~al.}(1993){Andre}, {Ward-Thompson}, \&
  {Barsony}}]{Andre1993}
{Andre}, P., {Ward-Thompson}, D., \& {Barsony}, M. 1993, \apj, 406, 122

\bibitem[{{Andr\'{e}} {et~al.}(2000){Andr\'{e}}, {Ward-Thompson}, \&
  {Barsony}}]{Andre2000}
{Andr\'{e}}, P., {Ward-Thompson}, D., \& {Barsony}, M. 2000, Protostars and
  Planets IV, 59

\bibitem[{{Anglada} {et~al.}(1996){Anglada}, {Rodriguez}, \&
  {Torrelles}}]{Anglada1996}
{Anglada}, G., {Rodriguez}, L.~F., \& {Torrelles}, J.~M. 1996, \apj, 473, L123

\bibitem[{{Bate} \& {Lor{\'e}n-Aguilar}(2017)}]{Bate2017}
{Bate}, M.~R. \& {Lor{\'e}n-Aguilar}, P. 2017, \mnras, 465, 1089

\bibitem[{{Bazell} \& {Dwek}(1990)}]{Bazell1990}
{Bazell}, D. \& {Dwek}, E. 1990, \apj, 360, 142

\bibitem[{{Beckwith} \& {Sargent}(1991)}]{BeckwithSargent1991}
{Beckwith}, S.~V.~W. \& {Sargent}, A.~I. 1991, \apj, 381, 250

\bibitem[{{Beltr{\'a}n} {et~al.}(2004){Beltr{\'a}n}, {Gueth}, {Guilloteau}, \&
  {Dutrey}}]{Beltran2004}
{Beltr{\'a}n}, M.~T., {Gueth}, F., {Guilloteau}, S., \& {Dutrey}, A. 2004,
  \aap, 416, 631

\bibitem[{{Bontemps} {et~al.}(1996){Bontemps}, {Andre}, {Terebey}, \&
  {Cabrit}}]{Bontemps1996}
{Bontemps}, S., {Andre}, P., {Terebey}, S., \& {Cabrit}, S. 1996, \aap, 311,
  858

\bibitem[{{Bracco} {et~al.}(2017){Bracco}, {Palmeirim}, {Andr{\'e}}, {Adam},
  {Ade}, {Bacmann}, {Beelen}, {Beno{\^\i}t}, {Bideaud}, {Billot}, {Bourrion},
  {Calvo}, {Catalano}, {Coiffard}, {Comis}, {D'Addabbo}, {D{\'e}sert},
  {Didelon}, {Doyle}, {Goupy}, {K{\"o}nyves}, {Kramer}, {Lagache}, {Leclercq},
  {Mac{\'\i}as-P{\'e}rez}, {Maury}, {Mauskopf}, {Mayet}, {Monfardini}, {Motte},
  {Pajot}, {Pascale}, {Peretto}, {Perotto}, {Pisano}, {Ponthieu},
  {Rev{\'e}ret}, {Rigby}, {Ritacco}, {Rodriguez}, {Romero}, {Roy}, {Ruppin},
  {Schuster}, {Sievers}, {Triqueneaux}, {Tucker}, \& {Zylka}}]{Bracco2017}
{Bracco}, A., {Palmeirim}, P., {Andr{\'e}}, P., {et~al.} 2017, \aap, 604, A52

\bibitem[{{Chac{\'o}n-Tanarro} {et~al.}(2017){Chac{\'o}n-Tanarro}, {Caselli},
  {Bizzocchi}, {Pineda}, {Harju}, {Spaans}, \&
  {D{\'e}sert}}]{Chacon-Tanarro2017}
{Chac{\'o}n-Tanarro}, A., {Caselli}, P., {Bizzocchi}, L., {et~al.} 2017, \aap,
  606, A142

\bibitem[{{Chac{\'o}n-Tanarro} {et~al.}(2019){Chac{\'o}n-Tanarro}, {Pineda},
  {Caselli}, {Bizzocchi}, {Gutermuth}, {Mason}, {Gomez-Ruiz}, {Harju},
  {Devlin}, {Dicker}, {Mroczkowski}, {Romero}, {Sievers}, {Stanchfield},
  {Offner}, \& {Sanchez-Arguelles}}]{Chacon-Tanarro2019}
{Chac{\'o}n-Tanarro}, A., {Pineda}, J.~E., {Caselli}, P., {et~al.} 2019, arXiv
  e-prints, arXiv:1901.02476

\bibitem[{{Chen} {et~al.}(2016){Chen}, {Di Francesco}, {Johnstone}, {Sadavoy},
  {Hatchell}, {Mottram}, {Kirk}, {Buckle}, {Berry}, {Broekhoven- Fiene},
  {Currie}, {Fich}, {Jenness}, {Nutter}, {Pattle}, {Pineda}, {Quinn}, {Salji},
  {Tisi}, {Hogerheijde}, {Ward-Thompson}, {Bastien}, {Bresnahan}, {Butner},
  {Chrysostomou}, {Coude}, {Davis}, {Drabek-Maunder}, {Duarte-Cabral}, {Fiege},
  {Friberg}, {Friesen}, {Fuller}, {Graves}, {Greaves}, {Gregson}, {Holland},
  {Joncas}, {Kirk}, {Knee}, {Mairs}, {Marsh}, {Matthews}, {Moriarty-Schieven},
  {Mowat}, {Pezzuto}, {Rawlings}, {Richer}, {Robertson}, {Rosolowsky},
  {Rumble}, {Schneider-Bontemps}, {Thomas}, {Tothill}, {Viti}, {White},
  {Wouterloot}, {Yates}, \& {Zhu}}]{Chen2016}
{Chen}, M. C.-Y., {Di Francesco}, J., {Johnstone}, D., {et~al.} 2016, \apj,
  826, 95

\bibitem[{{Chiang} {et~al.}(2012){Chiang}, {Looney}, \& {Tobin}}]{Chiang2012}
{Chiang}, H.-F., {Looney}, L.~W., \& {Tobin}, J.~J. 2012, \apj, 756, 168

\bibitem[{{Chiang} {et~al.}(2010){Chiang}, {Looney}, {Tobin}, \&
  {Hartmann}}]{Chiang2010}
{Chiang}, H.-F., {Looney}, L.~W., {Tobin}, J.~J., \& {Hartmann}, L. 2010, \apj,
  709, 470

\bibitem[{{Chini} {et~al.}(1997){Chini}, {Reipurth}, {Sievers},
  {Ward-Thompson}, {Haslam}, {Kreysa}, \& {Lemke}}]{Chini1997}
{Chini}, R., {Reipurth}, B., {Sievers}, A., {et~al.} 1997, \aap, 325, 542

\bibitem[{{Coupeaud} {et~al.}(2011){Coupeaud}, {Demyk}, {Meny}, {Nayral},
  {Delpech}, {Leroux}, {Depecker}, {Creff}, {Brubach}, \& {Roy}}]{Coupeaud2011}
{Coupeaud}, A., {Demyk}, K., {Meny}, C., {et~al.} 2011, \aap, 535, A124

\bibitem[{{Curiel} {et~al.}(1990){Curiel}, {Raymond}, {Moran}, {Rodriguez}, \&
  {Canto}}]{Curiel1990}
{Curiel}, S., {Raymond}, J.~C., {Moran}, J.~M., {Rodriguez}, L.~F., \& {Canto},
  J. 1990, \apjl, 365, L85

\bibitem[{{D'Alessio} {et~al.}(2001){D'Alessio}, {Calvet}, \&
  {Hartmann}}]{DAlessio2001}
{D'Alessio}, P., {Calvet}, N., \& {Hartmann}, L. 2001, \apj, 553, 321

\bibitem[{{Demyk} {et~al.}(1999){Demyk}, {Jones}, {Dartois}, {Cox}, \&
  {D'Hendecourt}}]{Demyk1999}
{Demyk}, K., {Jones}, A.~P., {Dartois}, E., {Cox}, P., \& {D'Hendecourt}, L.
  1999, \aap, 349, 267

\bibitem[{{Demyk} {et~al.}(2017{\natexlab{a}}){Demyk}, {Meny}, {Leroux},
  {Depecker}, {Brubach}, {Roy}, {Nayral}, {Ojo}, \& {Delpech}}]{Demyk2017_b}
{Demyk}, K., {Meny}, C., {Leroux}, H., {et~al.} 2017{\natexlab{a}}, \aap, 606,
  A50

\bibitem[{{Demyk} {et~al.}(2017{\natexlab{b}}){Demyk}, {Meny}, {Lu},
  {Papatheodorou}, {Toplis}, {Leroux}, {Depecker}, {Brubach}, {Roy}, {Nayral},
  {Ojo}, {Delpech}, {Paradis}, \& {Gromov}}]{Demyk2017}
{Demyk}, K., {Meny}, C., {Lu}, X.~H., {et~al.} 2017{\natexlab{b}}, \aap, 600,
  A123

\bibitem[{{Draine}(2006)}]{Draine2006}
{Draine}, B.~T. 2006, \apj, 636, 1114

\bibitem[{{Dupac} {et~al.}(2003){Dupac}, {Bernard}, {Boudet}, {Giard},
  {Lamarre}, {M{\'e}ny}, {Pajot}, {Ristorcelli}, {Serra}, {Stepnik}, \&
  {Torre}}]{Dupac2003}
{Dupac}, X., {Bernard}, J.-P., {Boudet}, N., {et~al.} 2003, \aap, 404, L11

\bibitem[{{Eiroa} {et~al.}(2005){Eiroa}, {Torrelles}, {Curiel}, \&
  {Djupvik}}]{Eiroa2005}
{Eiroa}, C., {Torrelles}, J.~M., {Curiel}, S., \& {Djupvik}, A.~A. 2005, \aj,
  130, 643

\bibitem[{{Enoch} {et~al.}(2011){Enoch}, {Corder}, {Duch{\^e}ne}, {Bock},
  {Bolatto}, {Culverhouse}, {Kwon}, {Lamb}, {Leitch}, {Marrone}, {Muchovej},
  {P{\'e}rez}, {Scott}, {Teuben}, {Wright}, \& {Zauderer}}]{Enoch2011}
{Enoch}, M.~L., {Corder}, S., {Duch{\^e}ne}, G., {et~al.} 2011, \apjs, 195, 21

\bibitem[{{Enoch} {et~al.}(2009){Enoch}, {Evans}, {Sargent}, \&
  {Glenn}}]{Enoch2009}
{Enoch}, M.~L., {Evans}, II, N.~J., {Sargent}, A.~I., \& {Glenn}, J. 2009,
  \apj, 692, 973

\bibitem[{{Facchini} {et~al.}(2017){Facchini}, {Birnstiel}, {Bruderer}, \& {van
  Dishoeck}}]{Facchini2017}
{Facchini}, S., {Birnstiel}, T., {Bruderer}, S., \& {van Dishoeck}, E.~F. 2017,
  \aap, 605, A16

\bibitem[{{Froebrich}(2005)}]{Froebrich2005}
{Froebrich}, D. 2005, \apjs, 156, 169

\bibitem[{{Fromang} {et~al.}(2006){Fromang}, {Hennebelle}, \&
  {Teyssier}}]{Fromang2006}
{Fromang}, S., {Hennebelle}, P., \& {Teyssier}, R. 2006, \aap, 457, 371

\bibitem[{{Galv{\'a}n-Madrid} {et~al.}(2018){Galv{\'a}n-Madrid}, {Liu},
  {Izquierdo}, {Miotello}, {Zhao}, {Carrasco-Gonz{\'a}lez}, {Lizano}, \&
  {Rodr{\'\i}guez}}]{GalvanMadrid2018}
{Galv{\'a}n-Madrid}, R., {Liu}, H.~B., {Izquierdo}, A.~F., {et~al.} 2018, \apj,
  868, 39

\bibitem[{{Gaudel} {et~al.}(2019){Gaudel}, {Maury}, {Belloche}, {Maret},
  {Andr\'{e}}, {Hennebelle}, {Galametz}, {Testi}, {Palmeirim}, {Ladjelate},
  {Cabrit}, {Codella}, \& {Podio}}]{Gaudel2019}
{Gaudel}, M., {Maury}, A.~J., {Belloche}, A., {et~al.} 2019, submitted to A\&A

\bibitem[{{Gueth} {et~al.}(1997){Gueth}, {Guilloteau}, {Dutrey}, \&
  {Bachiller}}]{Gueth1997}
{Gueth}, F., {Guilloteau}, S., {Dutrey}, A., \& {Bachiller}, R. 1997, \aap,
  323, 943

\bibitem[{{Guillet} {et~al.}(2018){Guillet}, {Fanciullo}, {Verstraete},
  {Boulanger}, {Jones}, {Miville-Desch{\^e}nes}, {Ysard}, {Levrier}, \&
  {Alves}}]{Guillet2018}
{Guillet}, V., {Fanciullo}, L., {Verstraete}, L., {et~al.} 2018, \aap, 610, A16

\bibitem[{{Harvey} {et~al.}(2003){Harvey}, {Wilner}, {Myers}, \&
  {Tafalla}}]{Harvey2003}
{Harvey}, D.~W.~A., {Wilner}, D.~J., {Myers}, P.~C., \& {Tafalla}, M. 2003,
  \apj, 596, 383

\bibitem[{{Hsieh} {et~al.}(2017){Hsieh}, {Lai}, \& {Belloche}}]{Hsieh2017}
{Hsieh}, T.-H., {Lai}, S.-P., \& {Belloche}, A. 2017, \aj, 153, 173

\bibitem[{{Jones} {et~al.}(2013){Jones}, {Fanciullo}, {K{\"o}hler},
  {Verstraete}, {Guillet}, {Bocchio}, \& {Ysard}}]{Jones2013}
{Jones}, A.~P., {Fanciullo}, L., {K{\"o}hler}, M., {et~al.} 2013, \aap, 558,
  A62

\bibitem[{{J{\o}rgensen} {et~al.}(2007){J{\o}rgensen}, {Bourke}, {Myers}, {Di
  Francesco}, {van Dishoeck}, {Lee}, {Ohashi}, {Sch{\"o}ier}, {Takakuwa},
  {Wilner}, \& {Zhang}}]{Jorgensen2007}
{J{\o}rgensen}, J.~K., {Bourke}, T.~L., {Myers}, P.~C., {et~al.} 2007, \apj,
  659, 479

\bibitem[{{Juvela} {et~al.}(2015){Juvela}, {Demyk}, {Doi}, {Hughes},
  {Lef{\`e}vre}, {Marshall}, {Meny}, {Montillaud}, {Pagani}, {Paradis},
  {Ristorcelli}, {Malinen}, {Montier}, {Paladini}, {Pelkonen}, \&
  {Rivera-Ingraham}}]{Juvela2015}
{Juvela}, M., {Demyk}, K., {Doi}, Y., {et~al.} 2015, \aap, 584, A94

\bibitem[{{Kaas} {et~al.}(2004){Kaas}, {Olofsson}, {Bontemps}, {Andr{\'e}},
  {Nordh}, {Huldtgren}, {Prusti}, {Persi}, {Delgado}, {Motte}, {Abergel},
  {Boulanger}, {Burgdorf}, {Casali}, {Cesarsky}, {Davies}, {Falgarone},
  {Montmerle}, {Perault}, {Puget}, \& {Sibille}}]{Kaas2004}
{Kaas}, A.~A., {Olofsson}, G., {Bontemps}, S., {et~al.} 2004, \aap, 421, 623

\bibitem[{{Karska} {et~al.}(2013){Karska}, {Herczeg}, {van Dishoeck},
  {Wampfler}, {Kristensen}, {Goicoechea}, {Visser}, {Nisini}, {San
  Jos{\'e}-Garc{\'{\i}}a}, {Bruderer}, {{\'S}niady}, {Doty}, {Fedele},
  {Y{\i}ld{\i}z}, {Benz}, {Bergin}, {Caselli}, {Herpin}, {Hogerheijde},
  {Johnstone}, {J{\o}rgensen}, {Liseau}, {Tafalla}, {van der Tak}, \&
  {Wyrowski}}]{Karska2013}
{Karska}, A., {Herczeg}, G.~J., {van Dishoeck}, E.~F., {et~al.} 2013, \aap,
  552, A141

\bibitem[{{Kataoka} {et~al.}(2014){Kataoka}, {Okuzumi}, {Tanaka}, \&
  {Nomura}}]{Kataoka2014}
{Kataoka}, A., {Okuzumi}, S., {Tanaka}, H., \& {Nomura}, H. 2014, \aap, 568,
  A42

\bibitem[{{Kern} {et~al.}(2016){Kern}, {Keown}, {Tobin}, {Mead}, \&
  {Gutermuth}}]{Kern2016}
{Kern}, N.~S., {Keown}, J.~A., {Tobin}, J.~J., {Mead}, A., \& {Gutermuth},
  R.~A. 2016, \aj, 151, 42

\bibitem[{{Kessler-Silacci} {et~al.}(2005){Kessler-Silacci}, {Hillenbrand},
  {Blake}, \& {Meyer}}]{Kessler-Silacci2005}
{Kessler-Silacci}, J.~E., {Hillenbrand}, L.~A., {Blake}, G.~A., \& {Meyer},
  M.~R. 2005, \apj, 622, 404

\bibitem[{{K{\"o}hler} {et~al.}(2015){K{\"o}hler}, {Ysard}, \&
  {Jones}}]{Kohler2015}
{K{\"o}hler}, M., {Ysard}, N., \& {Jones}, A.~P. 2015, \aap, 579, A15

\bibitem[{{Kristensen} {et~al.}(2012){Kristensen}, {van Dishoeck}, {Bergin},
  {Visser}, {Y{\i}ld{\i}z}, {San Jose-Garcia}, {J{\o}rgensen}, {Herczeg},
  {Johnstone}, {Wampfler}, {Benz}, {Bruderer}, {Cabrit}, {Caselli}, {Doty},
  {Harsono}, {Herpin}, {Hogerheijde}, {Karska}, {van Kempen}, {Liseau},
  {Nisini}, {Tafalla}, {van der Tak}, \& {Wyrowski}}]{Kristensen2012}
{Kristensen}, L.~E., {van Dishoeck}, E.~F., {Bergin}, E.~A., {et~al.} 2012,
  \aap, 542, A8

\bibitem[{{Kruegel} \& {Siebenmorgen}(1994)}]{Kruegel1994}
{Kruegel}, E. \& {Siebenmorgen}, R. 1994, \aap, 288, 929

\bibitem[{{Kwon} {et~al.}(2015){Kwon}, {Fern{\'a}ndez-L{\'o}pez}, {Stephens},
  \& {Looney}}]{Kwon2015}
{Kwon}, W., {Fern{\'a}ndez-L{\'o}pez}, M., {Stephens}, I.~W., \& {Looney},
  L.~W. 2015, \apj, 814, 43

\bibitem[{{Kwon} {et~al.}(2009){Kwon}, {Looney}, {Mundy}, {Chiang}, \&
  {Kemball}}]{Kwon2009}
{Kwon}, W., {Looney}, L.~W., {Mundy}, L.~G., {Chiang}, H.-F., \& {Kemball},
  A.~J. 2009, \apj, 696, 841

\bibitem[{{Lebreuilly} {et~al.}(2019){Lebreuilly}, {Commer{\c{c}}on}, \&
  {Laibe}}]{Lebreuilly2019}
{Lebreuilly}, U., {Commer{\c{c}}on}, B., \& {Laibe}, G. 2019, \aap, 626, A96

\bibitem[{{Lef{\`e}vre} {et~al.}(2014){Lef{\`e}vre}, {Pagani}, {Juvela},
  {Paladini}, {Lallement}, {Marshall}, {Andersen}, {Bacmann}, {McGehee},
  {Montier}, {Noriega-Crespo}, {Pelkonen}, {Ristorcelli}, \&
  {Steinacker}}]{Lefevre2014}
{Lef{\`e}vre}, C., {Pagani}, L., {Juvela}, M., {et~al.} 2014, \aap, 572, A20

\bibitem[{{LeGouellec} {et~al.}(2019){LeGouellec}, {Hull}, {Maury}, {Girart},
  {Tychoniec}, {Kristensen}, {Li}, {Louvet}, {Cortes}, \&
  {Rao}}]{LeGouellec2019}
{LeGouellec}, V. J.~M., {Hull}, C. L.~H., {Maury}, A.~J., {et~al.} 2019, arXiv
  e-prints, arXiv:1909.00046

\bibitem[{{Li} {et~al.}(2017){Li}, {Liu}, {Hasegawa}, \& {Hirano}}]{Li2017}
{Li}, J.~I., {Liu}, H.~B., {Hasegawa}, Y., \& {Hirano}, N. 2017, \apj, 840, 72

\bibitem[{{Lugo} {et~al.}(2004){Lugo}, {Lizano}, \& {Garay}}]{Lugo2004}
{Lugo}, J., {Lizano}, S., \& {Garay}, G. 2004, \apj, 614, 807

\bibitem[{{Martin} {et~al.}(2012){Martin}, {Roy}, {Bontemps},
  {Miville-Desch{\^e}nes}, {Ade}, {Bock}, {Chapin}, {Devlin}, {Dicker},
  {Griffin}, {Gundersen}, {Halpern}, {Hargrave}, {Hughes}, {Klein}, {Marsden},
  {Mauskopf}, {Netterfield}, {Olmi}, {Patanchon}, {Rex}, {Scott}, {Semisch},
  {Truch}, {Tucker}, {Tucker}, {Viero}, \& {Wiebe}}]{Martin2012}
{Martin}, P.~G., {Roy}, A., {Bontemps}, S., {et~al.} 2012, \apj, 751, 28

\bibitem[{{Mason} {et~al.}(2019){Mason}, {Dicker}, {Sadavoy}, {Stanchfield},
  {Mroczkowski}, {Romero}, {Friesen}, {Sarazin}, {Sievers}, \&
  {Stanke}}]{Mason2019}
{Mason}, B., {Dicker}, S., {Sadavoy}, S., {et~al.} 2019, arXiv e-prints,
  arXiv:1905.05221

\bibitem[{{Masson} {et~al.}(2016){Masson}, {Chabrier}, {Hennebelle}, {Vaytet},
  \& {Commer{\c{c}}on}}]{Masson2016}
{Masson}, J., {Chabrier}, G., {Hennebelle}, P., {Vaytet}, N., \&
  {Commer{\c{c}}on}, B. 2016, \aap, 587, A32

\bibitem[{{Mathis} {et~al.}(1977){Mathis}, {Rumpl}, \& {Nordsieck}}]{MRN1977}
{Mathis}, J.~S., {Rumpl}, W., \& {Nordsieck}, K.~H. 1977, \apj, 217, 425

\bibitem[{{Maury} {et~al.}(2011){Maury}, {Andr{\'e}}, {Men'shchikov},
  {K{\"o}nyves}, \& {Bontemps}}]{Maury2011}
{Maury}, A.~J., {Andr{\'e}}, P., {Men'shchikov}, A., {K{\"o}nyves}, V., \&
  {Bontemps}, S. 2011, \aap, 535, A77

\bibitem[{{Maury} {et~al.}(2019){Maury}, {Andr{\'e}}, {Testi}, {Maret},
  {Belloche}, {Hennebelle}, {Cabrit}, {Codella}, {Gueth}, {Podio}, {Anderl},
  {Bacmann}, {Bontemps}, {Gaudel}, {Ladjelate}, {Lef{\`e}vre}, {Tabone}, \&
  {Lefloch}}]{Maury2019}
{Maury}, A.~J., {Andr{\'e}}, P., {Testi}, L., {et~al.} 2019, \aap, 621, A76

\bibitem[{{Maury} {et~al.}(2018){Maury}, {Girart}, {Zhang}, {Hennebelle},
  {Keto}, {Rao}, {Lai}, {Ohashi}, \& {Galametz}}]{Maury2018}
{Maury}, A.~J., {Girart}, J.~M., {Zhang}, Q., {et~al.} 2018, \mnras, 477, 2760

\bibitem[{{McMullin} {et~al.}(1994){McMullin}, {Mundy}, {Wilking}, {Hezel}, \&
  {Blake}}]{McMullin1994}
{McMullin}, J.~P., {Mundy}, L.~G., {Wilking}, B.~A., {Hezel}, T., \& {Blake},
  G.~A. 1994, \apj, 424, 222

\bibitem[{{Meehan} {et~al.}(1998){Meehan}, {Wilking}, {Claussen}, {Mundy}, \&
  {Wootten}}]{Meehan1998}
{Meehan}, L.~S.~G., {Wilking}, B.~A., {Claussen}, M.~J., {Mundy}, L.~G., \&
  {Wootten}, A. 1998, \aj, 115, 1599

\bibitem[{{Melis} {et~al.}(2011){Melis}, {Duch{\^e}ne}, {Chomiuk}, {Palmer},
  {Perrin}, {Maddison}, {M{\'e}nard}, {Stapelfeldt}, {Pinte}, \&
  {Duvert}}]{Melis2011}
{Melis}, C., {Duch{\^e}ne}, G., {Chomiuk}, L., {et~al.} 2011, \apjl, 739, L7

\bibitem[{{Miotello} {et~al.}(2014){Miotello}, {Testi}, {Lodato}, {Ricci},
  {Rosotti}, {Brooks}, {Maury}, \& {Natta}}]{Miotello2014}
{Miotello}, A., {Testi}, L., {Lodato}, G., {et~al.} 2014, \aap, 567, A32

\bibitem[{{Miyake} \& {Nakagawa}(1993)}]{Miyake1993}
{Miyake}, K. \& {Nakagawa}, Y. 1993, \icarus, 106, 20

\bibitem[{{Motte} \& {Andr{\'e}}(2001)}]{Motte2001}
{Motte}, F. \& {Andr{\'e}}, P. 2001, \aap, 365, 440

\bibitem[{{Motte} {et~al.}(1998){Motte}, {Andre}, \& {Neri}}]{Motte1998}
{Motte}, F., {Andre}, P., \& {Neri}, R. 1998, \aap, 336, 150

\bibitem[{{Natta} {et~al.}(2007){Natta}, {Testi}, {Calvet}, {Henning},
  {Waters}, \& {Wilner}}]{Natta2007}
{Natta}, A., {Testi}, L., {Calvet}, N., {et~al.} 2007, in Protostars and
  Planets V, ed. B.~{Reipurth}, D.~{Jewitt}, \& K.~{Keil}, 767

\bibitem[{{Ormel} {et~al.}(2009){Ormel}, {Paszun}, {Dominik}, \&
  {Tielens}}]{Ormel2009}
{Ormel}, C.~W., {Paszun}, D., {Dominik}, C., \& {Tielens}, A.~G.~G.~M. 2009,
  \aap, 502, 845

\bibitem[{{Ortiz-Le{\'o}n} {et~al.}(2018{\natexlab{a}}){Ortiz-Le{\'o}n},
  {Loinard}, {Dzib}, {Galli}, {Kounkel}, {Mioduszewski}, {Rodr{\'{\i}}guez},
  {Torres}, {Hartmann}, {Boden}, {Evans}, {Brice{\~n}o}, \&
  {Tobin}}]{Ortiz-Leon2018}
{Ortiz-Le{\'o}n}, G.~N., {Loinard}, L., {Dzib}, S.~A., {et~al.}
  2018{\natexlab{a}}, \apj, 865, 73

\bibitem[{{Ortiz-Le{\'o}n} {et~al.}(2018{\natexlab{b}}){Ortiz-Le{\'o}n},
  {Loinard}, {Dzib}, {Kounkel}, {Galli}, {Tobin}, {Evans}, {Hartmann},
  {Rodr{\'{\i}}guez}, {Brice{\~n}o}, {Torres}, \&
  {Mioduszewski}}]{Ortiz-Leon2018_2}
{Ortiz-Le{\'o}n}, G.~N., {Loinard}, L., {Dzib}, S.~A., {et~al.}
  2018{\natexlab{b}}, \apjl, 869, L33

\bibitem[{{Ossenkopf}(1993)}]{Ossenkopf1993}
{Ossenkopf}, V. 1993, \aap, 280, 617

\bibitem[{{Ossenkopf} \& {Henning}(1994)}]{Ossenkopf1994}
{Ossenkopf}, V. \& {Henning}, T. 1994, \aap, 291, 943

\bibitem[{{Pagani} {et~al.}(2010){Pagani}, {Steinacker}, {Bacmann}, {Stutz}, \&
  {Henning}}]{Pagani2010}
{Pagani}, L., {Steinacker}, J., {Bacmann}, A., {Stutz}, A., \& {Henning}, T.
  2010, Science, 329, 1622

\bibitem[{{Pascucci} {et~al.}(2012){Pascucci}, {Gorti}, \&
  {Hollenbach}}]{Pascucci2012}
{Pascucci}, I., {Gorti}, U., \& {Hollenbach}, D. 2012, \apj, 751, L42

\bibitem[{{P{\'e}rez} {et~al.}(2012){P{\'e}rez}, {Carpenter}, {Chandler},
  {Isella}, {Andrews}, {Ricci}, {Calvet}, {Corder}, {Deller}, {Dullemond},
  {Greaves}, {Harris}, {Henning}, {Kwon}, {Lazio}, {Linz}, {Mundy}, {Sargent},
  {Storm}, {Testi}, \& {Wilner}}]{Perez2012}
{P{\'e}rez}, L.~M., {Carpenter}, J.~M., {Chandler}, C.~J., {et~al.} 2012, \apj,
  760, L17

\bibitem[{{P{\'e}rez} {et~al.}(2015){P{\'e}rez}, {Chandler}, {Isella},
  {Carpenter}, {Andrews}, {Calvet}, {Corder}, {Deller}, {Dullemond}, {Greaves},
  {Harris}, {Henning}, {Kwon}, {Lazio}, {Linz}, {Mundy}, {Ricci}, {Sargent},
  {Storm}, {Tazzari}, {Testi}, \& {Wilner}}]{Perez2015}
{P{\'e}rez}, L.~M., {Chandler}, C.~J., {Isella}, A., {et~al.} 2015, \apj, 813,
  41

\bibitem[{{Planck Collaboration} {et~al.}(2014){Planck Collaboration},
  {Abergel}, {Ade}, {Aghanim}, {Alves}, {Aniano}, {Armitage-Caplan}, {Arnaud},
  {Ashdown}, {Atrio-Barand ela}, {Aumont}, {Baccigalupi}, {Banday}, {Barreiro},
  {Bartlett}, {Battaner}, {Benabed}, {Beno{\^\i}t}, {Benoit-L{\'e}vy},
  {Bernard}, {Bersanelli}, {Bielewicz}, {Bobin}, {Bock}, {Bonaldi}, {Bond},
  {Borrill}, {Bouchet}, {Boulanger}, {Bridges}, {Bucher}, {Burigana}, {Butler},
  {Cardoso}, {Catalano}, {Chamballu}, {Chary}, {Chiang}, {Chiang},
  {Christensen}, {Church}, {Clemens}, {Clements}, {Colombi}, {Colombo},
  {Combet}, {Couchot}, {Coulais}, {Crill}, {Curto}, {Cuttaia}, {Danese},
  {Davies}, {Davis}, {de Bernardis}, {de Rosa}, {de Zotti}, {Delabrouille},
  {Delouis}, {D{\'e}sert}, {Dickinson}, {Diego}, {Dole}, {Donzelli},
  {Dor{\'e}}, {Douspis}, {Draine}, {Dupac}, {Efstathiou}, {En{\ss}lin},
  {Eriksen}, {Falgarone}, {Finelli}, {Forni}, {Frailis}, {Fraisse},
  {Franceschi}, {Galeotta}, {Ganga}, {Ghosh}, {Giard}, {Giardino},
  {Giraud-H{\'e}raud}, {Gonz{\'a}lez-Nuevo}, {G{\'o}rski}, {Gratton},
  {Gregorio}, {Grenier}, {Gruppuso}, {Guillet}, {Hansen}, {Hanson}, {Harrison},
  {Helou}, {Henrot-Versill{\'e}}, {Hern{\'a}ndez-Monteagudo}, {Herranz},
  {Hildebrand t}, {Hivon}, {Hobson}, {Holmes}, {Hornstrup}, {Hovest},
  {Huffenberger}, {Jaffe}, {Jaffe}, {Jewell}, {Joncas}, {Jones}, {Juvela},
  {Keih{\"a}nen}, {Keskitalo}, {Kisner}, {Knoche}, {Knox}, {Kunz},
  {Kurki-Suonio}, {Lagache}, {L{\"a}hteenm{\"a}ki}, {Lamarre}, {Lasenby},
  {Laureijs}, {Lawrence}, {Leonardi}, {Le{\'o}n-Tavares}, {Lesgourgues},
  {Levrier}, {Liguori}, {Lilje}, {Linden-V{\o}rnle}, {L{\'o}pez-Caniego},
  {Lubin}, {Mac{\'\i}as-P{\'e}rez}, {Maffei}, {Maino}, {Mand olesi}, {Maris},
  {Marshall}, {Martin}, {Mart{\'\i}nez-Gonz{\'a}lez}, {Masi}, {Massardi},
  {Matarrese}, {Matthai}, {Mazzotta}, {McGehee}, {Melchiorri}, {Mendes},
  {Mennella}, {Migliaccio}, {Mitra}, {Miville-Desch{\^e}nes}, {Moneti},
  {Montier}, {Morgante}, {Mortlock}, {Munshi}, {Murphy}, {Naselsky}, {Nati},
  {Natoli}, {Netterfield}, {N{\o}rgaard-Nielsen}, {Noviello}, {Novikov},
  {Novikov}, {Osborne}, {Oxborrow}, {Paci}, {Pagano}, {Pajot}, {Paladini},
  {Paoletti}, {Pasian}, {Patanchon}, {Perdereau}, {Perotto}, {Perrotta},
  {Piacentini}, {Piat}, {Pierpaoli}, {Pietrobon}, {Plaszczynski},
  {Pointecouteau}, {Polenta}, {Ponthieu}, {Popa}, {Poutanen}, {Pratt},
  {Pr{\'e}zeau}, {Prunet}, {Puget}, {Rachen}, {Reach}, {Rebolo}, {Reinecke},
  {Remazeilles}, {Renault}, {Ricciardi}, {Riller}, {Ristorcelli}, {Rocha},
  {Rosset}, {Roudier}, {Rowan-Robinson}, {Rubi{\~n}o-Mart{\'\i}n}, {Rusholme},
  {Sandri}, {Santos}, {Savini}, {Scott}, {Seiffert}, {Shellard}, {Spencer},
  {Starck}, {Stolyarov}, {Stompor}, {Sudiwala}, {Sunyaev}, {Sureau}, {Sutton},
  {Suur-Uski}, {Sygnet}, {Tauber}, {Tavagnacco}, {Terenzi}, {Toffolatti},
  {Tomasi}, {Tristram}, {Tucci}, {Tuovinen}, {T{\"u}rler}, {Umana},
  {Valenziano}, {Valiviita}, {Van Tent}, {Verstraete}, {Vielva}, {Villa},
  {Vittorio}, {Wade}, {Wandelt}, {Welikala}, {Ysard}, {Yvon}, {Zacchei}, \&
  {Zonca}}]{Planck2014}
{Planck Collaboration}, {Abergel}, A., {Ade}, P.~A.~R., {et~al.} 2014, \aap,
  571, A11

\bibitem[{{Plunkett} {et~al.}(2018){Plunkett}, {Fern{\'a}ndez-L{\'o}pez},
  {Arce}, {Busquet}, {Mardones}, \& {Dunham}}]{Plunkett2018}
{Plunkett}, A.~L., {Fern{\'a}ndez-L{\'o}pez}, M., {Arce}, H.~G., {et~al.} 2018,
  \aap, 615, A9

\bibitem[{{Poteet}(2012)}]{Poteet2012}
{Poteet}, C.~A. 2012, PhD thesis, The University of Toledo

\bibitem[{{Rawlings} {et~al.}(2013){Rawlings}, {Williams}, {Viti},
  {Cecchi-Pestellini}, \& {Duley}}]{Rawlings2013}
{Rawlings}, J.~M.~C., {Williams}, D.~A., {Viti}, S., {Cecchi-Pestellini}, C.,
  \& {Duley}, W.~W. 2013, \mnras, 430, 264

\bibitem[{{Reipurth} {et~al.}(2002){Reipurth}, {Rodr{\'{\i}}guez}, {Anglada},
  \& {Bally}}]{Reipurth2002}
{Reipurth}, B., {Rodr{\'{\i}}guez}, L.~F., {Anglada}, G., \& {Bally}, J. 2002,
  \aj, 124, 1045

\bibitem[{{Reissl} {et~al.}(2016){Reissl}, {Wolf}, \& {Brauer}}]{Reissl2016}
{Reissl}, S., {Wolf}, S., \& {Brauer}, R. 2016, \aap, 593, A87

\bibitem[{{Ricci} {et~al.}(2010){Ricci}, {Testi}, {Natta}, \&
  {Brooks}}]{Ricci2010}
{Ricci}, L., {Testi}, L., {Natta}, A., \& {Brooks}, K.~J. 2010, \aap, 521, A66

\bibitem[{{Robitaille} {et~al.}(2006){Robitaille}, {Whitney}, {Indebetouw},
  {Wood}, \& {Denzmore}}]{Robitaille2006}
{Robitaille}, T.~P., {Whitney}, B.~A., {Indebetouw}, R., {Wood}, K., \&
  {Denzmore}, P. 2006, \apjs, 167, 256

\bibitem[{{Rodr{\'{\i}}guez} {et~al.}(1997){Rodr{\'{\i}}guez}, {Anglada}, \&
  {Curiel}}]{Rodriguez1997}
{Rodr{\'{\i}}guez}, L.~F., {Anglada}, G., \& {Curiel}, S. 1997, \apjl, 480,
  L125

\bibitem[{{Sadavoy} {et~al.}(2014){Sadavoy}, {Di Francesco}, {Andre},
  {Pezzuto}, {Bernard}, {Maury}, {Men'shchikov}, {Motte}, {Nguyen-Lu'o'ng},
  {Schneider}, {Arzoumanian}, {Benedettini}, {Bontemps}, {Elia}, {Hennemann},
  {Hill}, {Konyves}, {Louvet}, {Peretto}, {Roy}, \& {White}}]{Sadavoy2014}
{Sadavoy}, S.~I., {Di Francesco}, J., {Andre}, P., {et~al.} 2014, \apjl, 787,
  L18

\bibitem[{{Sadavoy} {et~al.}(2016){Sadavoy}, {Stutz}, {Schnee}, {Mason}, {Di
  Francesco}, \& {Friesen}}]{Sadavoy2016}
{Sadavoy}, S.~I., {Stutz}, A.~M., {Schnee}, S., {et~al.} 2016, \aap, 588, A30

\bibitem[{{Schnee} {et~al.}(2014){Schnee}, {Mason}, {Di Francesco}, {Friesen},
  {Li}, {Sadavoy}, \& {Stanke}}]{Schnee2014}
{Schnee}, S., {Mason}, B., {Di Francesco}, J., {et~al.} 2014, \mnras, 444, 2303

\bibitem[{{Steinacker} {et~al.}(2015){Steinacker}, {Andersen}, {Thi},
  {Paladini}, {Juvela}, {Bacmann}, {Pelkonen}, {Pagani}, {Lef{\`e}vre},
  {Henning}, \& {Noriega-Crespo}}]{Steinacker2015}
{Steinacker}, J., {Andersen}, M., {Thi}, W.~F., {et~al.} 2015, \aap, 582, A70

\bibitem[{{Steinacker} {et~al.}(2010){Steinacker}, {Pagani}, {Bacmann}, \&
  {Guieu}}]{Steinacker2010}
{Steinacker}, J., {Pagani}, L., {Bacmann}, A., \& {Guieu}, S. 2010, \aap, 511,
  A9

\bibitem[{{Tazzari} {et~al.}(2016){Tazzari}, {Testi}, {Ercolano}, {Natta},
  {Isella}, {Chandler}, {P{\'e}rez}, {Andrews}, {Wilner}, {Ricci}, {Henning},
  {Linz}, {Kwon}, {Corder}, {Dullemond}, {Carpenter}, {Sargent}, {Mundy},
  {Storm}, {Calvet}, {Greaves}, {Lazio}, \& {Deller}}]{Tazzari2016}
{Tazzari}, M., {Testi}, L., {Ercolano}, B., {et~al.} 2016, \aap, 588, A53

\bibitem[{{Terebey} {et~al.}(1993){Terebey}, {Chandler}, \&
  {Andre}}]{Terebey1993}
{Terebey}, S., {Chandler}, C.~J., \& {Andre}, P. 1993, \apj, 414, 759

\bibitem[{{Testi} {et~al.}(2014){Testi}, {Birnstiel}, {Ricci}, {Andrews},
  {Blum}, {Carpenter}, {Dominik}, {Isella}, {Natta}, {Williams}, \&
  {Wilner}}]{Testi2014}
{Testi}, L., {Birnstiel}, T., {Ricci}, L., {et~al.} 2014, in Protostars and
  Planets VI, ed. H.~{Beuther}, R.~S. {Klessen}, C.~P. {Dullemond}, \&
  T.~{Henning}, 339

\bibitem[{{Testi} {et~al.}(2001){Testi}, {Natta}, {Shepherd}, \&
  {Wilner}}]{Testi2001}
{Testi}, L., {Natta}, A., {Shepherd}, D.~S., \& {Wilner}, D.~J. 2001, \apj,
  554, 1087

\bibitem[{{Testi} {et~al.}(2003){Testi}, {Natta}, {Shepherd}, \&
  {Wilner}}]{Testi2003}
{Testi}, L., {Natta}, A., {Shepherd}, D.~S., \& {Wilner}, D.~J. 2003, \aap,
  403, 323

\bibitem[{{Teyssier}(2002)}]{Teyssier2002}
{Teyssier}, R. 2002, \aap, 385, 337

\bibitem[{{Tobin} {et~al.}(2015){Tobin}, {Dunham}, {Looney}, {Li}, {Chandler},
  {Segura-Cox}, {Sadavoy}, {Melis}, {Harris}, {Perez}, {Kratter},
  {J{\o}rgensen}, {Plunkett}, \& {Hull}}]{Tobin2015_2}
{Tobin}, J.~J., {Dunham}, M.~M., {Looney}, L.~W., {et~al.} 2015, \apj, 798, 61

\bibitem[{{Tobin} {et~al.}(2013){Tobin}, {Hartmann}, {Chiang}, {Wilner},
  {Looney}, {Loinard}, {Calvet}, \& {D'Alessio}}]{Tobin2013_2}
{Tobin}, J.~J., {Hartmann}, L., {Chiang}, H.-F., {et~al.} 2013, \apj, 771, 48

\bibitem[{{Tobin} {et~al.}(2016){Tobin}, {Looney}, {Li}, {Chandler}, {Dunham},
  {Segura-Cox}, {Sadavoy}, {Melis}, {Harris}, {Kratter}, \&
  {Perez}}]{Tobin2016}
{Tobin}, J.~J., {Looney}, L.~W., {Li}, Z.-Y., {et~al.} 2016, \apj, 818, 73

\bibitem[{{Torres} {et~al.}(2009){Torres}, {Loinard}, {Mioduszewski}, \&
  {Rodr{\'{\i}}guez}}]{Torres2009}
{Torres}, R.~M., {Loinard}, L., {Mioduszewski}, A.~J., \& {Rodr{\'{\i}}guez},
  L.~F. 2009, \apj, 698, 242

\bibitem[{{Tychoniec} {et~al.}(2018{\natexlab{a}}){Tychoniec}, {Tobin},
  {Karska}, {Chand ler}, {Dunham}, {Li}, {Looney}, {Segura-Cox}, {Harris},
  {Melis}, \& {Sadavoy}}]{Tychoniec2018_III}
{Tychoniec}, {\L}., {Tobin}, J.~J., {Karska}, A., {et~al.} 2018{\natexlab{a}},
  \apj, 852, 18

\bibitem[{{Tychoniec} {et~al.}(2018{\natexlab{b}}){Tychoniec}, {Tobin},
  {Karska}, {Chandler}, {Dunham}, {Harris}, {Kratter}, {Li}, {Looney}, {Melis},
  {P{\'e}rez}, {Sadavoy}, {Segura-Cox}, \& {van Dishoeck}}]{Tychoniec2018_IV}
{Tychoniec}, {\L}., {Tobin}, J.~J., {Karska}, A., {et~al.} 2018{\natexlab{b}},
  \apjs, 238, 19

\bibitem[{{Valdivia} {et~al.}(2019){Valdivia}, {Maury}, {Brauer}, {Hennebelle},
  {Galametz}, {Guillet}, \& {Reissl}}]{Valdivia2019}
{Valdivia}, V., {Maury}, A., {Brauer}, R., {et~al.} 2019, \mnras, 488, 4897

\bibitem[{{Whitney} {et~al.}(2003){Whitney}, {Wood}, {Bjorkman}, \&
  {Wolff}}]{Whitney2003}
{Whitney}, B.~A., {Wood}, K., {Bjorkman}, J.~E., \& {Wolff}, M.~J. 2003, \apj,
  591, 1049

\bibitem[{{Wong} {et~al.}(2016){Wong}, {Hirashita}, \& {Li}}]{Wong2016}
{Wong}, Y. H.~V., {Hirashita}, H., \& {Li}, Z.-Y. 2016, \pasj, 68, 67

\bibitem[{{Ysard} {et~al.}(2019){Ysard}, {Koehler}, {Jimenez-Serra}, {Jones},
  \& {Verstraete}}]{Ysard2019}
{Ysard}, N., {Koehler}, M., {Jimenez-Serra}, I., {Jones}, A.~P., \&
  {Verstraete}, L. 2019, arXiv e-prints, arXiv:1909.05015

\bibitem[{{Zucker} {et~al.}(2019){Zucker}, {Speagle}, {Schlafly}, {Green},
  {Finkbeiner}, {Goodman}, \& {Alves}}]{Zucker2019}
{Zucker}, C., {Speagle}, J.~S., {Schlafly}, E.~F., {et~al.} 2019,
  arXiv:1902.01425

\end{thebibliography}

\appendix

\section{Free-free and synchrotron contributions to the 1.3 mm and 3.2 mm emission} \label{Nonthermaldust}

\begin{table*}[]
\caption{Upper limits on the contribution to the 1.3 mm and 3.2 mm emission not coming from thermal dust emission }
\centering
\begin{tabular}{ccccccc}
\hline
\vspace{-5pt}
&\\
Name            & $\alpha_\mathrm{ff+s}$ $^a$ 
& Flux @ 200 k$\lambda$     & Contribution 
& Flux @ 200 k$\lambda$     & Contribution & References $^b$ \\
&& at 1.3 mm & at 1.3 mm  
& at 3.2 mm  & at 3.2 mm    &  \\
&& (mJy) & (\%) & (mJy) & (\%) & \\
\vspace{-5pt}
&\\
\hline
\vspace{-5pt}
&\\
NGC1333-IRAS2A1     & 1.1               & 85      & 4.5       & 13       & 11         & 1\\
NGC1333-IRAS4A1     & 1.1           & 310     & 3.8        & 47      & 9.1       & 3 \\
NGC1333-IRAS4B      & 0.5               & 196     & 0.2       & 20       & 1.4       & 3 \\
L1448-2A                & 0.5           & 28      & 4.0       & 3.9      & 18        & 3 \\
L1448-C                 & 1.0               & 102     & 2.9       & 18       & 6.7       & 3 \\
L1448-NB1               & 0.3               & 82      & 3.6       & 13       & 14        & 2, 4 \\
SVS13B                  & 0.8               & 72      & 6.9       & 12       & 19        & 5 \\ 
L1527                   & 0.3           & 136     & 1.7       & 19       & 9.3       & 6 \\
SerpM-S68N              & 0.6           & 45      & 0.9       & 3.4      & 6.8       & 7,8 \\
SerpM-SMM4              & 1.1 {\it(f)}  & 46      & 9.6       & 8.7      & 19        & 8 \\
SerpS-MM18              & 1.1 {\it(f)}  & 97      & 3.6       & 8.2      & 16        & 9 \\
L1157                   & 0.3           & 103     & 0.5       & 12.6     & 3.2       & 10 \\
\hline
\end{tabular}
\begin{list}{}{}
\item[$^a$] As in \citet{Tobin2015_2}, the spectral slopes $\alpha_\mathrm{ff+s}$ are defined by F$_{\nu}$ = $F_0(\lambda/\lambda_0)^{-\alpha_\mathrm{ff+s}}$, with $\lambda_0$ = 1 mm. {\it(f)} indicates when $\alpha_\mathrm{ff+s}$ has been fixed to 1.1.
\item[$^b$] References for the centimeter fluxes used to estimate the free-free+synchrotron contribution: [1] \citet{Tobin2015_2}, [2] \citet{Reipurth2002}, [3] \cite{Tychoniec2018_IV}, [4] \citet{Curiel1990}, [5] \citet{Rodriguez1997}, [6] \citet{Melis2011}, [7] \citet{McMullin1994}, [8] \citet{Eiroa2005}, [9] \citet{Kern2016}, [10] \citet{Meehan1998}.
\end{list}
\label{tab:freefree}
\end{table*}

In order to quantify the contribution to the PdBI observations not coming from thermal dust emission, we assume that the $\geq$ 2 cm emission can be considered as free from thermal dust emission \citep[as observed in IRAS2A by][]{Tobin2015_2} and considered as compact. We gathered fluxes beyond 2 cm published in the literature. For sources with more than two observational constraints, we estimated the free-free + synchrotron spectral slope $\alpha_\mathrm{ff+s}$ as
\begin{equation}
    F_{\lambda} \propto (\lambda/\lambda_0)^{-\alpha_\mathrm{ff+s}}
\label{equafree}
,\end{equation}

\noindent where $\lambda_0$ = 1 mm. We then extrapolated the fitted spectrum to the millimeter regime and compare it with the 1.3 mm and 3.2 mm PdBI fluxes at 200 k$\lambda$ (so on our smallest scales) to estimate the contamination at both wavelengths. The references of the used radio fluxes, the derived $\alpha_\mathrm{ff+s}$ slopes, and the contamination to the 1.3 mm and 3.2 mm emission are reported in Table~\ref{tab:freefree}. For SerpM-SMM4, where only the 3.6 cm flux was found, we fixed $\alpha_\mathrm{ff+s}$ to 1.1, which is the $\alpha_\mathrm{ff+s}$ value of IRAS4A1, i.e., the source in the sample that is the closest in terms of envelope mass, temperature, and internal luminosity. For SerpS-MM18, the $\alpha_\mathrm{ff+s}$ value was derived by \citet{Kern2016} from observations at 4.75 and 7.25 GHz. We note that several millimeter sources reside in the 4\arcsec\ beam, as observed in \citet{Plunkett2018}. The high $\alpha_\mathrm{ff+s}$ found by \citet{Kern2016} ($\sim$2-2.3) suggests that thermal dust emission might still contribute to the centimeter emission. As in 
SerpM-SMM4, we decided to fix $\alpha_\mathrm{ff+s}$ to 1.1 for this source.
The contribution is systematically lower than 10\% at 1.3 mm, but it can become non-negligible at 3.2 mm. Calculated using the radio sources of \citet{Kern2016} for SerpS-MM18, the contamination accounts for all the 200 k$\lambda$ 1.3 mm and 3.2mm fluxes; these fluxes have a large $\alpha_\mathrm{ff+s}$ ($>$ 2) usually associated with optically thick emission.

\section{Predictions from synthetic observations: Noticeable impact of the central region} \label{SyntheticObservations}

\begin{table}[]
\centering
\caption{Average amplitude for $>$200 k$\lambda$ baselines} 
\begin{tabular}{ccc}
\hline
\vspace{-5pt}
&\\
Name    & F$^{1.3mm}_{>200 k\lambda}$   & F$^{3.2mm}_{>200 k\lambda}$     \\ 
                & (mJy)         &       (mJy)                           \\
\vspace{-5pt}
&\\
\hline
\vspace{-5pt}
&\\
IRAS2A1                         & 85.4 & 13.3   \\
IRAS4A1                         &309.6 & 47.2 \\
IRAS4B                          &196.1 & 20.0 \\
SVS13B                          & 71.8 & 12.4 \\
L1527                           &135.6 & 18.9 \\
SerpM-S68N                      & 45.2 & 3.4\\
Serp-SMM4                       & 83.8 & 8.7 \\
SerpS-MM18                      & 97.4 & 8.2 \\
L1157                           & 103.0 & 12.6 \\
\hline
\end{tabular}
\label{FluxCorrected}
\end{table}

\begin{figure}
\begin{tabular}{m{8.5cm}}
\includegraphics[width=8.5cm]{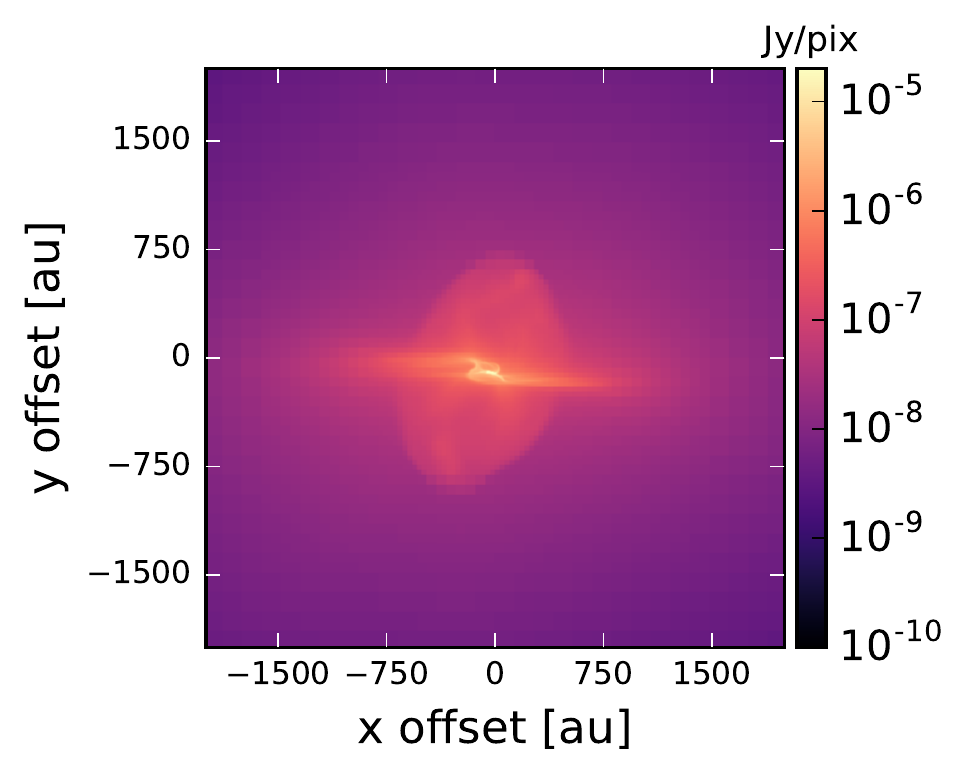} \\
\vspace{-105pt}\hspace{58pt}\color{white}{\bf \Large @ 1.3 mm} \\ 
\vspace{-20pt}\includegraphics[width=8.5cm]{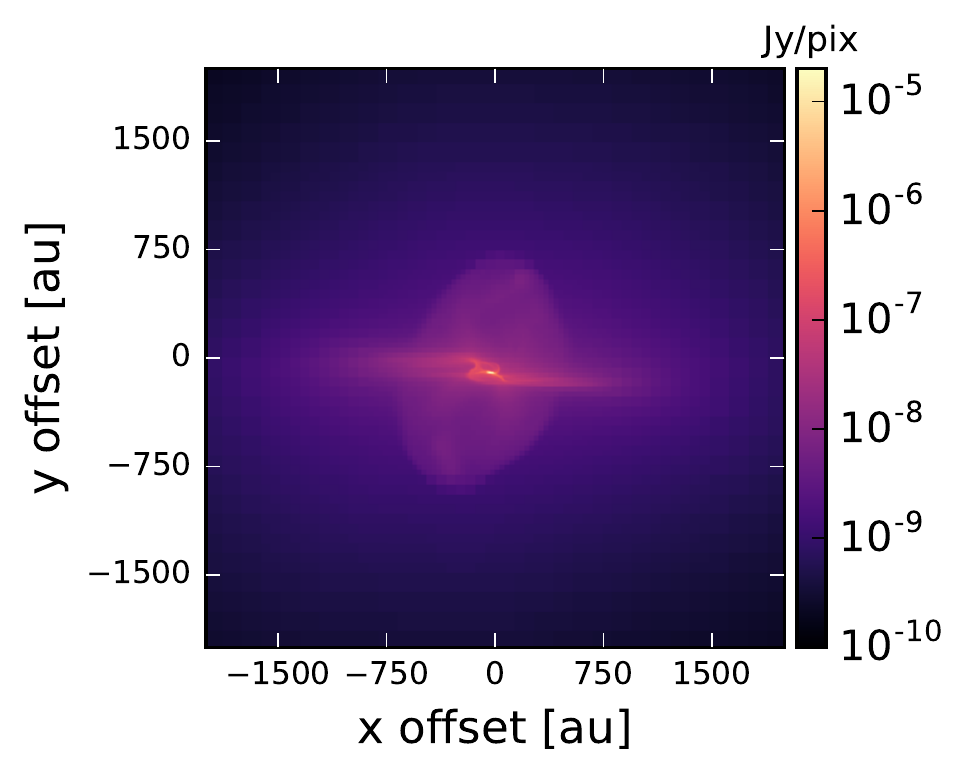} \\
\vspace{-105pt}\hspace{58pt}\color{white}{\bf \Large @ 3.2 mm} \\
\vspace{-15pt}\includegraphics[width=8.5cm]{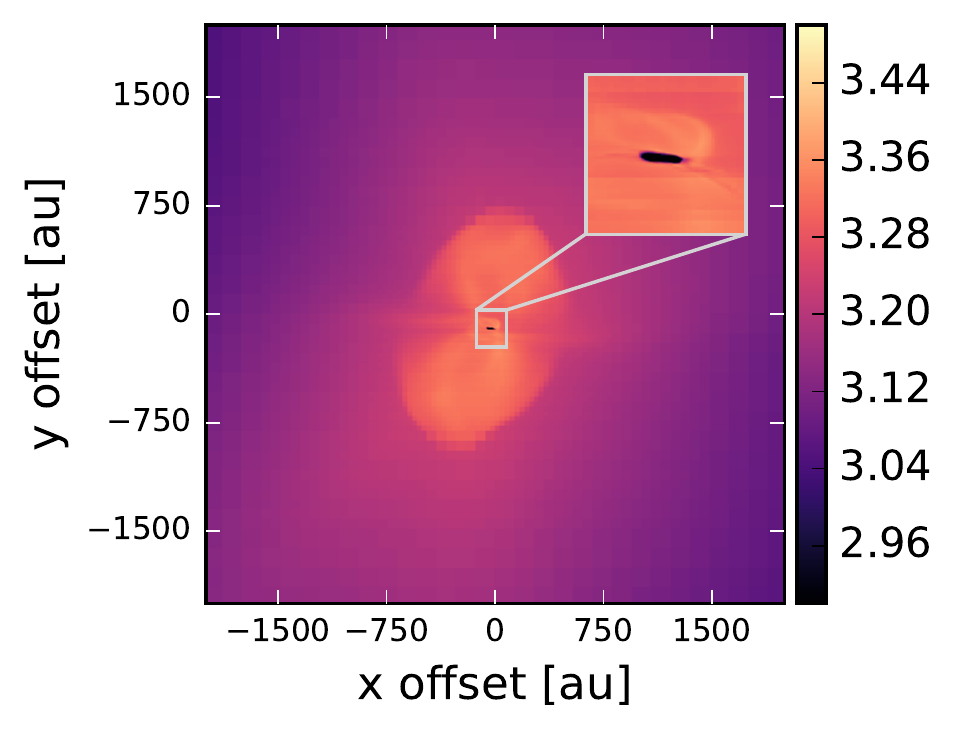} \\
\vspace{-105pt}\hspace{50pt}\color{white}{\bf \huge ~~$\alpha_\mathrm{1-3mm}$} \\
\end{tabular}
 \vspace{-20pt}
\caption{{\it Top two panels:} Synthetic observations at 1.3 mm and 3.2 mm of the core-collapse simulation used in this analysis. The radiative transfer is performed using the POLARIS code. The flux is in Jy/pixel. The pixel size is 1 au. 
{\it Bottom:} Spectral indices $\alpha_\mathrm{1-3mm}$ derived from the simulation. The parameter $\alpha_\mathrm{1-3mm}$ is defined as log (F$_{\nu0}$ / F$_{\nu1}$)~/~log ($\nu0$ / $\nu1$), where $\nu0$ and $\nu1$ are 231 and 94 GHz, respectively.}The inset shows a zoom on the central 200 au region. 
\label{Simulations1}
\end{figure}

\begin{figure}
\begin{tabular}{m{8.5cm}}
\hspace{5pt}\includegraphics[width=8.5cm]{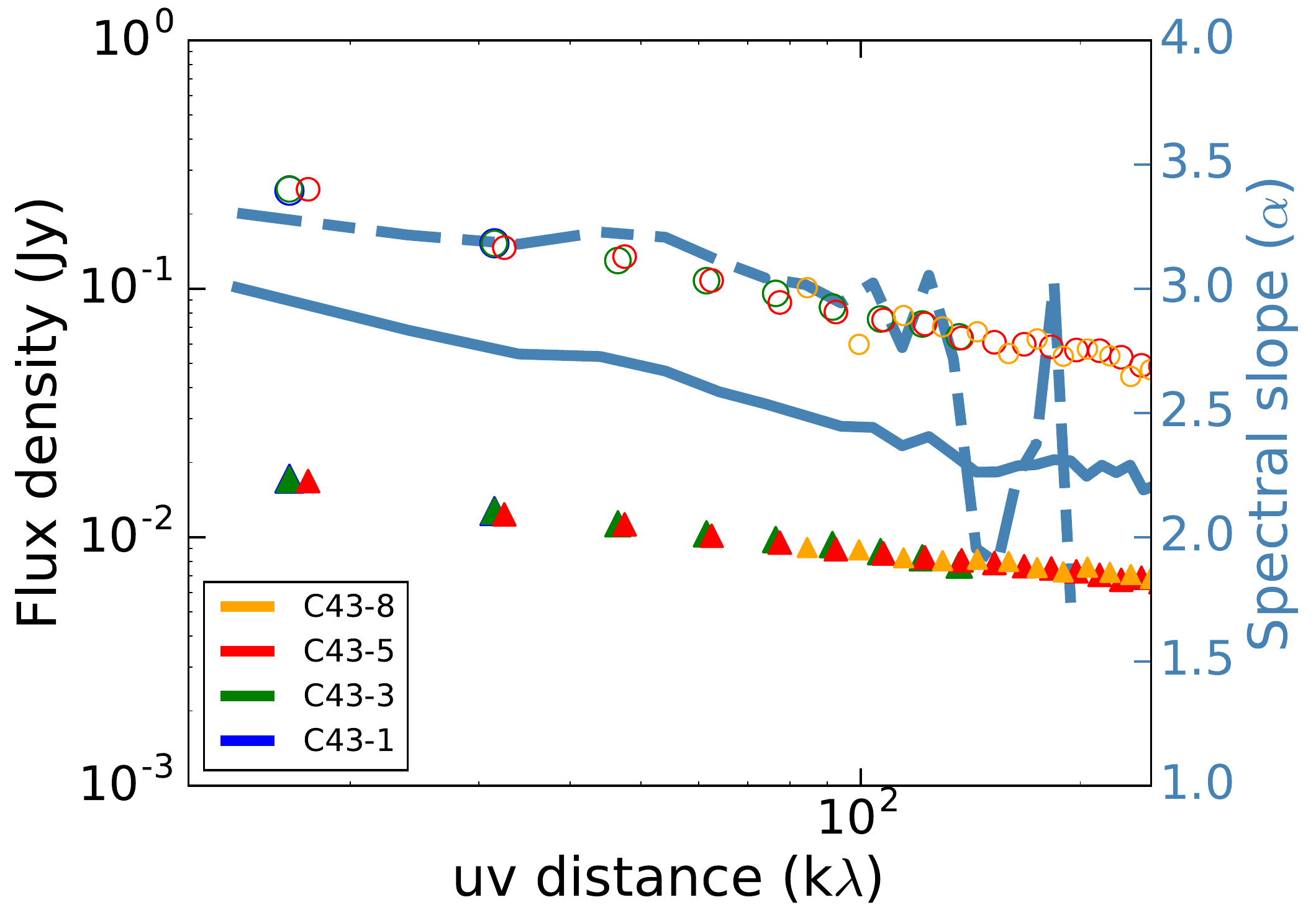} \\
\end{tabular}
\vspace{-5pt}
\caption{Predicted flux density of the dust continuum at 1.3 mm (open circles) and 3.2 mm (filled triangles) as a function of the uv distance averaged every 25 k$\lambda$ for different ALMA antenna configurations. 
The corresponding spectral index $\alpha$ is shown with the blue line. Values are given on the right y-axis. The dashed blue line indicates $\alpha$ values corrected for the central regions (average flux around 200 k$\lambda$ removed).}
\label{Simulations2}
\end{figure}

The observations have shown that the analysis of the spectral index performed in the uv domain can be biased by unresolved optically thick emission or the presence of central compact components. We subtracted the $>$200 k$\lambda$ average amplitude to correct for these contaminations. These average amplitudes are tabulated in Table~\ref{FluxCorrected}. In order to further test the effects of this correction on the 1.3 mm and 3.2 mm visibility profiles and deduced $\beta_\mathrm{1-3mm}$, we generated synthetic observations starting from a nonideal MHD simulation of the gravitational collapse of a pre-stellar core. The simulation is generated using the RAMSES code \citep{Teyssier2002,Fromang2006}. It was carried out by P. Hennebelle in a fashion similar to the non-ideal MHD models of protostar formation shown in \citet{Masson2016}. The dust radiative transfer output from this MHD model that we use here was already presented in \citet{Valdivia2019}. The core has a mass of $2.5~\mathrm{M_\odot}$, a mass-to-flux ratio $\mu=6,$ and a rotational to gravitational energy ratio of $10^{-2}$. The effective resolution is $\sim 1~\mathrm{au}$. More details on this simulation are provided in \citet{Valdivia2019} \citep[see also][]{Maury2018}. We selected a snapshot in which the central embryo has accreted $0.2~\mathrm{M_\odot}$ and we post-processed a zoomed-in region of $8000~\mathrm{au}$ using the radiative transfer code POLArized RadIation Simulator  \citep[POLARIS]{Reissl2016}. We assumed a gas-to-dust mass ratio of 100 and a MRN-like \citep{MRN1977} dust grain size distribution (n(a) $\propto$ a$^{-3.5}$; 5 nm $\le$ a $\le$ 1 \mic). The composition is 62.5\% astronomical silicates and 37.5\% carbonaceous grains: the simulation includes neither dust evolution nor dust dynamics. The chosen accretion luminosity and distance are 1\lsun\ and 120 pc, respectively. Figure~\ref{Simulations1} shows the integrated dust emission of the protostellar envelope at 1.3 mm and 3.2 mm, with a density distribution is close to a simple power law, a pseudo-disk, and bipolar outflow bubbles extending to up to 800 au. We show how the predicted spectral index (Eq.~\ref{equalpha}) varies locally in the synthetic envelope in the bottom panel, where $\alpha$ $\sim$ 3 in the outer part of the envelopes, increasing up to 3.3 in the outflow bubble, then decreasing drastically below 2 in the central 50 au where an optically thick disk component resides.

The 1.3 mm and 3.2 mm emission maps are used as inputs for the ALMA simulator of the \texttt{CASA}\footnote{https://casa.nrao.edu/} software. We ran the {\it simobserve} task to produce four simulated measurement sets for compact and extended ALMA antenna configurations, C43-1, C43-3, C43-5, and C43-8\footnote{The configuration files are retrieved from https://almascience.nrao.edu/tools/casa-simulator.}. 
We used integration times of 7200s per configuration
and assumed a PWV of 0.7 mm (i.e., under good atmospheric conditions). Thermal noise (default ATM model) was added to the simulated data. We then extracted the concatenated visibilities using \texttt{GILDAS/MAPPING}. Figure~\ref{Simulations2} shows the predicted observations in the uv domain at 1.3 mm and 3.2 mm and their associated spectral indices $\alpha$. The $\alpha$ values plunging at small scales can be explained by the central ($\sim$50 au) region that, with its weak ($\alpha<3$) values, affects the whole visibility profile. To correct for this contribution, we masked the inner region: we calculated the fluxes at 1.3 m and 3.2 mm at 200 k$\lambda$ (to be sure to encompass the 50 au central region) and removed them from the visibility amplitudes at all shortest baselines. The corrected trend for $\alpha$ is overlaid in Fig.~\ref{Simulations2} (dashed blue line). The $\alpha\geq$ 3 values observed in the $\alpha$ map (Fig.~\ref{Simulations1} bottom) are recovered for most of the envelope; there is a nearly constant $\alpha$ after the correction. We also reran the tests using a simulation in which the temperature is fixed to 10 K during the post-processing, but this does not affect our results: we conclude that temperature gradients along the line of sight should only have a minor effect on our results.

\section{Maps of $\beta_\mathrm{1-3mm}$} \label{BetaMaps}

\begin{table*}
\centering
\caption{Characteristics of the dust continuum maps shown in Fig.\ref{Maps}}.
\begin{tabular}{cccccccc}
\hline
\hline
\vspace{-5pt}
&\\
Name            &       \multicolumn{2}{c}{Synthesized beams} 
& &     \multicolumn{2}{c}{rms (Jy/beam)} \\
\vspace{-5pt}
&\\
\cline{2-3} 
\cline{5-7} 
\vspace{-5pt}
&\\
& 94 GHz        & 231 GHz-tapered       &&  94 GHz      & 231 GHz-tapered       \\         
\vspace{-5pt}
&\\
\hline
\vspace{-5pt}
&\\
IRAS2A1
&1\farcs45 $\times$ 1\farcs02   (38$^{\circ}$)  
&1\farcs40 $\times$ 1\farcs06   (35$^{\circ}$)          
&& 1.71 $\times$ 10$^{-4}$      &  4.04 $\times$ 10$^{-3}$\\
IRAS4A1
&1\farcs58 $\times$ 1\farcs02   (-155$^{\circ}$)        
&1\farcs59 $\times$ 0\farcs93   (26$^{\circ}$)          
&& 1.03 $\times$ 10$^{-3}$      &  1.26 $\times$ 10$^{-2}$\\
IRAS4B
&1\farcs57 $\times$ 1\farcs01   (25$^{\circ}$)  
&1\farcs53 $\times$ 0\farcs96   (25$^{\circ}$)          
&& 1.33 $\times$ 10$^{-3}$      &  9.31 $\times$ 10$^{-3}$\\
L1448-2A
&1\farcs47 $\times$ 0\farcs96   (43$^{\circ}$)  
&1\farcs43 $\times$ 0\farcs99   (44$^{\circ}$)          
&& 6.27 $\times$ 10$^{-5}$      &  8.23 $\times$ 10$^{-4}$\\
L1448-C
&1\farcs52 $\times$ 0\farcs99   (42$^{\circ}$)  
&1\farcs52 $\times$ 0\farcs99   (45$^{\circ}$)          
&& 9.79 $\times$ 10$^{-5}$      &  1.82 $\times$ 10$^{-3}$\\
L1448-NB1
&1\farcs49 $\times$ 1\farcs02   (35$^{\circ}$)  
&1\farcs46 $\times$ 1\farcs00   (37$^{\circ}$)          
&& 2.60 $\times$ 10$^{-4}$      &  6.37 $\times$ 10$^{-3}$\\
SVS13B
&1\farcs58 $\times$ 1\farcs00   (24$^{\circ}$)  
&1\farcs58 $\times$ 0\farcs98   (25$^{\circ}$)          
&& 2.13 $\times$ 10$^{-4}$      &  6.61 $\times$ 10$^{-3}$\\
L1527
&1\farcs56 $\times$ 1\farcs05   (35$^{\circ}$)  
&1\farcs54 $\times$ 1\farcs06   (36$^{\circ}$)          
&& 7.52 $\times$ 10$^{-5}$      &  1.89 $\times$ 10$^{-3}$\\
SerpM-S68N
&1\farcs69 $\times$ 0\farcs89   (29$^{\circ}$)
&1\farcs69 $\times$ 0\farcs91   (30$^{\circ}$)
&& 8.28 $\times$ 10$^{-5}$      &   1.42 $\times$ 10$^{-3}$ \\
SerpM-SMM4
&1\farcs75 $\times$ 0\farcs88   (28$^{\circ}$)  
&1\farcs79 $\times$ 0\farcs98   (29$^{\circ}$)          
&& 2.02 $\times$ 10$^{-4}$      &  5.16 $\times$ 10$^{-3}$\\
SerpS-MM18
&1\farcs71 $\times$ 0\farcs84   (27$^{\circ}$)  
&1\farcs71 $\times$ 0\farcs92   (24$^{\circ}$)          
&& 2.18 $\times$ 10$^{-4}$      &  5.43 $\times$ 10$^{-3}$\\
L1157
&1\farcs41 $\times$ 1\farcs09   (62$^{\circ}$)  
&1\farcs47 $\times$ 1\farcs01   (61$^{\circ}$)          
&& 9.23 $\times$ 10$^{-5}$      &  1.80 $\times$ 10$^{-3}$\\
\vspace{-5pt}
&&\\
\hline
\end{tabular}
\label{BeamSizes}
\end{table*}

\begin{figure*}
\begin{tabular}{m{5.6cm}m{5.6cm}m{5.6cm}}
 \includegraphics[width=5.8cm]{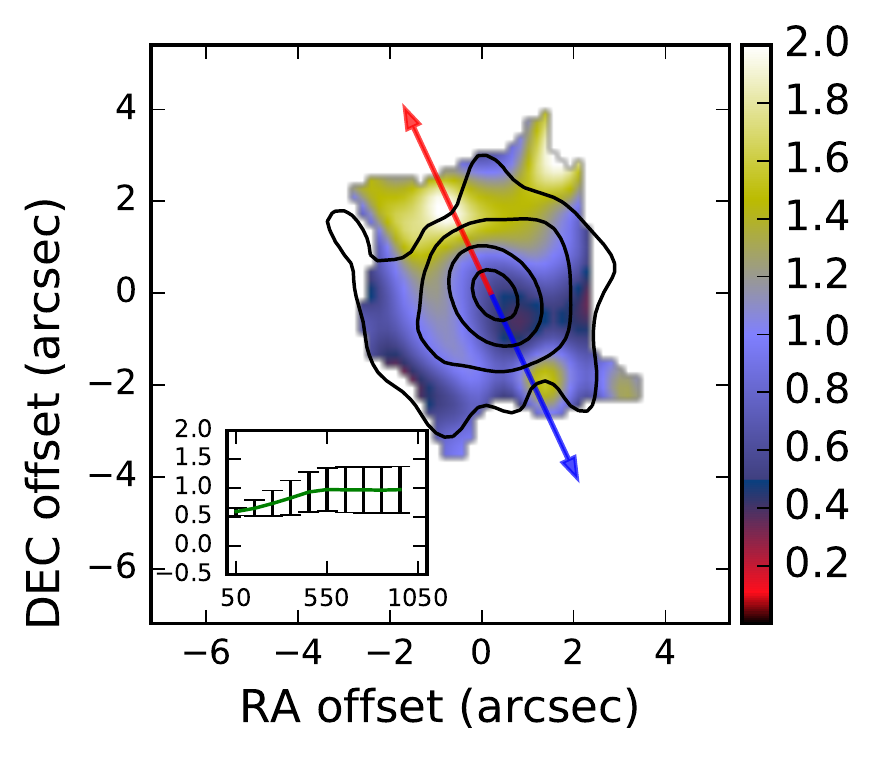}      &
 \includegraphics[width=5.8cm]{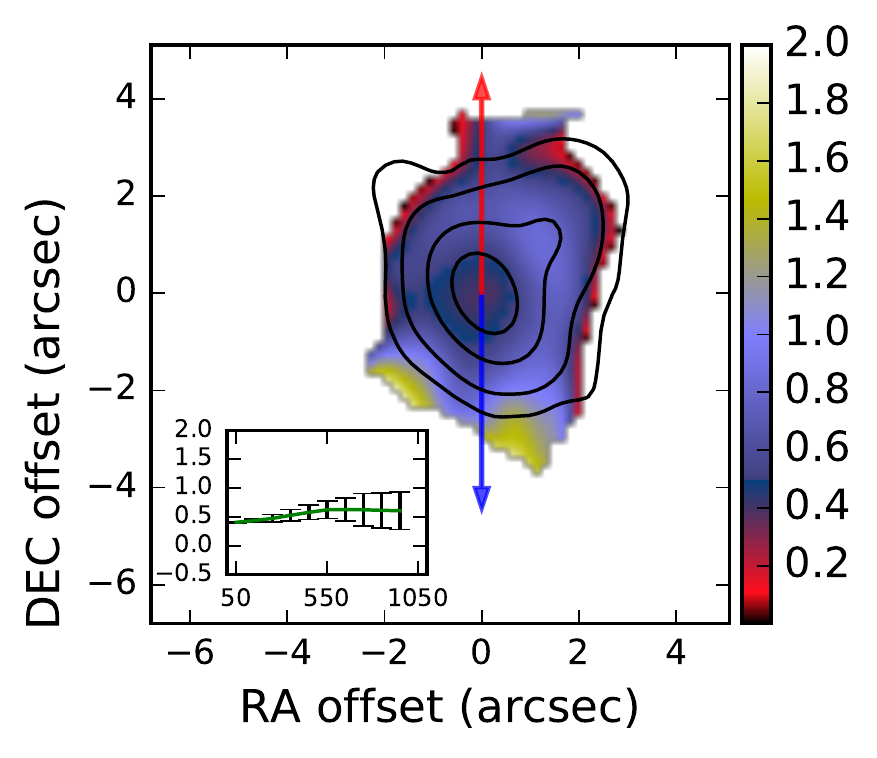}      &
 \includegraphics[width=5.8cm]{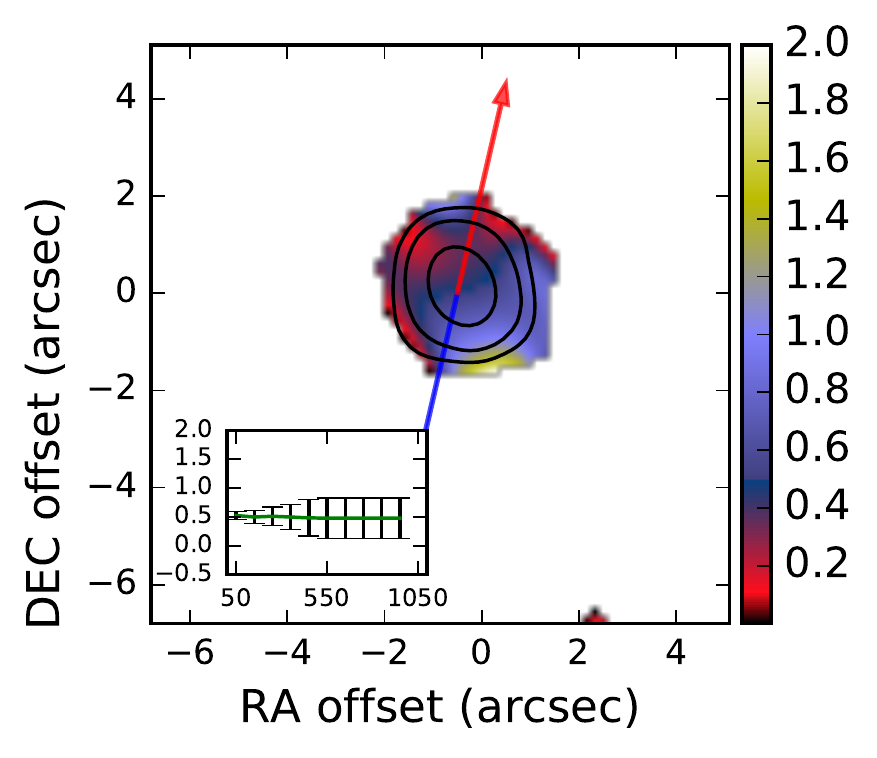}      \\
\vspace{-260pt}\hspace{34pt}{\bf IRAS2A1} & 
\vspace{-260pt}\hspace{34pt}{\bf IRAS4A1} & 
\vspace{-260pt}\hspace{34pt}{\bf IRAS4B} \\
\vspace{-15pt}\includegraphics[width=5.8cm]{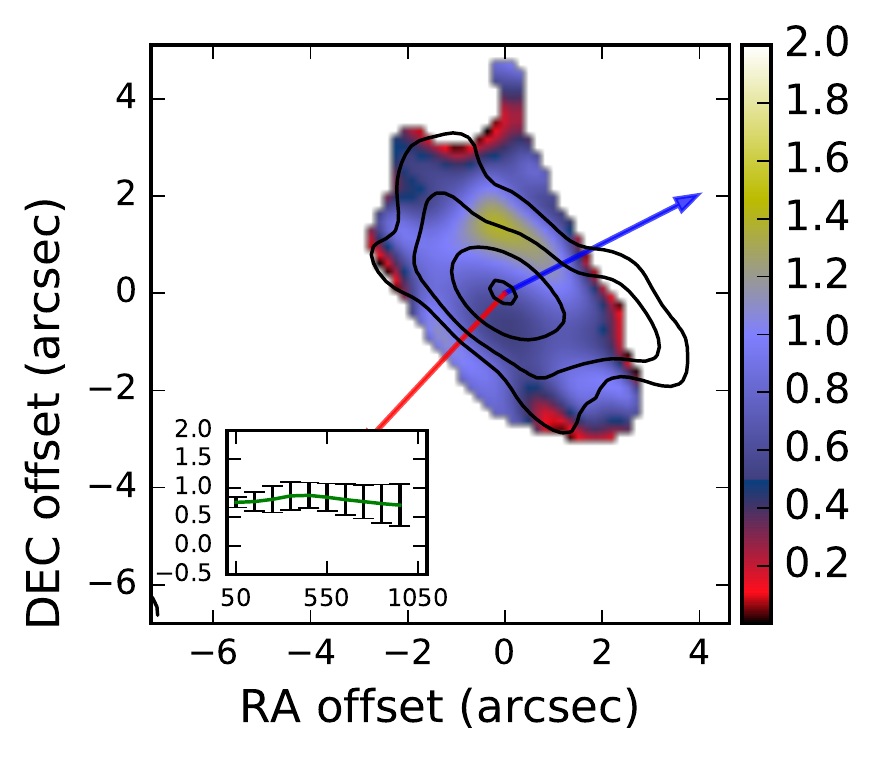} &
\vspace{-15pt}\includegraphics[width=5.8cm]{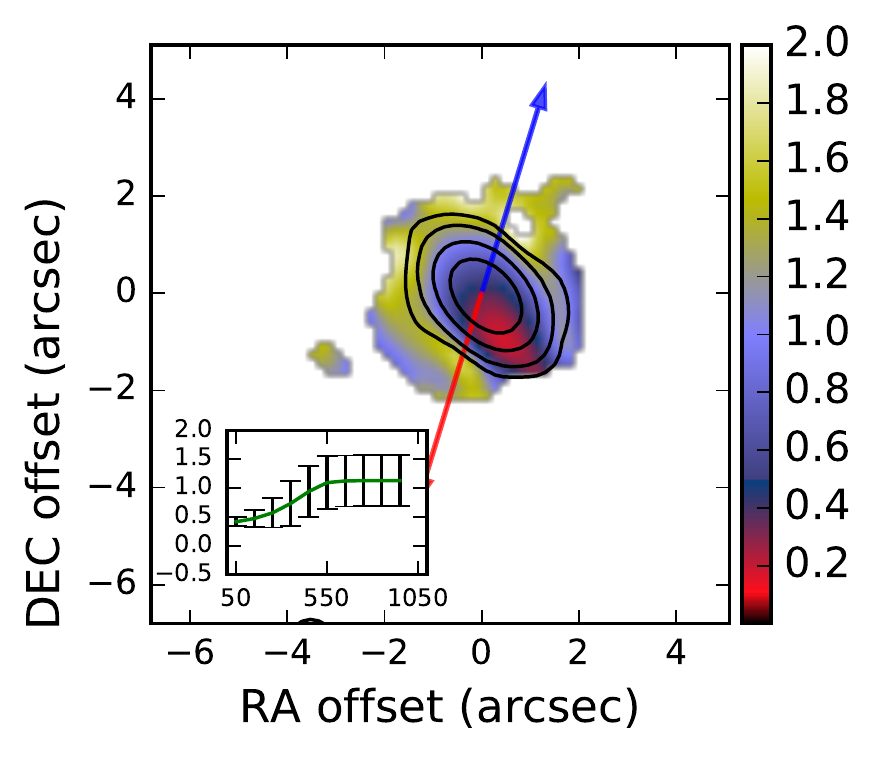} &
\vspace{-15pt}\includegraphics[width=5.8cm]{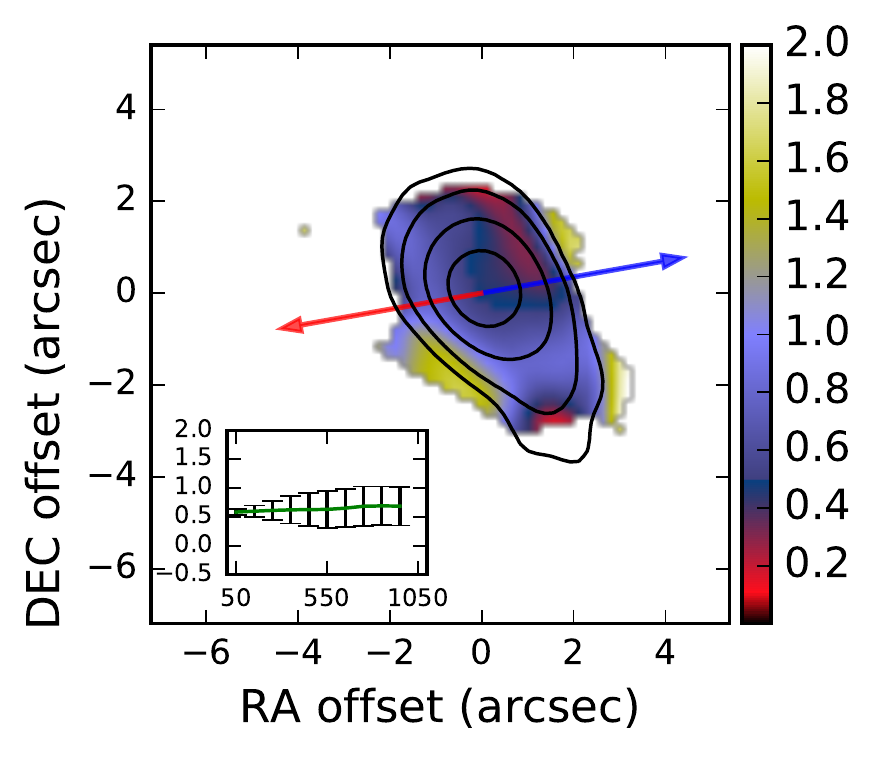}         \\
 \vspace{-260pt}\hspace{34pt}{\bf L1448-2A} & 
\vspace{-260pt}\hspace{34pt}{\bf L1448-C} & 
\vspace{-260pt}\hspace{34pt}{\bf L1448-NB1} \\
 \vspace{-15pt}\includegraphics[width=5.8cm]{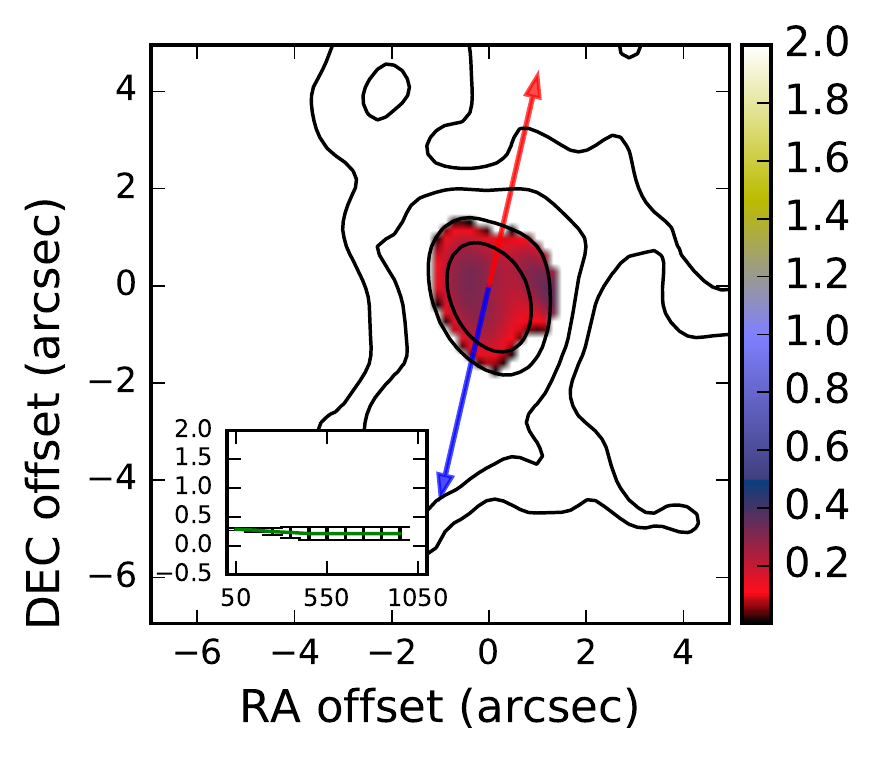}  &
 \vspace{-15pt}\includegraphics[width=5.8cm]{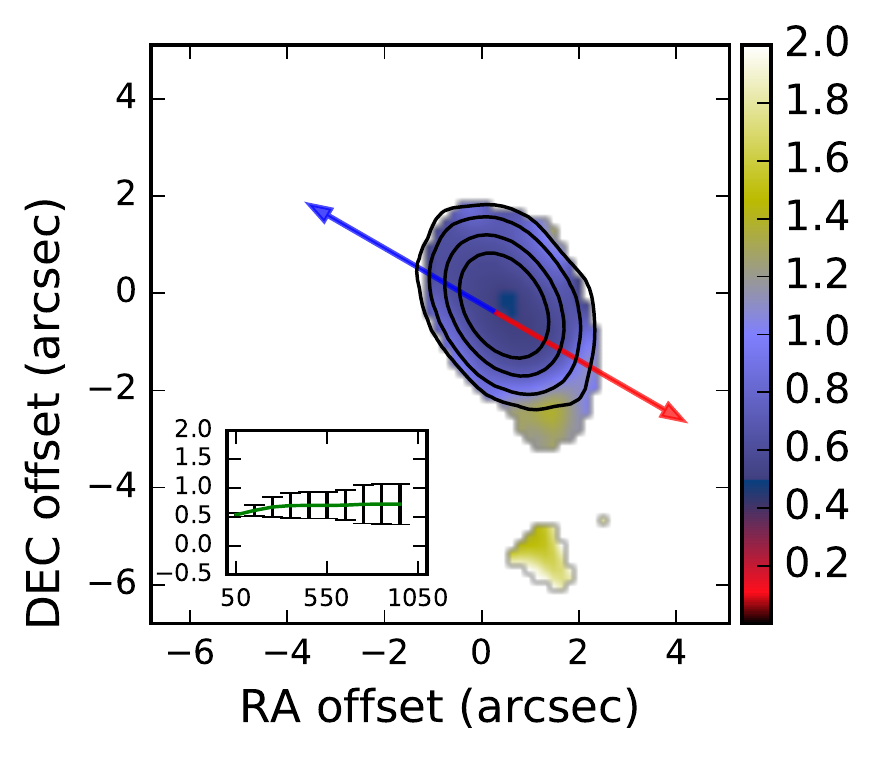}         &  
 \vspace{-15pt}\includegraphics[width=5.8cm]{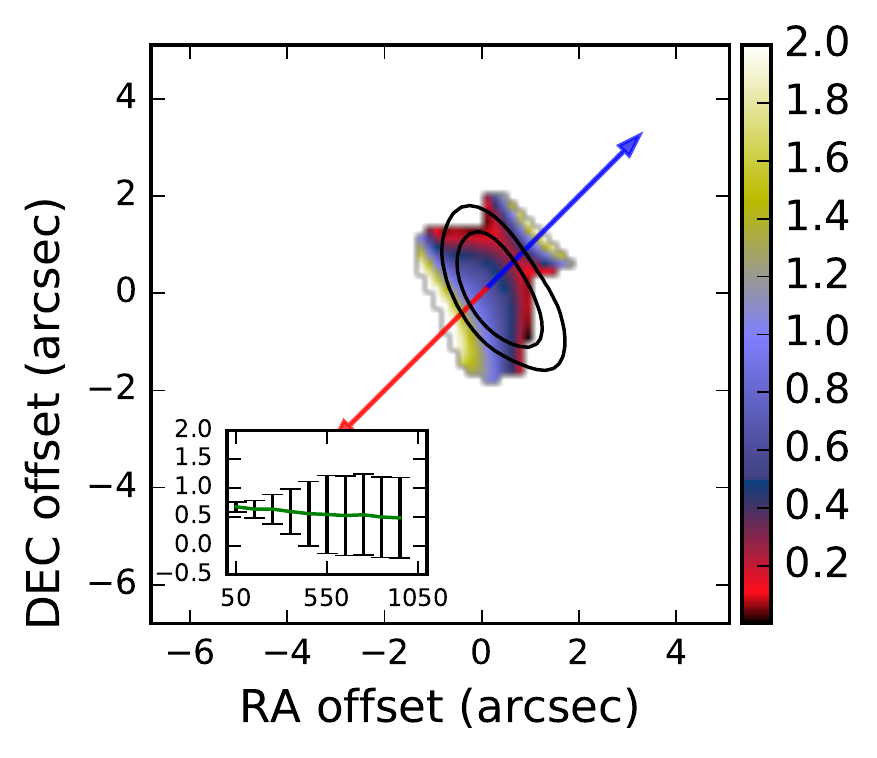}    \\
\vspace{-260pt}\hspace{34pt}{\bf SVS13B} & 
\vspace{-260pt}\hspace{34pt}{\bf L1527} & 
\vspace{-260pt}\hspace{34pt}{\bf SerpM-S68N} \\
\vspace{-15pt} \includegraphics[width=5.8cm]{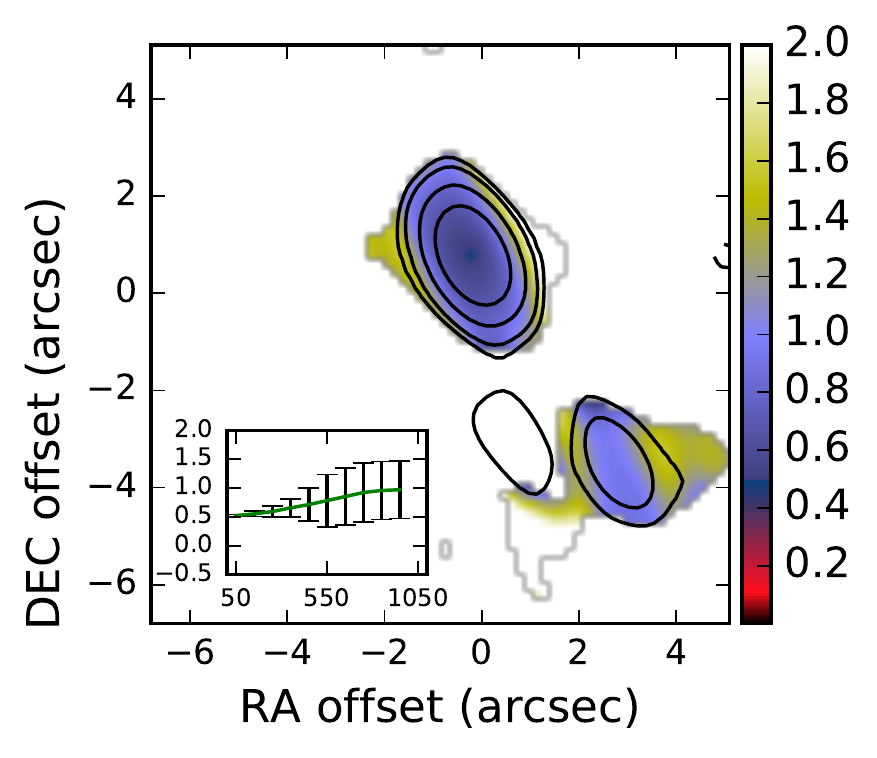}  &
\vspace{-15pt} \includegraphics[width=5.8cm]{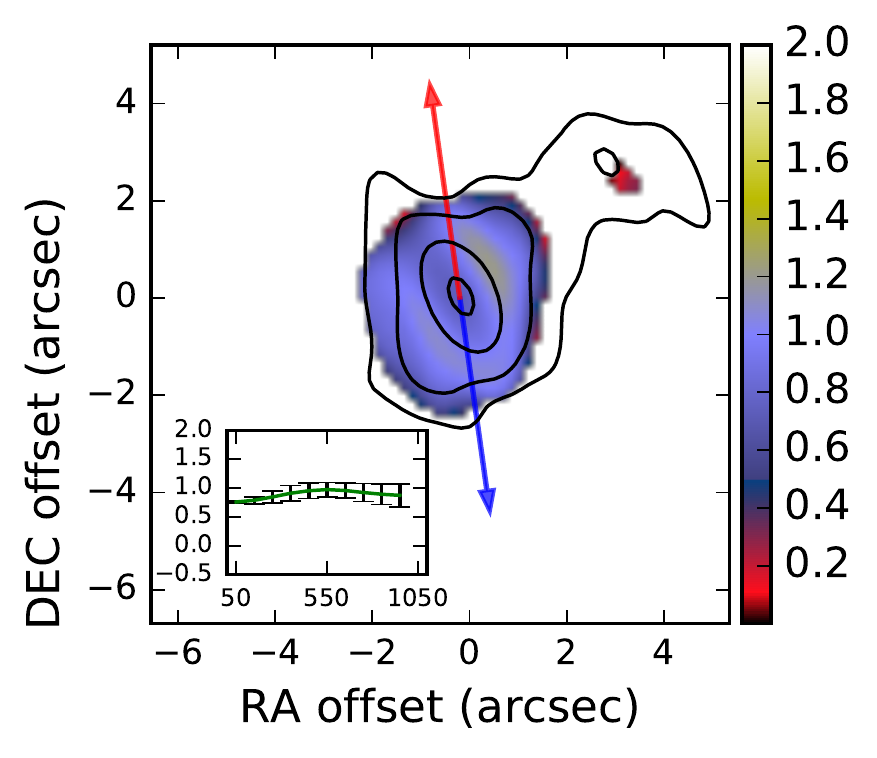}        &
\vspace{-15pt} \includegraphics[width=5.8cm]{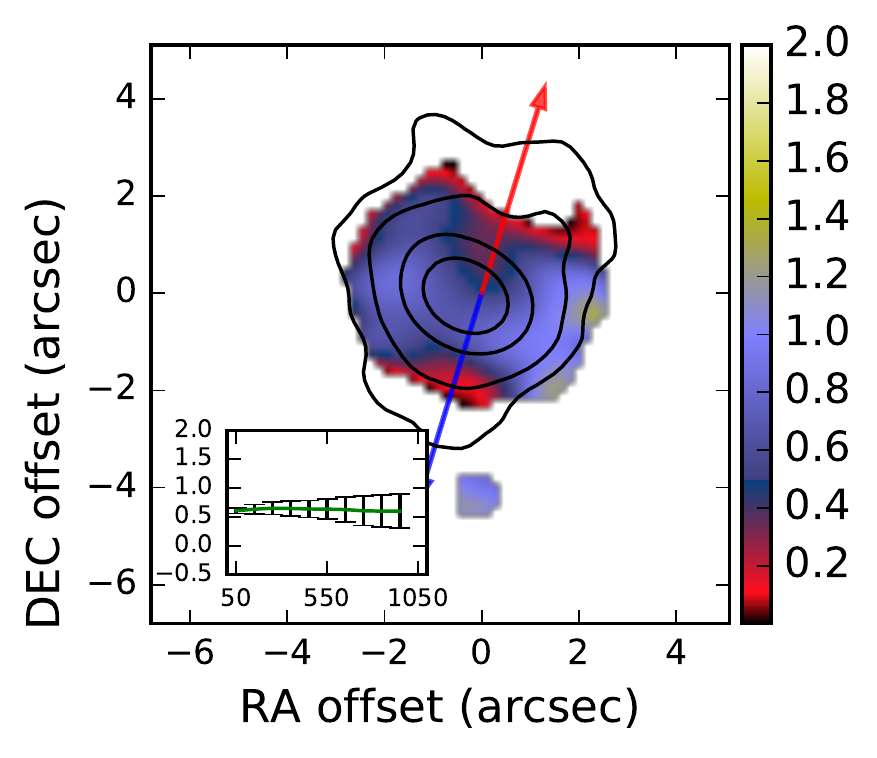}   \\
\vspace{-260pt}\hspace{34pt}{\bf SerpM-SMM4} & 
\vspace{-260pt}\hspace{34pt}{\bf SerpS-MM18} & 
\vspace{-260pt}\hspace{34pt}{\bf L1157} \\
\end{tabular}
\vspace{-10pt}
\caption{Maps of the dust emissivity index $\beta_\mathrm{1-3mm}$. Only pixels with a 2$\sigma$ detection at both wavelengths are shown. The contours indicate the 94 GHz emission detected at 3, 5, 10, 30, and 80$\sigma$. The blue and red arrows indicate the direction of the bipolar jets. The insets show how $\beta_\mathrm{1-3mm}$ evolves as values get integrated to larger and larger scales. The x-axis of the inset plots is labeled in astronomical units.  } 
\label{Maps}
\end{figure*}

For a qualitative assessment of the spatial variations of the dust emissivity index $\beta_\mathrm{1-3mm}$, we performed an imaging of the dust continuum emission (including the secondary sources) using \texttt{GILDAS/MAPPING}. We applied a robust weighting scheme, compromise between the natural weighting and the uniform weighting that minimizes the synthesized beam sides lobes in particular. Table~\ref{BeamSizes} provides the synthesized beam and the rms of the obtained 94 and 231 GHz maps. In order to build maps of $\beta_\mathrm{1-3mm}$, we need to have both 94 and 231 GHz maps tracing the same physical scales at the same resolution. We thus apply an additional uv tapering on the visibilities at 231 GHz  to obtain beams at 231 GHz similar (in size and orientation) to those at 94 GHz. Maps are produced with the same pixel size. The `tapered' sizes and rms of these new maps are also given in Table~\ref{BeamSizes} (231 GHz-tapered columns). We finally used the temperature profiles estimated for each source to apply Eq.~\ref{equbeta} on a pixel-by-pixel basis. Figure~\ref{Maps} shows the $\beta_\mathrm{1-3mm}$ maps of the sample. $\beta_\mathrm{1-3mm}$ is shown at locations where the emission at 1.3 mm and 3.2 mm is detected with a signal-to-noise ratio higher than two. We observe that the dust emissivity index changes in the envelopes of most of the sources as observed in the uv domain. However, when $\beta_\mathrm{1-3mm}$ values are averaged over various aperture sizes, we observe that $\beta_\mathrm{1-3mm}$ gradients are not as clearly observed as we probe smaller and smaller scales compared to the results obtained from the uv domain analysis. This demonstrates the difficulty of analyzing $\beta_\mathrm{1-3mm}$ spatial variations directly from interferometric maps affected by filtering.

\end{document}